\pdfoutput = 1

\documentclass[12pt,a4paper]{article}

\usepackage{amssymb,epsf,epsfig,amsmath,color}
\usepackage{multirow}
\usepackage{accents}
\usepackage{braket}
\usepackage{float}
\usepackage{multicol,lipsum}
\usepackage{simplewick}
\usepackage[noadjust]{cite}
\usepackage[toc,page]{appendix}

\newcommand{\lsim}{
\mathrel{\hbox{\rlap{\hbox{\lower4pt\hbox{$\sim$}}}\hbox{$<$}}}}
\newcommand{\gsim}{
\mathrel{\hbox{\rlap{\hbox{\lower4pt\hbox{$\sim$}}}\hbox{$>$}}}}

\begin{document}
\begin{titlepage}
\vspace*{-0.7truecm}
\begin{flushright}

SI-HEP-2021-31

P3H-21-092
\end{flushright}

\vspace{1.6truecm}

\begin{center}
\boldmath
{\Large{\bf Estimating QCD-factorization amplitudes  \\[0.4em] through $SU(3)$ symmetry in $B\rightarrow P P$ decays}
}
\unboldmath
\end{center}

\vspace{0.8truecm}

\begin{center}
{\bf Tobias Huber and 
Gilberto~Tetlalmatzi-Xolocotzi}

\vspace{0.5truecm}

{\sl Theoretische Physik 1, Naturwissenschaftlich-Technische Fakult\"{a}t,\\
Universit\"{a}t Siegen,Walter-Flex-Strasse 3, D-57068 Siegen, Germany.}

\end{center}

\vspace*{1.7cm}

\begin{abstract}
\noindent 
We estimate the potential size of the weak annihilation amplitudes in QCD factorization as allowed by experimental data. To achieve this goal, we establish a connection between the amplitudes in the QCD factorization and the so-called topological and $SU(3)$-invariant descriptions. Our approach is based purely on the analysis of the tensor structure of the decay amplitudes. By focusing on the processes $B\rightarrow P P$ and considering data from CP asymmetries and branching fractions, we perform a global fit to the $SU(3)$-irreducible quantities assuming minimal theoretical bias. Subsequently we translate the outcome to the QCD factorization decomposition, and find that the most constrained weak annihilation amplitudes are below $0.04$. However, in view of the large uncertainties in several of the experimental input parameters, values up to $\sim0.3$ are allowed in certain cases.
\end{abstract}
\end{titlepage}


\section{Introduction}

Charmless non-leptonic $B_{(s)}$-meson decays play a prominent role in testing the CKM mechanism of quark flavour mixing, in determining the angles of the unitarity triangle and -- closely related -- in quantifying the amount of CP violation in the quark flavour sector of the Standard Model (SM), which represents the only established source of CP violation to date. Moreover, these decays offer numerous observables such as branching ratios, direct CP asymmetries, and polarization fractions, which entail a rich and interesting phenomenology, in particular if precision can be achieved both on the experimental and the theoretical sides.

On the experimental side numerous non-leptonic decay channels have been scrutinized at the B-factories~\cite{BaBar:2014omp} and at hadron collider experiments in the past years. Ongoing and future programs at the upgraded LHC and at Belle II will further increase the experimental precision of non-leptonic $B$ decays~\cite{LHCb:2018roe,Belle-II:2018jsg} which will be particularly important since quite a number of decay channels and observables -- especially direct CP asymmetries -- are not yet measured or are only poorly constrained.

On the theoretical side the phenomenology of charmless non-leptonic $B_{(s)}$-decays is governed by the interplay between CKM factors (weak phases), Wilson coefficients, and non-perturbative hadronic matrix elements, where the computation of the latter has been the bottleneck to precision predictions for quite some time due to the appearance of QCD effects from many different scales that arise in the purely hadronic initial and final states, and due to the fact that these matrix elements cannot be directly computed with non-perturbative methods such as lattice QCD at present (however, see~\cite{RBC:2015gro} for an interesting study on kaon decays). The hadronic matrix element is oftentimes split up into the so-called topological amplitudes such as the colour-allowed, the colour-suppressed tree, annihilation, the QCD and electroweak penguin amplitudes,  etc.~\cite{Chau:1986jb,Chau:1987tk,Chau:1990ay,Gronau:1994rj,Gronau:1995hm,Cheng:2014rfa}. 

There are various strategies that have been developed to get a handle on the hadronic matrix elements, each having its virtues and drawbacks. These approaches are mainly based on flavour symmetries of the light quarks (see e.g.~\cite{Zeppenfeld:1980ex} for an early work) or on factorization, like PQCD~\cite{Keum:2000ph,Lu:2000em} or QCD factorization (QCDF)~\cite{Beneke:1999br,Beneke:2000ry,Beneke:2001ev}. The latter provides a rigorous and systematic framework to disentangle short- from long-distance physics in the heavy quark limit, and at leading power in $\Lambda_{\rm QCD}/m_b$ has arrived at a highly sophisticated level~\cite{Beneke:2003zv,Beneke:2006hg,Beneke:2005vv,Kivel:2006xc,Pilipp:2007mg,Beneke:2006mk,Bell:2007tv,Bell:2009nk,Bell:2009fm,Beneke:2009ek,Kim:2011jm,Bell:2014zya,Bell:2015koa,Bell:2020qus,Beneke:2020vnb,Beneke:2021jhp}. One of the shortcomings of the QCDF approach in its present use is the failure to compute sub-leading power corrections in the heavy-quark expansion from first field-theoretical principles (see~\cite{Khodjamirian:2005wn} for a determination of annihilation topologies for 
$B\rightarrow \pi \pi$ using QCD-sum
rules). Unfortunately, this results in sizeable uncertainties that in many observables spoil the precision achieved at leading power (see e.g.~\cite{Bell:2015koa}). However, it is fair to say that combinations of observables can be built that are robust against hadronic uncertainties such as the combination $\Delta A_{\rm CP}$ in the $K\pi$ channel~\cite{Gronau:2005kz}, which still persists as the so-called $K\pi$ puzzle (see e.g.~\cite{Buras:2003yc,Chiang:2004nm,Mishima:2004um,Cheng:2004ru,Fleischer:2007mq,Bell:2015koa}).

Flavour symmetries based on the approximate $SU(3)$ invariance in the $(u,d,s)$ flavour space or one of its SU(2) subgroups isospin, U-spin and V-spin have been used extensively in the Literature and have been applied both to non-leptonic bottom~\cite{Savage:1989ub,Gronau:1990ka,Deshpande:1994ii,Gronau:1995hm,He:1998rq,He:2000ys,Fu:2003fy,Chiang:2006ih,Chiang:2008zb,Chiang:2008vc,Grossman:2013lya,Cheng:2014rfa,Hsiao:2015iiu} and charm decays~\cite{Rosner:1999xd,Lipkin:2000sf,Gronau:2000ru,Bhattacharya:2008ss,Hiller:2012xm,Feldmann:2012js,Muller:2015lua,Cheng:2016ejf}. The advantages of this approach are the fact that hardly any assumption about the scales of QCD effects are needed, and that it relates different decay channels, thereby reducing the number of parameters, for instance in a global analysis. While the isospin subgroup of the full flavour $SU(3)$ can be considered unbroken, the U and V-spin subgroups as well as the full flavour $SU(3)$ are severely broken by the splitting between the down/up and the strange quark masses. Although some progress has been made in quantifying the amount of flavour breaking~\cite{Gronau:1995hm,Jung:2009pb,Cheng:2012xb,Grossman:2012ry,Muller:2015lua}, the failure of its rigorous implementation can still be regarded as the main drawback of this approach.

One obvious idea is to combine the different approaches, with the goal to benefit from the various advantages which each one of them offers, while at the same time minimizing the sensitivities to their individual drawbacks. Aspects of factorization combined with the topological amplitudes  were put forward in the so-called factorization-assisted topological amplitude approach (FAT)~\cite{Li:2012cfa,Qin:2013tje,Zhou:2015jba,Wang:2017hxe,Jiang:2017zwr}, while combinations of factorization and flavour symmetries were studied e.g.\ in~\cite{Gronau:1995hn,Descotes-Genon:2006spp,Cheng:2011qh,Hsiao:2015iiu}. Also numerous other phenomenological studies with aspects from several approaches can be found in the literature~\cite{Fleischer:1993gr,Buras:1995pz,Gronau:1995hm,Fleischer:1997um,Ciuchini:1997hb,Buras:1997cv,Neubert:1998pt,Neubert:1998jq,Gronau:1998fn,Buras:1998ra,Ali:1998eb,Ali:1998gb,Fleischer:1999jv,Ciuchini:2001gv,Grossman:2003qp,Zenczykowski:2004tw,Buras:2004th,Grossman:2005jb,Lipkin:2005pb,Cheng:2005bg,Imbeault:2006ss,Ali:2007ff,Cheng:2009cn,Cheng:2009mu,Imbeault:2011jz,Bobeth:2014rra,AlvarengaNogueira:2016qyt,Fleischer:2016jbf,Fleischer:2017vrb,Fleischer:2018bld,Bediaga:2020qxg,Fleischer:2021cct,Fleischer:2021cwb,Wang:2020gmn}.

Recently the amplitudes of the topological and $SU(3)$-invariant methods were revisited in~\cite{He:2018php}, whose main findings are that the number of amplitudes in these two descriptions is equal and that the relations between them are linear. In the present work, we make use of the relations between the topological, the $SU(3)$-irreducible decomposition and the factorization approach. We start from the $SU(3)$-irreducible amplitude which we first fit to current experimental data on branching ratios and direct CP asymmetries by performing a $\chi^2$-fit with minimal theoretical bias.  We determine our best fit point using random sampling and then evaluate the two-dimensional $1\sigma$ confidence regions for each of the relevant complex amplitudes in the $SU(3)$-irreducible approach. Subsequently we translate the $SU(3)$ amplitudes into those from QCD factorization, with the goal of obtaining the corresponding confidence regions, in particular for the annihilation amplitudes which are currently inaccessible from first QFT principles. Our $\chi^2$-fit is performed under the assumption of exact flavour $SU(3)$. We leave the implementation of symmetry breaking and further theoretical input, e.g.\ from QCDF at leading power, to future work.

This paper is organised as follows. After specifying the effective Hamiltonian in section~\ref{sec:Heff}, we give the decay amplitudes in the topological and $SU(3)$-irreducible frameworks in section~\ref{sec:TopIrr}. Section~\ref{sec:QCDF} deals with the QCD factorization amplitudes, while their connection to the topological ones are given in section~\ref{sec:QCDFTop}. In sections~\ref{sec:prepfit} and~\ref{sec:fit} we give details on the input parameters and statistical analysis, specifically on the $\chi^2$-fit. In section~\ref{sec:results} we present the results, including best-fit points, confidence regions and the value and uncertainty bands for branching fractions and CP asymmetries. We conclude in section~\ref{sec:conclusion}.


\section{Effective Hamiltonian basis}
\label{sec:Heff}
Our starting point is the effective weak Hamiltonian~\cite{Buchalla:1995vs,Chetyrkin:1997gb}. Since our analysis is to a large extent independent of the precise operator basis we adopt the convention of~\cite{Beneke:2003zv}.
\begin{align}
\mathcal{H}^q_{\rm eff}&= \frac{G_F}{\sqrt{2}}\sum_{p=u,c}\lambda^{(q)}_p\Biggl[C_1 Q^{q,p}_1 + C_2 Q^{q,p}_2  +  \sum^{10}_{i=3} C_i Q^q_i \Biggl]+\rm{h.c.} \nonumber\\
&=\frac{G_F}{\sqrt{2}}\Biggl[\lambda^{(q)}_u \bigl[ C_1 (Q^{q,u}_1-Q^{q,c}_1) + C_2 (Q^{q,u}_2-Q^{q,c}_2) \bigl] \nonumber\\
& \hspace*{40pt}-\lambda^{(q)}_t \biggl( C_1 Q^{q,c}_1 + C_2 Q^{q,c}_2 + \sum^{10}_{i=3} C_i Q^q_i \biggl)\Biggl]+\rm{h.c.}, \label{eq:EffH}
\end{align}
where $q\in\{d,s\}$, $\lambda^{(q)}_p=V_{pb} V^*_{pq}$ and we have moreover made use of the unitarity relation
\begin{equation}
\lambda^{(q)}_u+\lambda^{(q)}_c+\lambda^{(q)}_t=0.
\label{eq:CKMrel}
\end{equation}
The four-quark operators are given as follows
\begin{align}
\allowdisplaybreaks
Q_{1}^{q, \, p}&=\bigl(\bar{p}_{\beta}b_{\alpha}\bigl)_{V-A}
                      \bigl(\bar{ q }_{\alpha}   p_{\beta}\bigl)_{V-A}
                      \, , & 
Q_{2}^{q, \, p}=& \bigl(\bar{p} b\bigl)_{V-A} \bigl(\bar{q} p\bigl)_{V-A}\, ,
\nonumber    \\
Q^q_{3}&=\bigl(\bar{q}b\bigl)_{V-A}            \sum_k\bigl(\bar{k}k\bigl)_{V-A}  \, ,
& 
Q^q_{4}=&\bigl(\bar{q}_\alpha b_\beta\bigl)_{V-A}\sum_k\bigl(\bar{k}_\beta k_\alpha\bigl)_{V-A}\, ,
\nonumber\\
Q^q_{5}&=\bigl(\bar{q} b\bigl)_{V-A}           \sum_k\bigl(\bar{k}k\bigl)_{V+A} \, ,
& Q^q_{6}=&\bigl(\bar{q}_\alpha b_\beta\bigl)_{V-A}\sum_k\bigl(\bar{k}_\beta k_\alpha\bigl)_{V+A}\, ,
\nonumber\\
Q^q_{7}&=\bigl(\bar{q} b\bigl)_{V-A}\sum_k\frac{3}{2}e_k\bigl(\bar{k} k\bigl)_{V+A} \, ,
& Q^q_{8}=&\bigl(\bar{q}_\alpha b_\beta\bigl)_{V-A}\sum_k\frac{3}{2}e_k\bigl(\bar{k}_\beta k_\alpha\bigl)_{V+A}\, ,
\nonumber\\
 Q^q_{9 }&=\bigl(\bar{q}       b      \bigl)_{V-A}\sum_{k}\frac{3}{2}e_k\bigl(\bar{k}       k    \bigl)_{V-A}\, ,
&Q^q_{10}=&\bigl(\bar{q}_\alpha b_\beta \bigl)_{V-A}\sum_{k}\frac{3}{2}e_k\bigl(\bar{k}_\beta k_{\alpha} \bigl)_{V-A}\,.
\label{eq:mainbasis}
\end{align}
Colour indices in singlet currents are suppressed to ease the reading. The sum over $k$ includes the five lightest quark flavours, and $e_k$ is the electric charge of quark $k$ in units of the positron charge. Currently, the Wilson Coefficients $C_i$ have been computed to high orders in different bases~\cite{Buchalla:1995vs,Gorbahn:2004my}.


\section{Topological and $SU(3)$-irreducible representation}
\label{sec:TopIrr}
The physical amplitudes of the processes $B\rightarrow P P$, $P$ being a light pseudoscalar meson, are traditionally written using two representations, the topological and the $SU(3)$-irreducible one. The equivalence between both approaches was recently discussed in \cite{He:2018php}. To make the present paper self-consistent, we summarize the most important relations below.

\subsection{Topological representation}
\label{sec:Top}

\begin{figure}[tp]
\begin{center}
\includegraphics[width=1.0\textwidth]{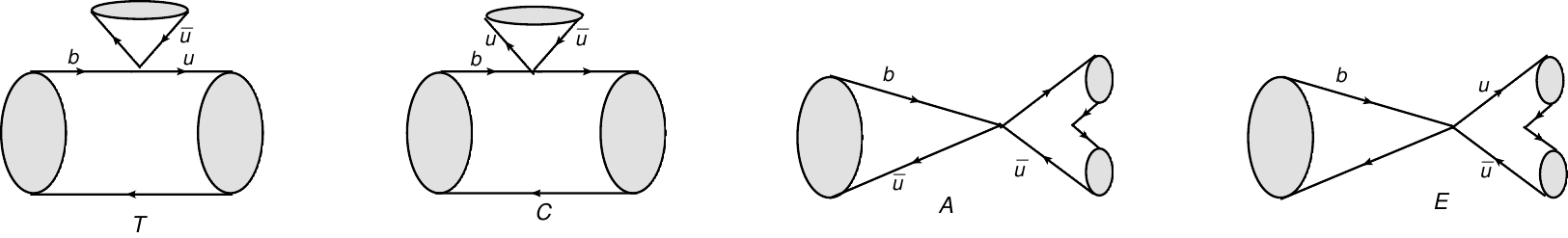}
\includegraphics[width=1.0\textwidth]{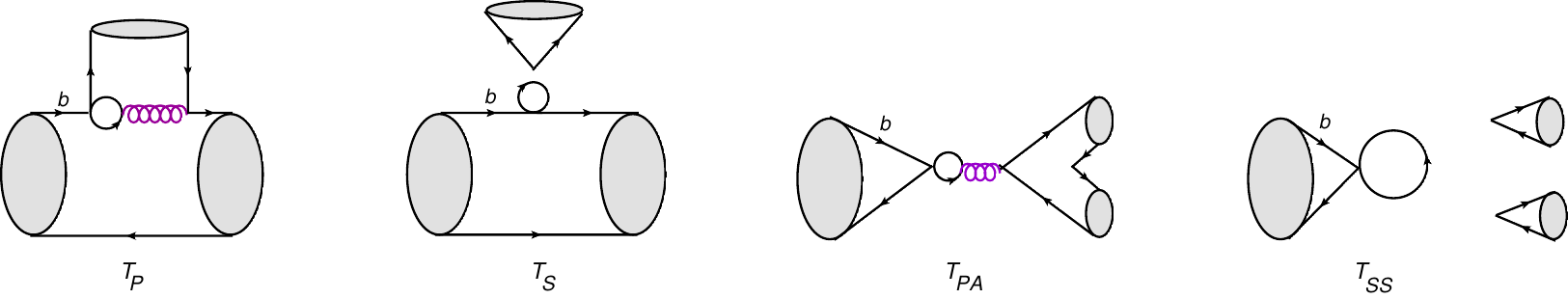}
\includegraphics[width=0.5\textwidth]{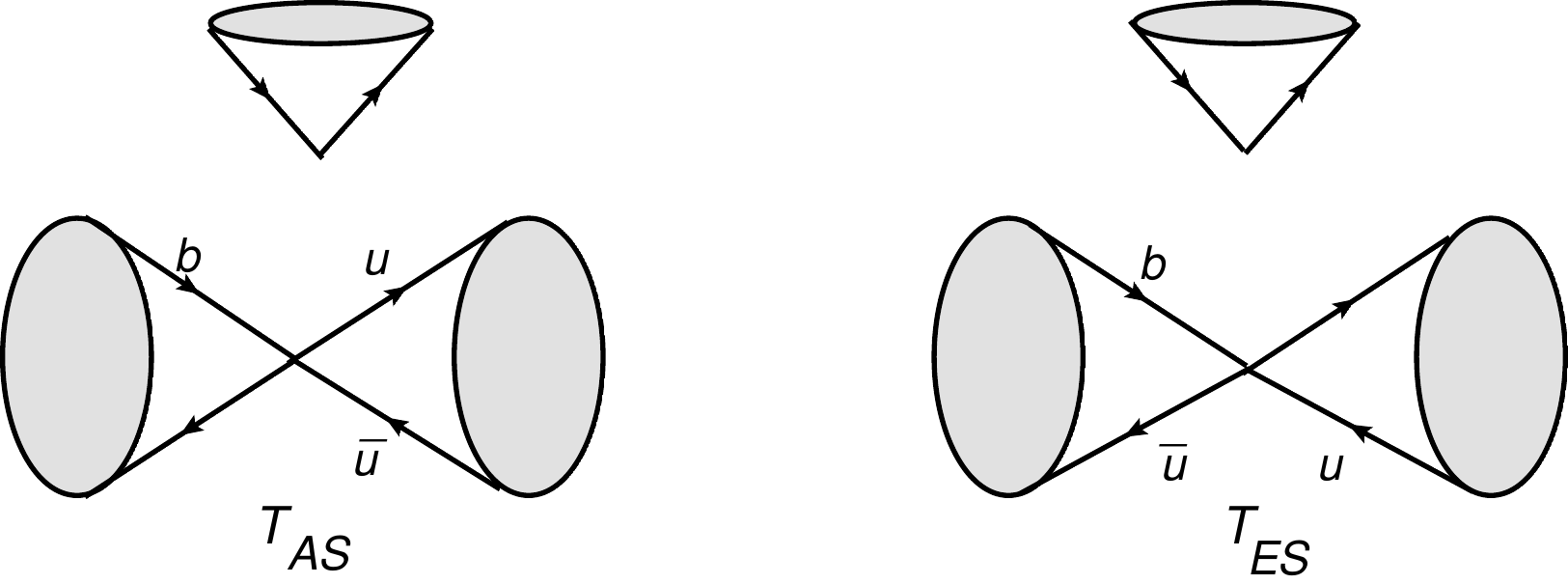}
\caption{Diagrammatic representation of the tree-like topological amplitudes in Eq.~(\ref{eq:new_tree}).}
\label{fig:Prop}
\end{center}
\end{figure}

Usually, the following notation is used to describe the different topological amplitudes
\begin{itemize}
    \item $T$: referring to the color-allowed tree amplitude.
    \item $C$: denoting the color-suppressed tree diagram.
    \item $E$: referring to the $W$-exchange diagram. 
    \item $P$, denoting QCD penguin contributions.
    \item $S$, representing the flavor QCD singlet penguin.
    \item $A$, denoting annihilation diagrams.
\end{itemize}
Using the notation above, the labels for other topologies can be written, for example $T_{AS}$ is used to denote a tree-annihilation singlet topology. Then, the topological decomposition of the physical amplitudes reads
\begin{eqnarray}
{\cal A}^{TDA} = i\frac{G_{F}}{\sqrt{2}}\Bigl[\mathcal{T}^{TDA}  +  \mathcal{P}^{TDA}\Bigl],
\label{eq:TDA}
\end{eqnarray}
with
\begin{eqnarray}
{\cal T}^{TDA} &=&  T~B_i (M)^{i}_j   \bar H^{jl}_k  (M)^k_l   + C~B_i (M)^{i}_j \bar H^{lj}_k  (M)^k_l
+ A~B_i \bar H^{il}_j   (M)^j_k (M)^{k}_l\nonumber \\
&&+ E~B_i  \bar H^{li}_j (M)^j_k (M)^{k}_l
+ T_{ES} B_i  \bar H^{ij}_{l}   (M)^{l}_j    (M)^k_k
+T_{AS} B_i \bar H^{ji}_{l}  (M)^{l}_j  
(M)^k_k\nonumber\\
&&+ T_{S} B_i  (M)^{i}_j  \bar H^{lj}_{l}  (M)^k_k + T_{PA} B_i \bar H^{li}_{l}  (M)^j_k (M)^{k}_j +T_{P} B_i (M)^{i}_j   (M)^j_k \bar H^{lk}_{l}\nonumber\\ 
&&    + T_{SS} B_i \bar H^{li}_{l}  (M)^j_j (M)^{k}_{k},\label{eq:new_tree}
\end{eqnarray}
and
\begin{eqnarray}
{\cal P}^{TDA}&=& P~B_i (M)^{i}_{j} (M)^{j}_k \tilde H^k
+ P_{T}~B_i (M)^{i}_j   \tilde H^{jl}_k  (M)^k_l
+ S~B_i (M)^{i}_{j}  \tilde H^j (M)^{k}_k\nonumber\\
&&+ P_C~B_i (M)^{i}_j \tilde H^{lj}_k  (M)^k_l
+ P_{TA} B_i \tilde H^{il}_j   (M)^j_k (M)^{k}_l
+ P_A~B_i \tilde H^i  (M)^{j}_{k} (M)^{k}_j \nonumber\\
&&  + P_{TE} B_i  \tilde H^{ji}_k   (M)^k_l (M)^{l}_j 
 +P_{AS} B_i \tilde H^{ji}_{l}  (M)^{l}_j  (M)^k_k
+P_{SS} B_i \tilde H^i  (M)^{j}_{j} (M)^{k}_k\nonumber\\
&& +P_{ES} B_i  \tilde H^{ij}_{l}   (M)^{l}_j    (M)^k_k,
\label{eq:new_peng}
\end{eqnarray} 
where all the indices can assume the values $1,2,3$. 
The $B$-meson vector with components $(B_i)$ reads $B=(B^+,B^0_d,B^0_s)$ and the light-meson matrix $(M)^i_j$ has the structure
\begin{equation}
\label{eq:Mmatrix}
\begin{aligned}
   M &= \left( \begin{array}{ccc} 
    \frac{\pi^0}{\sqrt 2} + \frac{\eta_q}{\sqrt 2}
     + \frac{\eta'_q}{\sqrt 2} & \pi^- & K^- \\
    \pi^+ & - \frac{\pi^0}{\sqrt2} + \frac{\eta_q}{\sqrt 2}
     + \frac{\eta'_q}{\sqrt 2} & \bar K^0 \\
    K^+ & K^0 & \eta_s + \eta'_s \end{array} \right).
\end{aligned}
\end{equation}
The states $\eta_{q}$, $\eta'_{q}$, $\eta_{s}$ and $\eta'_{s}$ appearing in the diagonal of the meson matrix in Eq.~(\ref{eq:Mmatrix}) correspond to the flavor states used to describe $\eta$~--~$\eta'$ mixing in the Feldmann–Kroll–Stech scheme, for more details see section~\ref{Sec:etamixing}.
The graphical representation of the topological amplitudes presented in  Eq.~(\ref{eq:new_tree}) is given in Fig.~\ref{fig:Prop}.

Here, we have used the symbols  $\tilde{H}$ and $\bar{H}$ to denote the flavour tensors, which allow to construct flavour singlets when contracted with the meson vector and matrices. Unlike the standard conventions here we are absorbing the different CKM-components inside the flavour tensors, thus we are using the symbol $``-"$ to indicate the presence of the component $\lambda^{(q)}_u$  and the symbol $``\sim"$  to indicate the presence of $\lambda^{(q)}_t$, for $q=d, s$.
The non-zero elements of $\tilde{H}^{ij}_{k}$ and $\bar{H}^{ij}_{k}$ are then
\begin{eqnarray}
\bar{H}^{12}_1=\lambda^{(d)}_{u},&~&
\bar{H}^{13}_1=\lambda^{(s)}_{u},\nonumber\\
\tilde{H}^{12}_1=\lambda^{(d)}_{t},&~&
\tilde{H}^{13}_1=\lambda^{(s)}_{t}.
\end{eqnarray}
Analogously, for the vectors $\bar{H}^i$ and $\tilde{H}^i$ the non-zero components are
\begin{eqnarray}
\bar{H}^{2}=\lambda^{(d)}_{u},&~&
\bar{H}^{3}=\lambda^{(s)}_{u},\nonumber\\
\tilde{H}^{2}=\lambda^{(d)}_{t},&~&
\tilde{H}^{3}=\lambda^{(s)}_{t}.
\end{eqnarray}

\subsection{$SU(3)$-irreducible representation}
\label{sec:Irr}

Alternatively to the topological decomposition in Eq.~(\ref{eq:TDA}), the $SU(3)$-irreducible decomposition of the physical amplitudes reads
\begin{eqnarray}
{\cal A}^{IRA} =i\frac{G_{F}}{\sqrt{2}}\Bigl[ \mathcal{T}^{IRA}  +  \mathcal{P}^{IRA}\Bigl].
\label{eq:IRA}
\end{eqnarray}
The irreducible representation for the ``tree'' contribution to the amplitude for the $B\rightarrow P P$ decay processes is expressed as
\begin{eqnarray}\label{eq:su3basis}
 \mathcal{T}^{IRA} &=&A_3^T B_i (\bar{H}_{\bar 3})^i (M)_k^j(M)_j^k +C_3^T B_i (M)^i_j (M)^j_k (\bar{H}_{\bar 3})^k+B_3^T   B_i (\bar{H}_3)^i (M)_k^k(M)_j^j \nonumber\\
  &&    
  +D_3^T B_i (M)^i_j  (\bar{H}_{\bar 3})^j (M)^k_k+A_6^T B_i (H_{ 6})^{ij}_k (M)_j^l(M)_l^k 
  +C_6^T B_i (M)^i_j (\bar{H}_{ 6})^{jl}_k (M)_l^k  \nonumber\\
  && 
  +B_6^T B_i (\bar{H}_{ 6})^{ij}_k (M)_j^k(M)_l^l+A_{15}^T B_i (\bar{H}_{\overline{15}})^{ij}_k (M)_j^l(M)_l^k 
  +C_{15}^T B_i (M)^i_j (\bar{H}_{\overline{15}})^{jk}_l (M)_k^l\nonumber\\
  &&+B_{15}^T B_i (\bar{H}_{\overline{15}})^{ij}_k (M)_j^k(M)_l^l.
\end{eqnarray}
with 
\begin{eqnarray}
(\bar{H}_3)^2=\lambda^{(d)}_u,\quad
(\bar{H}_6)^{12}_1=-(\bar{H}_6)^{21}_1=(\bar{H}_6)^{23}_3=-(H_6)^{32}_3&=&\lambda^{(d)}_u \, , \nonumber\\
2(\bar{H}_{15})^{12}_1=2(\bar{H}_{15})^{21}_1=-3(\bar{H}_{15})^{22}_2=-6(\bar{H}_{15})^{23}_3=-6(\bar{H}_{15})^{32}_3&=&6\lambda^{(d)}_u \, ,
\end{eqnarray}
and
\begin{eqnarray}
(\bar{H}_3)^3=\lambda^{(s)}_u,\quad
(\bar{H}_6)^{13}_1=-(\bar{H}_6)^{31}_1=(\bar{H}_6)^{32}_3=-(H_6)^{23}_3&=&\lambda^{(s)}_u \, , \nonumber\\
2(\bar{H}_{15})^{13}_1=2(\bar{H}_{15})^{31}_1=-3(\bar{H}_{15})^{33}_3=-6(\bar{H}_{15})^{32}_3=-6(\bar{H}_{15})^{23}_3&=&6\lambda^{(s)}_u \, ,
\end{eqnarray}
where the subindices in the different tensors denote the $SU(3)$-multiplet under consideration. The penguin components of the amplitude $\mathcal{P}^{IRA}$ can be obtained from  $\mathcal{T}^{IRA}$ via the replacements 
\begin{eqnarray}
A^T_{j}\rightarrow A^P_{j},\quad C^T_{j}\rightarrow C^P_{j}, \quad B^T_{j}\rightarrow B^P_{j} \quad D^T_{j}\rightarrow D^P_{j}\label{eq:treepengSU3}
\end{eqnarray}
and
\begin{eqnarray}
\bar{H}_3\rightarrow 
\tilde{H}_3,\quad
\bar{H}_6\rightarrow 
\tilde{H}_{6},\quad
\bar{H}_{15}\rightarrow 
\tilde{H}_{15}\, ,
\end{eqnarray}
with the non-zero components given by
\begin{eqnarray}
(\tilde{H}_3)^2=\lambda^{(d)}_t,\quad
(\tilde{H}_6)^{12}_1=-(\tilde{H}_6)^{21}_1=(\tilde{H}_6)^{23}_3=-(\tilde{H}_6)^{32}_3&=&\lambda^{(d)}_t \, ,\nonumber\\
2(\tilde{H}_{15})^{12}_1=2(\tilde{H}_{15})^{21}_1=-3(\tilde{H}_{15})^{22}_2=-6(\tilde{H}_{15})^{23}_3=-6(\tilde{H}_{15})^{32}_3&=&6\lambda^{(d)}_t \, ,
\end{eqnarray}
and
\begin{eqnarray}
(\tilde{H}_3)^3=\lambda^{(s)}_t,\quad
(\tilde{H}_6)^{13}_1=-(\bar{H}_6)^{31}_1=(\bar{H}_6)^{32}_3=-(H_6)^{23}_3&=&\lambda^{(s)}_t \, , \nonumber\\
2(\tilde{H}_{15})^{13}_1=2(\tilde{H}_{15})^{31}_1=-3(\tilde{H}_{15})^{33}_3=-6(\tilde{H}_{15})^{32}_3=-6(\tilde{H}_{15})^{23}_3&=&6\lambda^{(s)}_t.
\end{eqnarray}

Interestingly, once the physical amplitudes are computed,  the component $A^T_6$ appears always in the combinations $C^T_6-A^T_6$ and $B^T_6 + A^T_6$ and consequently it can always be absorbed according to the following rules \cite{Paz:2002ev},
\begin{eqnarray}
C^T_6-A^T_6\rightarrow C^T_6,\nonumber\\
B^T_6 + A^T_6 \rightarrow B^T_6,
\label{eq:reparameterization}
\end{eqnarray}
and analogously for the case of the corresponding penguin amplitudes,
\begin{eqnarray}
C^P_6-A^P_6\rightarrow C^P_6,\nonumber\\
B^P_6 + A^P_6 \rightarrow B^P_6.
\label{eq:reparameterizationpeng}
\end{eqnarray}
In other words, we redefine $C_6$ and $B_6$ to include $A_6$ for ``trees'' and ``penguins'', respectively. Here, we follow~\cite{He:2018php} in the decomposition of the color allowed and color suppressed electroweak penguin operators $P_{EW}$ and $P^{C}_{EW}$ according to the rule
\begin{eqnarray}
\bar{q}b\sum_{q'} e_{q'}\bar{q}'q'=\bar{q}b\bar{u}u+\bar{q}b\bar{c}c-\frac{1}{3}\bar{q}b\sum_{q'}\bar{q}'q',\label{eq:splitEW}
\end{eqnarray}
where the first two terms and their color suppressed analog lead to the amplitudes $P_T$ and $P_C$, respectively. The remaining part of Eq.~(\ref{eq:splitEW}) is then absorbed into the penguin amplitude $P$. Thus, we have
\begin{eqnarray}
P_{EW}&=&P_T -\frac{1}{3}P',\\
P^{C}_{EW}&=&P_{C}-\frac{1}{3}P^{'C}.
\end{eqnarray}

We proceed by counting the total number of real degrees of freedom. In total we have 20 complex coefficients which appear in the tree and penguin contributions. This translates into a total of 40 real quantities. Once the parameter shifts in Eq.~(\ref{eq:reparameterization}) and  Eq.~(\ref{eq:reparameterizationpeng})  are performed we are left with 36 real quantities. However, an overall phase can always be absorbed and is irrelevant for the modulus of the physical amplitudes. Hence we are left with 35 real coefficients. Once we include the $\eta-\eta'$ mixing angle, the total count increases to 36.

\begin{table}[tp]
  \begin{center}
    \scalebox{0.85}{\begin{tabular}{|c|c|c|c|c|c|c|c|c|c|c|}
      \hline 
      \textbf{Channel} & $A_{3}^T$& $C_{3}^T$&$A_{6}^T$&$C_{6}^T$&$A_{15}^T$&$C_{15}^T$&$B_{3}^T$&$B_{6}^T$&$B_{15}^T$ &$D_{3}^T$ \\
      \hline
      $B^-\rightarrow \pi^0\pi^-$ & $0$ & $0$& $0$&$0$&$0$&$4\sqrt{2}$&$0$&$0$&$0$&$0$\\
      $B^-\rightarrow K^0 K^-$ & $0$ &$1$ &$1$&$-1$&$3$&$-1$&$0$&$0$&$0$&$0$\\
      $B^0\rightarrow \pi^+ \pi^-$& $2$ & $1$ &$-1$ &$1$ &$1$&$3$&$0$&$0$&$0$&$0$\\
      $B^0\rightarrow \pi^0 \pi^0$& $2$&$1$&$-1$&$1$&$1$&$-5$&$0$&$0$&$0$&$0$ \\
      $B^0\rightarrow K^+ K^-$& $2$&$0$&$0$&$0$&$2$&$0$&$0$&$0$&$0$&$0$\\
     $B^0\rightarrow K^0 \bar{K}^0$& $2$& $1$& $1$& $-1$& $-3$& $-1$& $0$& $0$&$0$&$0$\\
      $B_s\rightarrow \pi^0 K^0$& 0&$-\frac{1}{\sqrt{2}}$&$\frac{1}{\sqrt{2}}$&$-\frac{1}{\sqrt{2}}$&
      $\frac{1}{\sqrt{2}}$&$\frac{5}{\sqrt{2}}$&$0$&$0$&$0$&$0$ \\
      $B_s\rightarrow \pi^- K^+$&$0$&$1$&$-1$&$1$&$-1$&$3$&$0$&$0$&$0$&$0$\\
      $B^-\rightarrow \pi^0 K^-$& 0&$\frac{1}{\sqrt{2}}$& $\frac{1}{\sqrt{2}}$& $-\frac{1}{\sqrt{2}}$&$\frac{3}{\sqrt{2}}$&$\frac{7}{\sqrt{2}}$&$0$&$0$&$0$&$0$\\
      $B^-\rightarrow \pi^- K^0$& $0$&$1$&$1$&$-1$&$3$&$-1$&$0$&$0$&$0$&$0$\\
      $B^0\rightarrow \pi^+ K^-$&$0$&$1$&$-1$&$1$&$-1$&$3$&$0$&$0$&$0$&$0$\\
      $B^0\rightarrow \pi^0 K^0$&0&$-\frac{1}{\sqrt{2}}$&
      $\frac{1}{\sqrt{2}}$&$-\frac{1}{\sqrt{2}}$&$\frac{1}{\sqrt{2}}$&$\frac{5}{\sqrt{2}}$&$0$&$0$&$0$&$0$\\
      $B_s\rightarrow \pi^+ \pi^-$&$2$&$0$&$0$&$0$&$2$&$0$&$0$&$0$&$0$&$0$\\
      $B_s\rightarrow \pi^0\pi^0$&$2$&$0$&$0$&$0$&$2$&$0$&$0$&$0$&$0$&$0$\\
      $B_s\rightarrow K^+ K^-$&$2$&$1$&$-1$&$1$&$1$&$3$&$0$&$0$&$0$&$0$\\
      $B_s\rightarrow K^0 \bar{K}^0$& $2$&$1$&$1$&$-1$&$-3$&$-1$&$0$&$0$&$0$&$0$\\
      \hline
    \end{tabular}}
     \caption{$B\rightarrow P P$ amplitudes decomposition in the $SU(3)$-basis. By virtue of Eq.~(\ref{eq:treepengSU3}) the corresponding numbers for the penguin components coincide with the tree ones. Where applicable, the coefficients agree with~\cite{He:2018php}. \label{tab:tableAmplBPP}}
  \end{center}
\end{table}

To make the present paper self-contained we give in Tables~\ref{tab:tableAmplBPP} and \ref{tab:tableAmplBPPeta} the decay amplitudes for $B\rightarrow P P$ transitions in the $SU(3)$-de\-composition. 

\begin{table}[tp]
  \begin{center}
    \scalebox{0.85}{\begin{tabular}{|c|c|c|c|c|c|c|c|c|c|c|}
      \hline 
      \textbf{Channel} & $A_{3}^T$& $C_{3}^T$&$A_{6}^T$&$C_{6}^T$&$A_{15}^T$&$C_{15}^T$&$B_{3}^T$&$B_{6}^T$&$B_{15}^T$ &$D_{3}^T$\\
      \hline
      $B^-\rightarrow \eta_q \pi^-$&0& $\sqrt{2}$& $\sqrt{2}$& 0& $3\sqrt{2}$& $2\sqrt{2}$&
      0& $\sqrt{2}$& $3\sqrt{2}$& $\sqrt{2}$ \\
      $B^-\rightarrow \eta_s \pi^-$&$0$& $0$& $0$& $1$& $0$& $-1$& $0$& $1$& $3$& $1$\\
      $B^0\rightarrow \eta_q \pi^0$&0& $-1$& $-1$& $0$& $5$& $2$& $0$& $-1$& $5$& $-1$\\
      $B^0\rightarrow \eta_s \pi^0$&$0$& $0$& $0$& $-\frac{1}{\sqrt{2}}$& 0& $\frac{1}{\sqrt{2}}$& 0&$-\frac{1}{\sqrt{2}}$&$\frac{5}{\sqrt{2}}$& $-\frac{1}{\sqrt{2}}$\\
      $B_s\rightarrow \eta_q K^0$&0& $\frac{1}{\sqrt{2}}$& $-\frac{1}{\sqrt{2}}$& $-\frac{1}{\sqrt{2}}$&$-\frac{1}{\sqrt{2}}$& $\frac{1}{\sqrt{2}}$& 0& $-\sqrt{2}$&$-\sqrt{2}$& $\sqrt{2}$\\
      $B_s\rightarrow \eta_s K^0$&$0$& $1$& $-1$& $0$& $-1$& $-2$& $0$& $-1$& $-1$& $1$\\
      $B^-\rightarrow \eta_q K^-$&$0$& $\frac{1}{\sqrt{2}}$& $\frac{1}{\sqrt{2}}$& $\frac{1}{\sqrt{2}}$&$\frac{3}{\sqrt{2}}$& $\frac{5}{\sqrt{2}}$& $0$& $\sqrt{2}$& $3\sqrt{2}$&$\sqrt{2}$\\
      $B^-\rightarrow \eta_s K^-$&$0$& $1$& $1$& $0$& $3$& $-2$& $0$& $1$& $3$& $1$\\
      $B^0\rightarrow \eta_q K^0$&$0$& $\frac{1}{\sqrt{2}}$& $-\frac{1}{\sqrt{2}}$& $-\frac{1}{\sqrt{2}}$&$-\frac{1}{\sqrt{2}}$& $\frac{1}{\sqrt{2}}$& $0$& $-\sqrt{2}$&
      $-\sqrt{2}$& $\sqrt{2}$\\
      $B^0\rightarrow \eta_s K^0$&$0$&$1$&$-1$& $0$& $-1$& $-2$& $0$& $-1$& $-1$& $1$\\
      $B_s\rightarrow \eta_q \pi^0$&$0$& $0$& $-2$& $0$& $4$& $0$& $0$& $-2$& $4$& $0$\\
      $B_s\rightarrow \eta_s \pi^0$&$0$& $0$& $0$& $-\sqrt{2}$& $0$& $2\sqrt{2}$& $0$&$-\sqrt{2}$& $2\sqrt{2}$& $0$ \\
      $B^0\rightarrow \eta_q \eta_q$&1&$\frac{1}{2}$& -$\frac{1}{2}$& $-\frac{1}{2}$& $\frac{1}{2}$&$\frac{1}{2}$& $2$& $-1$& $1$& $1$\\
      $B^0\rightarrow \eta_q \eta_s$&$0$& $0$& $0$& $\frac{1}{\sqrt{2}}$& $0$& $-\frac{1}{\sqrt{2}}$& $2\sqrt{2}$& $\frac{1}{\sqrt{2}}$&-$\frac{1}{\sqrt{2}}$& $\frac{1}{\sqrt{2}}$ \\
      $B^0\rightarrow \eta_s\eta_s$&$1$& $0$& $1$& $0$& $-1$& $0$& $1$& $1$& $-1$& $0$\\
      $B_s\rightarrow \eta_q\eta_q$&$1$& $0$& $0$& $0$& $1$& $0$& $2$& $0$& $2$& $0$\\
      $B_s\rightarrow \eta_q\eta_s$&$0$& $0$& $0$& $0$& $0$& $\sqrt{2}$& $2\sqrt{2}$& $0$&$-\sqrt{2}$& $\sqrt{2}$\\
      $B_s\rightarrow \eta_s\eta_s$&$1$& $1$& $0$& $0$& $-2$& $-2$& $1$& $0$& $-2$& $1$\\
 \hline
    \end{tabular}}
     \caption{$B\rightarrow P P$ amplitudes decomposition in the $SU(3)$-basis including singlets. Also here the corresponding numbers for the penguin components coincide with the tree ones.\label{tab:tableAmplBPPeta}}
  \end{center}
\end{table}

\subsection{Connecting $SU(3)$-invariant and topological decomposition}
\label{sec:connect}
The transformation rules between the topological basis and the $SU(3)$-invariant one for trees are given by~\cite{He:2018php}
\begin{align}
A_3^T&= -\frac{A}{8} + \frac{3E}{8}+T_{PA}, &
B_3^T&=  T_{SS} +\frac{3T_{AS}-T_{ES}}{8},\nonumber\\
C_3^T&=  \frac{1}{8} ({3A-C-E+3T})+T_P, &  D_3^T&=  T_{S} +\frac{1}{8} (3C-T_{AS}+3T_{ES}-T), \nonumber\\
A_6^T&= \frac{1}{4}(A-E),  &
B_6^T&=  \frac{1}{4}(T_{ES}-T_{AS}),\nonumber\\
C_6^T&=  \frac{1}{4}(-C+T),  &
A_{15}^T&=  \frac{A+E}{8},  \nonumber\\
B_{15}^T&= \frac{T_{ES}+T_{AS}}{8}, &
C_{15}^T&=  \frac{C+T}{8},
\end{align}
and for penguins by
\begin{align}
A_3^P&= -\frac{P_{TA}}{8} + \frac{3 P_{TE}}{8}+P_{A}, & 
B_3^P&=  P_{SS} +\frac{3P_{AS}-P_{ES}}{8},\nonumber\\
C_3^P&=  \frac{1}{8} ({3 P_{TA}-P_{C}-P_{TE}+3P_{T}})+ P, &  D_3^P&=  P_{S} +\frac{1}{8} (3P_{C}-P_{AS}+3P_{ES}-P), \nonumber\\
A_6^P&=  \frac{1}{4}(P_{A}-P_{E}),  &
B_6^P&=  \frac{1}{4}(P_{ES}-P_{AS}),\nonumber\\
C_6^P&=  \frac{1}{4}(-P_{C}+P_{T}),  &
A_{15}^P&=  \frac{P_A+P_E}{8},  \nonumber\\
B_{15}^P&= \frac{P_{ES}+P_{AS}}{8}, &
C_{15}^P&=  \frac{P_C+P_T}{8}. 
\end{align}
For convenience we also give the inverse rules which allow to convert the topological basis into the $SU(3)$-invariant one. For the tree-like amplitudes, they read,
\begin{eqnarray}
\label{eq:ToptoSU3}
T&=& 2 C_6^T +4C_{15}^T, \;\; C= 4C_{15}^T -2C_{6}^T, \;\; A= 2A_6^T +4A_{15}^T,\;\; E= 4A_{15}^T-2A_6^T,\nonumber\\
T_{P}&=& -A_{6}^T-A_{15}^T+C_{3}^T-C_{6}^T-C_{15}^T, \;\; T_{PA}= A_{3}^T+A_{6}^T -A_{15}^T,\;\; T_{AS}= 4B_{15}^T -2B_{6}^T, \nonumber\\
T_{ES}&=& 2B_6^T +4B_{15}^T, \;\; T_{SS}=B_{3}^T+B_{6}^T -B_{15}^T,\;\;T_{S}=-B_{6}^T-B_{15}^T+C_{6}^T -C_{15}^T +D_{3}^T.\nonumber\\ 
\end{eqnarray}
In the case of the penguins the rules can easily be derived if the following substitutions are applied
\begin{align}
T&\rightarrow P_{T}, & C & \rightarrow P_{C}, &
A&\rightarrow P_{TA},& T_{P}&\rightarrow P, &
E&\rightarrow P_{TE},\nonumber\\
T_{PA}&\rightarrow P_{A}, &
T_{AS}&\rightarrow P_{AS},&
T_{ES}&\rightarrow P_{ES}, &
T_{SS}&\rightarrow P_{SS}, &
T_{S}&\rightarrow P_{S}.
\end{align}


\section{Amplitudes in QCD factorization}
\label{sec:QCDF}

The calculation of the amplitude for $B\rightarrow M_1 M_2$ in QCD factorization can be obtained by applying the general formula~\cite{Beneke:2003zv}
\begin{eqnarray}
   &&\hspace{-1.0truecm}
   \mathcal{A}^{\rm{QCDF}}=i\frac{G_{F}}{\sqrt{2}}\sum_{p=u,c}\,A_{M_1 M_2}\,\bigg\{
    B M_1 \left( \alpha_1\delta_{pu}\hat{U} +
    \alpha_4^p\hat{I}
     + \alpha_{4,EW}^p\hat{Q} \right) M_2\,
     \Lambda_p \nonumber\\
   &&\qquad\mbox{}+ B M_1 \Lambda_p\cdot
    \mbox{Tr}\left[\left( \alpha_2\delta_{pu} \hat{U} + \alpha_3^p\hat{I}
    + \alpha_{3,EW}^p\,\hat {Q} \right) M_2\right] \nonumber\\
   &&\qquad\mbox{}+ B \left( \beta_2\delta_{pu} \hat{U} + \beta_3^p\hat{I}
    + \beta_{3,EW}^p\,\hat{Q} \right)
    M_1 M_2\Lambda_p \nonumber\\
   &&\qquad\mbox{}+ B \Lambda_p\cdot
    \mbox{Tr}\left[\left( \beta_1 \delta_{pu} \hat{U} + \beta_4^p\hat{I}
    + b_{4,EW}^p\,\hat{Q} \right) M_1 M_2\right]
    \nonumber\\
   &&\qquad\mbox{}+ B \left( \beta_{S2}\delta_{pu}\hat{U} + \beta_{S3}^p\hat{I}
    + \beta_{S3,EW}^p\,\hat{Q} \right)
    M_1\Lambda_p\cdot\mbox{Tr}M_2 \nonumber\\
   &&\qquad\mbox{}+ B\Lambda_p\cdot
    \mbox{Tr}\left[\left( \beta_{S1}\delta_{pu}\hat{U} + \beta_{S4}^p\hat{I}
    + b_{S4,EW}^p\hat{Q} \right) M_1\right]\cdot
    \mbox{Tr}M_2 \bigg\} \,,
    \label{eq:mastereq}
\end{eqnarray}
where
\begin{align}
   \Lambda_p &= \left( \begin{array}{c}
    0 \\ \lambda_p^{(d)} \\ \lambda_p^{(s)}
   \end{array} \right), & \hat{U} &= \left( \begin{array}{ccc}
    1~ & 0~ & 0 \\
    0~ & 0~ & 0 \\
    0~ & 0~ & 0
   \end{array} \right) , \\[0.2em]
   \hat {Q} &= \frac{3}{2} Q = \left( \begin{array}{ccc}
    1 & 0 & 0 \\
    0 & -\frac12 & 0 \\
    0 & 0 & -\frac12
   \end{array} \right), &
    \hat{I} &=\left( \begin{array}{ccc}
    1~ & 0~ & 0 \\
    0~ & 1~ & 0 \\
    0~ & 0~ & 1
   \end{array} \right),
\end{align}
and the meson vector $B$ and matrix $M$ coincide with the ones introduced in section~\ref{sec:Top}. In view of the connection to the $SU(3)$-invariant decomposition that we establish below, we drop the dependence of the QCDF amplitudes $\{\alpha_i,\alpha^{p}_{i}, \beta_{i},b_i\}$ on the final-state mesons from now on. The coefficient $A_{M_1 M_2}$ corresponds to
\begin{eqnarray}
A_{M_1 M_2}&=& M^2_B \, F^{B\rightarrow M_1}_0(0) \, f_{M_2}.
\label{eq:AM1M2}
\end{eqnarray}

It is relevant for our purposes to note that the general formula  Eq.~(\ref{eq:mastereq}) consists of the contraction of the meson matrices $B$, $M_1$ and $M_2$ with the generic structure
\begin{eqnarray}
\tilde{\hat{C}}_{r}&=&\sum_{p=u,c}\Bigl[(\tilde{T}_{r}\delta_{pu} \hat{U} + \tilde{P}^{(1),p}_{r} \hat{I} +\tilde{P}^{(2),p}_{r}\hat{Q})\otimes\Lambda_p\Bigl]
\label{eq:amplstr}
\end{eqnarray}
where
\begin{eqnarray}
\tilde{T}_{k}&\in&\{\alpha_1,\alpha_2,\beta_2,\beta_1, \beta_{S2}, \beta_{S1}\},\nonumber\\
\tilde{P}^{(1), p}_{k}&\in&\{\alpha^p_4, \alpha^p_3, \beta^p_3, \beta^p_4, \beta^p_{S3},\beta^p_{S4}\},\nonumber\\
\tilde{P}^{(2), p}_{k} &\in& \{\alpha^p_{4, EW}, \alpha^p_{3, EW},\beta^p_{3, EW}, b^p_{4, EW},\beta^p_{S3, EW}, b^p_{S4,EW}\}.\quad\quad\quad
\end{eqnarray}
Thus, Eq.~(\ref{eq:mastereq}) can be written in terms of indices as
\begin{eqnarray}
   \mathcal{A}^{\rm{QCDF}}&=&i\frac{G_{F}}{\sqrt{2}}
   A_{M_1 M_2}\Bigl\{B_i M^{i}_{j}(\tilde{\hat{C}}_{1})^{j l}_{k}M^{k}_{l}
   +
   B_i M^{i}_{j}(\tilde{\hat{C}}_{2})^{lj }_{k}M^{k}_{l}
   +
   B_i(\tilde{\hat{C}}_{3})^{ij }_{k}M^{k}_{l}M^{l}_{j}\nonumber\\
   &&\qquad\qquad+B_i (\tilde{\hat{C}}_{4})^{li}_{k} M^{k}_{r}M^{r}_{l} + B_i (\tilde{\hat{C}}_{5})^{ij}_{k}M^{k}_{i}M^{l}_{l}+ B_i (\tilde{\hat{C}}_{6})^{ji}_{k})M^{k}_{j}M^{l}_{l}\Bigl\},
\end{eqnarray}
where we have omitted the indices $1$ and $2$ in the meson matrices and we have identified
\begin{eqnarray}
(\tilde{\hat{C}}_{r})^{ij}_{k}&=&\sum_{p=u,c}\Bigl[(\tilde{T}_{r}\delta_{pu} \hat{U} + \tilde{P}^{(1),p}_{r} \hat{I} +\tilde{P}^{(2),p}_{r}\hat{Q})^{i}_{k}\Lambda^{j}_p\Bigl].
\label{eq:Cmatrix}
\end{eqnarray}


\section{Equivalence of QCDF and the topological basis}
\label{sec:QCDFTop}

Our aim is to establish a connection between the amplitudes $\{\alpha_i,\alpha^{p}_{i}, \beta_{i},b_i\}$ appearing in the  QCDF formalism inside Eq.~(\ref{eq:mastereq}) and the topological/$SU(3)$-irreducible representations as presented in Eqs.~(\ref{eq:new_tree}), (\ref{eq:new_peng}) and (\ref{eq:su3basis}). The desired transformation formulas are most conveniently obtained first between the topological and the QCDF expressions. Once this is done, the relationships between the QCDF amplitudes and the $SU(3)$-irreducible ones can be easily calculated since we already know the transformation rules between the topological and the $SU(3)$ case. 

To obtain the desired conversion rules we
first decompose the matrix $\hat{Q}$ in terms of $\hat{U}$ and $\hat{I}$
\begin{eqnarray}
\hat{Q}&=&\frac{3}{2}\hat{U}-\frac{1}{2}\hat{I}.
\label{eq:Qdecomposition}
\end{eqnarray}
Then, we notice that one of the key differences between the topological and the QCDF decomposition are the CKM factors. In the first case the amplitudes are written in terms of $\lambda^{(q)}_u$ and $\lambda^{(q)}_t$, for $q=d,s$. However in the second case this is done through the factors $\lambda^{(q)}_u$ and $\lambda^{(q)}_c$ encoded inside the vectors $\Lambda_u$ and $\Lambda_c$. Therefore, we translate the unitarity rule in Eq.~(\ref{eq:CKMrel}) in terms of the CKM $\Lambda$-vectors and express  $\Lambda_t$ in terms of $\Lambda_u$ and $\Lambda_c$ via
\begin{eqnarray}
\Lambda_t=-\Lambda_u - \Lambda_c.
\label{eq:CKMLambda}
\end{eqnarray}
Finally,  we substitute Eq.~(\ref{eq:Qdecomposition}) and Eq.~(\ref{eq:CKMLambda}) inside the amplitude in Eq.~(\ref{eq:amplstr}) to arrive at
\begin{eqnarray}
\tilde{\hat{C}}_{r}&=&\Bigl[\tilde{T} +\frac{3}{2}\tilde{P}^u_2 - \frac{3}{2}\tilde{P}^c_2\Bigl]\hat{U}\otimes\Lambda_u + \Bigl[\tilde{P}^u_1-\tilde{P}^c_1-\frac{1}{2}\Bigl\{\tilde{P}^u_2-\tilde{P}^c_2\Bigl\}\Bigl]\hat{I}\otimes\Lambda_u\nonumber\\
&&-\frac{3}{2}\tilde{P}^c_2 \hat{U}\otimes\Lambda_t -\Bigl[ \tilde{P}^c_1 -\frac{\tilde{P}^c_2}{2}\Bigl]\hat{I}\otimes\Lambda_t,
\label{eq:amplstrexpanded}
\end{eqnarray}
or in terms of components
\begin{eqnarray}
(\tilde{\hat{C}}_{r})^{ij}_{k}&=&\Bigl[\tilde{T} +\frac{3}{2}\tilde{P}^u_2 - \frac{3}{2}\tilde{P}^c_2\Bigl]\hat{U}^{i}_{k}(\Lambda_u)^{j} + \Bigl[\tilde{P}^u_1-\tilde{P}^c_1-\frac{1}{2}\Bigl\{\tilde{P}^u_2-\tilde{P}^c_2\Bigl\}\Bigl]\delta^{i}_{k}(\Lambda_u)^{j}\nonumber\\
&&-\frac{3}{2}\tilde{P}^c_2 \hat{U}^{i}_{k}(\Lambda_t)^{j} -\Bigl[ \tilde{P}^c_1 -\frac{\tilde{P}^c_2}{2}\Bigl]\delta^{i}_{k}(\Lambda_t)^{j}.
\label{eq:amplstrexpandedcomp}
\end{eqnarray}

We are now ready to establish the equivalence between the QCD-factorization little amplitude basis and the topological decomposition basis discussed in Sec.~\ref{sec:TopIrr}. This can easily be done if we
make the following identifications with the different tensors appearing in Eq.~(\ref{eq:new_tree}) and Eq.~(\ref{eq:new_peng})
\begin{eqnarray}
U^i_k(\Lambda_u)^j=\bar{H}^{ij}_k,\quad\quad 
U^i_k(\Lambda_t)^j=\tilde{H}^{ij}_k,
\quad\quad
(\Lambda_t)^i=\tilde{H}^i.
\end{eqnarray}
Our set of transformation rules finally reads
\begin{align}
T&=A_{M_{1}M_{2}}\Bigl[\alpha_1 + \frac{3}{2}\alpha^{u}_{4,EW}
-\frac{3}{2}\alpha^{c}_{4,EW}\Bigl],&
C&=A_{M_{1}M_{2}}\Bigl[\alpha_2+\frac{3}{2}\alpha^{u}_{3,EW}
-\frac{3}{2}\alpha^{c}_{3,EW}\Bigl],\nonumber\\
E&=A_{M_1 M_2}\Bigl[\beta_1+\frac{3}{2}b^{u}_{4,EW}-\frac{3}{2}b^{c}_{4,EW}\Bigl], &
A&=A_{M_1 M_2}\Bigl[\beta_{2}+\frac{3}{2}\beta^{u}_{3,EW}-\frac{3}{2}\beta^{c}_{3,EW}\Bigl],\nonumber\\
T_{AS}&=A_{M_1 M_2}\Bigl[\beta_{S1}+\frac{3}{2}b^{u}_{S4,EW}-\frac{3}{2}b^{c}_{S4,EW} \Bigl],&
T_{ES}&=A_{M_1 M_2}\Bigl[\beta_{S2}+\frac{3}{2}\beta^u_{S3, EW}-\frac{3}{2}\beta^c_{S3, EW}\Bigl],\nonumber
\end{align}
\begin{eqnarray}
T_{PA}&=&A_{M_1 M_2}\Bigl[\beta^{u}_{4}-\beta^{c}_{4}-\Bigl(\frac{b^{u}_{4,EW}}{2}-\frac{b^{c}_{4,EW}}{2}\Bigl)\Bigl], \nonumber\\
T_{SS}&=&A_{M_1 M_2}\Bigl[\beta^{u}_{S4}-\beta^{c}_{S4}-\Bigl(\frac{b^{u}_{S4,EW}}{2}-\frac{b^{c}_{S4,EW}}{2}\Bigl)\Bigl],\nonumber\\
T_{S}&=&A_{M_1 M_2}\Bigl[\alpha^{u}_{3}-\alpha^{c}_{3}-\Bigl(\frac{\alpha^{u}_{3,EW}}{2}-\frac{\alpha^{c}_{3,EW}}{2}\Bigl)+\Bigl(\beta^{u}_{S3}-\beta^{c}_{S3}\Bigl)-\Bigl(\frac{\beta^{u}_{S3,EW}}{2}-\frac{\beta^{c}_{S3,EW}}{2}\Bigl)\Bigl],\nonumber\\
T_{P}&=&A_{M_1 M_2}\Bigl[\alpha^{u}_{4}-\alpha^{c}_{4}-\Bigl(\frac{\alpha^{u}_{4, EW}}{2}-\frac{\alpha^{c}_{4, EW}}{2}\Bigl)+\Bigl(\beta^{u}_{3}-\beta^{c}_{3}\Bigl)-\Bigl(\frac{\beta^{u}_{3,EW}}{2}-\frac{\beta^{c}_{3,EW}}{2}\Bigl)\Bigl],\nonumber\\
S&=&-A_{M_1 M_2}\Bigl[\alpha^{c}_{3}+\beta^{c}_{S3}-\frac{\alpha^{c}_{3, EW}}{2}-\frac{\beta_{S3, EW}}{2}\Bigl],\nonumber\\
P&=&-A_{M_1 M_2}\Bigl[\alpha^{c}_{4}+\beta^{c}_{3}-\frac{\alpha^{c}_{4,EW}}{2}-\frac{\beta^{c}_{3,EW}}{2}\Bigl],\nonumber
\end{eqnarray}
\begin{eqnarray}
P_{A}=-A_{M_1 M_2}\Bigl[\beta^{c}_{4}-\frac{b^{c}_{4, EW}}{2}\Bigl],&~~&P_{SS}=-A_{M_1 M_2}\Bigl[\beta^{c}_{S4}-\frac{b^{c}_{S4, EW}}{2}\Bigl],\nonumber\\
P_{C}=-\frac{3}{2}A_{M_1 M_2}\alpha^{c}_{3,EW},&~~&P_{T}=-\frac{3}{2}A_{M_1 M_2}\alpha^{c}_{4,EW},\nonumber\\
P_{TA}=-\frac{3}{2}A_{M_1M_2}\beta^{c}_{3,EW},&~~&P_{TE}=-\frac{3}{2}A_{M_1M_2}b^{c}_{4,EW},\nonumber\\
P_{AS}=-\frac{3}{2}A_{M_1M_2}b^{c}_{S4,EW},&~~&P_{ES}=-\frac{3}{2}A_{M_1M_2}\beta^{c}_{S3, EW}.
\label{eq:QCDFtotopological}
\end{eqnarray}

One of the interesting features of the formulas in Eq.~(\ref{eq:QCDFtotopological}) is the fact that the flavour structure combines amplitudes with different physical content. For instance, consider the case of the structures $(\alpha_4-\alpha_{4,EW}/2)$ and $(\beta_3-\beta_{3,EW}/2)$ which appear multiplying the same factor  $B_i M^{i}_{j} M^{j}_{k}\bar{H}^{lk}_{l}$,
where the first factor refers to penguin topologies, whereas the second to annihilation ones.\\

The formulas in Eqs.~(\ref{eq:QCDFtotopological}) can be simplified if we consider the following NLO results \cite{Beneke:2003zv}
\begin{eqnarray}
\alpha^{u}_{3}=\alpha^{c}_{3}=\alpha_{3},
\quad
\alpha^{u}_{3,EW}=\alpha^{c}_{3,EW}=\alpha_{3,EW},
\quad
\beta^{u}_{i}=\beta^{c}_{i}=\beta_{i},
\quad
b^{u}_i=b^{c}_i=b_i.
\label{eq:QCDFconditions}
\end{eqnarray}
Moreover, at NLO the amplitudes $\alpha^{p}_{4,EW}$ are at the permille level, such that the differences $|\alpha^{c}_{4,EW}-\alpha^{u}_{4,EW}|$ are at most $\mathcal{O}(10^{-3})$. Analogously for $\alpha^{p}_{4}$ the magnitudes are about $0.1$, and the differences $|\alpha^{c}_{4}-\alpha^{u}_{4}|$ are about $0.02$. Then, we simplify Eqs.~(\ref{eq:QCDFtotopological}) as follows
\begin{align}
T&=A_{M_1 M_2}\alpha_1, &
C&= A_{M_1 M_2} \alpha_2, &
E&=A_{M_1 M_2} \beta_1, \nonumber\\
A&=A_{M_1 M_2} \beta _2, &
T_{AS}&=A_{M_1 M_2}\beta_{S1},&
T_{ES}&=A_{M_1 M_2}\beta_{S2},
\nonumber
\end{align}
\begin{eqnarray}
S&=&-A_{M_1 M_2}\Bigl[\alpha_{3}+\beta_{S3}-\frac{\alpha_{3, EW}}{2}-\frac{\beta_{S3, EW}}{2}\Bigl],\nonumber\\
P&=&-A_{M_1 M_2}\Bigl[\alpha^{c}_{4}+\beta_{3}-\frac{\alpha^{c}_{4,EW}}{2}-\frac{\beta_{3,EW}}{2}\Bigl],\nonumber
\end{eqnarray}
\begin{align}
P_A&=-A_{M_1 M_2}(\beta_4-\frac{b_{4, EW}}{2}), &
P_{SS}&=-A_{M_1 M_2}(\beta_{S4}-\frac{b_{S4, EW}}{2}),\nonumber\\
P_C&=-\frac{3}{2}A_{M_1 M_2}\alpha_{3, EW}, &
P_{T}&=-\frac{3}{2}A_{M_1 M_2}\alpha^{c}_{4, EW},\nonumber\\
P_{TA}&=-\frac{3}{2}A_{M_1 M_2}\beta_{3, EW}, &
P_{TE}&=-\frac{3}{2}A_{M_1 M_2}b_{4, EW},\nonumber\\
P_{AS}&=-\frac{3}{2}A_{M_1 M_2}b_{S4, EW},&
P_{ES}&=-\frac{3}{2}A_{M_1 M_2}\beta_{S3, EW},\nonumber
\end{align}
\begin{align}
T_{PA}&=0, & T_{SS}&=0, &T_{S}&=0, & |T_{P}|&<0.02.
\label{eq:TopologicalQCDF}
\end{align}

We realize that the expressions in Eq.~(\ref{eq:QCDFconditions}) simplify drastically the relationships between the topological and the QCDF decomposition of the physical amplitudes. For instance there are multiple cancellations between the annihilation 
contributions $\beta_i$. Moreover, most of the extra contributions of the form
$\alpha^{c}_i-\alpha^{u}_i$ are expected to be subleading and after applying model dependent information can in most of the cases be neglected, with the only exception $\alpha^c_4-\alpha^u_4$.\\

We can now invert and solve for the QCDF little amplitudes in terms of the topological decomposition coefficients, and find the following results
\begin{align}
A_{M_1 M_2}\alpha_1&= T, & A_{M_1 M_2}\alpha_2&= C, & A_{M_1 M_2}\beta_1&= E, & A_{M_1 M_2}\beta_2 &= A,\nonumber\\
A_{M_1 M_2}\beta_{S1} &= T_{AS}, & A_{M_1 M_2}\beta_{S2}&=T_{ES}, &
A_{M_1 M_2}\alpha_{3, EW}&=-\frac{2}{3}P_C, &
A_{M_1 M_2}\alpha^{c}_{4, EW}&=-\frac{2}{3}P_{T},\nonumber
\end{align}
\begin{align}
A_{M_1 M_2}\beta_{3, EW}&=-\frac{2}{3}P_{TA},&
A_{M_1 M_2}b_{4, EW}&=-\frac{2}{3}P_{TE},&
A_{M_1 M_2}\beta_4&=-P_{A}+\frac{1}{3}P_{TE},\nonumber\\
A_{M_1 M_2}\beta_{S3, EW}&=-\frac{2}{3}P_{ES},&
A_{M_1 M_2}b_{S4, EW}&=-\frac{2}{3}P_{AS},&
A_{M_1 M_2}\beta_{S4}&=-P_{SS}+\frac{1}{3}P_{AS},\nonumber
\end{align}
\begin{align}
A_{M_1 M_2}\Bigl(\alpha_3 + \beta_{S3}\Bigl)&=-S-\frac{P_{C}}{3}-\frac{P_{ES}}{3},&
A_{M_1 M_2}\Bigl(\alpha^{c}_{4}+\beta_{3}\Bigl)&=-P-\frac{P_T}{3}-\frac{P_{TA}}{3}.
\label{eq:SU£toTop}
\end{align}

So far, we have decomposed the physical amplitudes in terms of the CKM factors $\lambda_u^{(q)}$ and $\lambda_t^{(q)}$, or equivalently $\Lambda_u$ and $\Lambda_t$. However, we can also try an alternative parameterization in terms of $\Lambda_u$ and $\Lambda_c$. Thus, we proceed in an analogous way as we did previously during the decomposition of the $\tilde{\hat{C}}_r$ factors in Eq.~(\ref{eq:amplstrexpanded}), but this time we do not use the CKM-unitary relationship. Our result is then

\begin{eqnarray}
\tilde{\hat{C}}_r&=&\Bigl[\tilde{T}_{r} + \frac{3}{2}\tilde{P}^{(2), u}_r\Bigl]\hat{U}\otimes \Lambda_{u}+\Bigl[\tilde{P}^{(1), u}_r-\frac{\tilde{P}^{(2), u}_r}{2}\Bigl]\hat{I}\otimes \Lambda_{u}\nonumber\\
&&+\frac{3}{2} \tilde{P}^{(2), c}_r \hat{U}\otimes \Lambda_{c}+ \Bigl[\tilde{P}^{(1), c}_r-\frac{\tilde{P}^{(2), c}_r}{2} \Bigl] \hat{I}\otimes \Lambda_{c}.
\end{eqnarray}
For the physical amplitude we write
\begin{eqnarray}
\mathcal{A}^{\rm QCDF}=i\frac{G_{F}}{\sqrt{2}} \, \sum_{p=u,c}&&
\Bigl(c^{{\rm QCDF},p}_{1}B M_1 \hat{U} M_2 \Lambda_p 
+ c^{{\rm QCDF},p}_{2}B \hat{U} M_1 M_2 \Lambda_p \nonumber\\
&&+c^{{\rm QCDF},p}_{3} B M_1 M_2 \Lambda_p + c^{{\rm QCDF},p}_{4}
B M_1 \Lambda_p \mbox{Tr}\Bigl[\hat{U} M_2\Bigl] \nonumber\\
&&+ c^{{\rm QCDF},p}_5 B\hat{U} M_1 \Lambda_p \mbox{Tr}\Bigl[M_2\Bigl]
+ c^{{\rm QCDF},p}_6 B M_1 \Lambda_p \mbox{Tr}\Bigl[M_2\Bigl]\nonumber\\
&&+c^{{\rm QCDF}, p}_7 B \Lambda_p \mbox{Tr}\Bigl[\hat{U} M_1 M_2\Bigl]
+c^{{\rm QCDF}, p}_8 B\Lambda_p \mbox{Tr}\Bigl[\hat{U} M_1\Bigl]
\mbox{Tr}\Bigl[M_2\Bigl]\nonumber\\
&&+c^{{\rm QCDF}, p}_{9} B\Lambda_p \mbox{Tr}\Bigl[M_1 M_2\Bigl]
+ c^{{\rm QCDF}, p}_{10} B\Lambda_p \mbox{Tr}\Bigl[M_1\Bigl] \mbox{Tr}\Bigl[M_2\Bigl]\Bigl) \, .
\end{eqnarray}
We can identify the following set of independent coefficients proportional to $\Lambda_{u}$,
\begin{align}
c^{{\rm QCDF}, u}_{1}&=A_{M_1 M_2}\Bigl(\alpha_{1}+\frac{3}{2}\alpha^{u}_{4, EW}\Bigl), &
c^{{\rm QCDF}, u}_{7}&=A_{M_1 M_2}\Bigl(\beta_{1}+\frac{3}{2}b^{u}_{4, EW}\Bigl),\nonumber\\ 
c^{{\rm QCDF}, u}_{2}&=A_{M_1 M_2}\Bigl(\beta_{2}+\frac{3}{2}\beta^{u}_{3, EW}\Bigl), &
c^{{\rm QCDF}, u}_{8}&=A_{M_1 M_2}\Bigl(\beta_{S1}+\frac{3}{2}b^{u}_{S4, EW}\Bigl),\nonumber\\
c^{{\rm QCDF}, u}_{3}&= A_{M_1 M_2}\Bigl(\alpha^{u}_{4}-\frac{\alpha^{u}_{4, EW}}{2}+\beta^{u}_3-\frac{\beta^{u}_{3, EW}}{2}\Bigl), &
c^{{\rm QCDF}, u}_{9}&=A_{M_1 M_2}\Bigl(\beta^{u}_{4}-\frac{b^u_{4, EW}}{2}\Bigl),\nonumber\\
c^{{\rm QCDF}, u}_{4}&=A_{M_1 M_2}\Bigl(\alpha_2 +\frac{3}{2}\alpha^{u}_{3, EW}\Bigl), &
c^{{\rm QCDF}, u}_{10}&=A_{M_1 M_2}\Bigl(\beta^u_{S4}-\frac{b^u_{S4, EW}}{2}\Bigl),\nonumber\\
c^{{\rm QCDF}, u}_{5}&=A_{M_1 M_2}\Bigl(\beta_{S2}+\frac{3}{2}\beta^{u}_{S3, EW}\Bigl), & &\nonumber\\
c^{{\rm QCDF}, u}_{6}&=A_{M_1 M_2}\Bigl(\alpha^{u}_{3}-\frac{\alpha^{u}_{3, EW}}{2}+\beta^{u}_{S3}-\frac{\beta^{u}_{S3, EW}}{2}\Bigl), & & \label{eq:QCDF1}
\end{align}
and analogously for $\Lambda_{c}$,
\begin{align}
\allowdisplaybreaks
c^{{\rm QCDF}, c}_{1}&=\frac{3}{2}A_{M_1 M_2}\alpha^{c}_{4, EW}, &
c^{{\rm QCDF}, c}_{7}&=\frac{3}{2}A_{M_1 M_2} b^c_{4, EW}, \nonumber\\
c^{{\rm QCDF}, c}_{2}&=\frac{3}{2}A_{M_1 M_2}\beta^c_{3, EW}, &
c^{{\rm QCDF}, c}_{8}&=\frac{3}{2}A_{M_1 M_2}b^c_{S4, EW}, \nonumber\\
c^{{\rm QCDF}, c}_{3}&=A_{M_1 M_2}\Bigl(\alpha^{c}_{4}-\frac{\alpha^{c}_{4, EW}}{2}+\beta_{3}-\frac{\beta_{3, EW}}{2}\Bigl), &
c^{{\rm QCDF}, c}_{9}&=A_{M_1 M_2}\Bigl(\beta^{c}_{4}-\frac{b^c_{4, EW}}{2}\Bigl),    \nonumber\\
c^{{\rm QCDF}, c}_{4}&=\frac{3}{2}A_{M_1 M_2}\alpha^{c}_{3, EW}, &
c^{{\rm QCDF}, c}_{10}&=A_{M_1 M_2}\Bigl(\beta^c_{S4}-\frac{b^c_{S4, EW}}{2}\Bigl).\nonumber\\
c^{{\rm QCDF}, c}_{5}&=\frac{3}{2}A_{M_1 M_2}\beta^c_{S3, EW}, && \nonumber\\
c^{{\rm QCDF}, c}_{6}&=A_{M_1 M_2}\Bigl(\alpha^{c}_{3}-\frac{\alpha^{c}_{3, EW}}{2}+\beta^c_{S3}-\frac{\beta^c_{S3, EW}}{2}\Bigl),  &&  \label{eq:QCDF2}
\end{align}

Without considering any further relations between the amplitudes $c^{{\rm QCDF}, u}_{i}$ and $c^{{\rm QCDF}, c}_{i}$ we have in total 20 complex quantities and in principle 40 real degrees of freedom. However, to describe the phenomenological processes we only need 18 complex amplitudes.
To prove this, we consider the coefficient matrix resulting from expanding each one of the 34  transitions for $B\rightarrow P P$ in terms of the numerical coefficients $c^{{\rm QCDF}, u}_{i}$ analogous to those in Tables \ref{tab:tableAmplBPP} and \ref{tab:tableAmplBPPeta} given for the coefficients in the $SU(3)$ irreducible representation. It is easy to see that the rank of this matrix is $9$, meaning that once the phenomenology is taken into account not all the $c^{{\rm QCDF}, u}_{i}$ coefficients are independent and only 9 linear combinations of them are required. Analogous considerations can be done for the set of coefficients $c^{{\rm QCDF}, c}_{i}$ with the same outcome. Thus the total number of complex amplitudes required is $18$, implying 36 real degrees of freedom. This is in line with~\cite{beneketalk} where the fact that QCDF describes $B\to PP$ transitions in terms of $18$ independent complex amplitudes was already pointed out. The linear relations between the QCDF and topological amplitudes, however, are to the best of our knowledge worked out for the first time in the present article.

 As argued before, an additional overall phase can be absorbed since it is not physically relevant. This can be done by taking one of the coefficients $c^{{\rm QCDF},u}_{i}$ or $c^{{\rm QCDF},c}_{i}$ to be real. Hence, if we consider Eqs.~(\ref{eq:QCDF1}) and~(\ref{eq:QCDF2}) and the absorption of the global phase,  we see that the total number of meaningful real parameters is $35$. The counting leads to 36 real values once the $\eta-\eta'$ mixing angle is added. This is the same result as was obtained when doing the analysis in the $\Lambda_{u}$,   $\Lambda_{t}$ decomposition. 
 
Using Eqs. (\ref{eq:QCDFconditions})  we can simplify Eqs.~(\ref{eq:QCDF1}) and (\ref{eq:QCDF2}),  we then obtain two extra conditions

\begin{eqnarray}
 c^{{\rm QCDF}, c}_{9}=c^{{\rm QCDF}, u}_{9} \, ,& &c^{{\rm QCDF}, c}_{10}=c^{{\rm QCDF}, u}_{10}.
\end{eqnarray}

For completeness we invert the simplified expressions derived from Eqs.~(\ref{eq:QCDF1}) and~(\ref{eq:QCDF2}) leading to

\begin{align}
A_{M_1 M_2}\alpha^{c}_{4, EW}&=\frac{2}{3}c^{{\rm QCDF},c}_{1}, &
A_{M_1 M_2}\beta_{3, EW}&=\frac{2}{3}c^{{\rm QCDF},c}_{2}, &
A_{M_1 M_2}\alpha^{c}_{3, EW}&=\frac{2}{3}c^{{\rm QCDF}, c}_{4}, \nonumber\\
A_{M_1 M_2}\beta_{S3, EW}&=\frac{2}{3}c^{{\rm QCDF}, c}_{5}, &
A_{M_1 M_2}b_{4, EW}&=\frac{2}{3}c^{{\rm QCDF}, c}_{7}, &
A_{M_1 M_2}b_{S4, EW}&=\frac{2}{3}c^{{\rm QCDF}, c}_{8}, \nonumber
\end{align}
\begin{align}
A_{M_1 M_2}\beta_1&=c^{{\rm QCDF}, u}_{7} - c^{{\rm QCDF}, c}_{7}, &
A_{M_1 M_2}\beta_{S1}&=c^{{\rm QCDF}, u}_{8} - c^{{\rm QCDF}, c}_{8}, \nonumber\\
A_{M_1 M_2}\beta_{2}&=c_{2}^{{\rm QCDF}, u}-c_{2}^{{\rm QCDF}, c}, &
A_{M_1 M_2}\beta_{4}&=c^{{\rm QCDF}, u}_{9}+\frac{1}{3}c^{{\rm QCDF}, c}_{7},\nonumber\\
A_{M_1 M_2}\beta_{S4}&=c^{{\rm QCDF}, u}_{10}+\frac{1}{3}c^{{\rm QCDF}, c}_{8}, &
A_{M_1 M_2}\Bigl(\alpha_{1}+\frac{3}{2}\alpha^{u}_{4, EW}\Bigl)&=c^{{\rm QCDF}, u}_{1},\nonumber
\end{align}
\begin{align}
A_{M_1 M_2}\Bigl(\alpha^{c}_{4}+\beta_3 \Bigl)&=\frac{1}{3}c^{{\rm QCDF}, c}_{1}+\frac{1}{3}c^{{\rm QCDF}, c}_{2}+c^{{\rm QCDF}, c}_{3}, \nonumber\\
A_{M_1 M_2}\Bigl(\alpha^{c}_3+\beta_{S3}\Bigl)&=
\frac{1}{3}c^{{\rm QCDF}, c}_{4}+\frac{1}{3}c^{{\rm QCDF}, c}_{5}+c^{{\rm QCDF}, c}_{6},\nonumber\\
A_{M_1 M_2}\Bigl(\alpha^{u}_3+\beta_{S3}-\frac{1}{2}\alpha^{c}_{3, EW}\Bigl)&=
c^{{\rm QCDF}, u}_{6}+\frac{1}{3}c^{{\rm QCDF}, c}_{5}, \nonumber\\
A_{M_1 M_2}\Bigl(\alpha^{u}_4+\beta_3-\frac{1}{2}\alpha^{u}_{4, EW}\Bigl)&=c^{{\rm QCDF}, u}_{3}+\frac{1}{3}c^{{\rm QCDF}, c}_{2}.
\end{align}

We can realize how the $\Lambda_u$~--~$\Lambda_t$ parameterization in Eq.~(\ref{eq:TopologicalQCDF}) has advantages with respect to the $\Lambda_u$~--~$\Lambda_c$ one in Eqs.~(\ref{eq:QCDF1}) and~(\ref{eq:QCDF2}) when it comes to disentangle the QCDF amplitudes $\{\alpha_i,\alpha^{p}_{i}, \beta_{i},b_i\}$. For instance, once we account for the expressions in Eq.~(\ref{eq:QCDFconditions}), there are multiple cancellations between the up and charm versions of a given little subamplitude, leading to simple relationships between the elements of the topological basis and the QCDF one as shown in Eq.~(\ref{eq:TopologicalQCDF}). On the other hand, in the $\Lambda_u$~--~$\Lambda_c$ decomposition in Eqs.~(\ref{eq:QCDF1})  and~(\ref{eq:QCDF2}) these cancellations do not arise, leaving the coefficients $c^{{\rm QCDF}, p}_{i}$ in terms of linear combinations of multiple $\{\alpha_i,\alpha^{p}_{i}, \beta_{i},b_i\}$ quantities.


\section{Setting the stage for the $\chi^2$-fit}
\label{sec:prepfit}

\subsection{Relevant physical quantities}

We extract the values for the $SU(3)$-irreducible parameters using branching fractions and CP-asymmetries for the different $B\rightarrow P P$ transitions, where $P$ is a pseudoscalar meson.  

The basic ingredient behind our phenomenological determinations is the decay rate 
\begin{eqnarray}
\Gamma(\bar{B}\rightarrow M_1 M_2)&=&\frac{S}{16\pi M_B}|\mathcal{A}_{B\rightarrow M_1 M_2}|^2,
\end{eqnarray}
where $S$ is a symmetry factor with values $S=1$ if $M_1$ and $M_2$ are different and $S=1/2$ if $M_1$ and $M_2$ are identical. Then, we determine the CP-averaged branching fraction of the process $\bar{B}\rightarrow \bar{f}$ via the formula
\begin{eqnarray}
\mathcal{B}(\bar{B}\rightarrow \bar{f})&=&\frac{1}{2}\tau_{B} \Bigl[\Gamma(\bar{B}\rightarrow \bar{f}) + \Gamma(B\rightarrow f)\Bigl].
\end{eqnarray}
In the case of the CP asymmetries, our theoretical evaluations are performed using
\begin{eqnarray}
\mathcal{A}_{\rm CP}(\bar{B}\rightarrow \bar{f})&=&\frac{\Gamma(\bar{B}
\rightarrow \bar{f})-\Gamma(B\rightarrow f)}{\Gamma(\bar{B}\rightarrow \bar{f})+\Gamma(B\rightarrow f)},
\end{eqnarray}
where we have followed the conventions introduced in \cite{Beneke:2003zv}.

The list of processes considered, together with the corresponding state of the art experimental status are shown in Tables~\ref{tab:tableBr_combined} and~\ref{tab:tableAs_combined}, whose presentation we relegate to section~\ref{sec:results} in order to have the experimental input and the $\chi^2$-fit result at a glance. 

\subsection{The $\eta$-$\eta'$ system}
\label{Sec:etamixing}

During our analysis, we account for $\eta$-$\eta'$ mixing. In particular, we adopt the Feldmann–Kroll–Stech (FKS) scheme~\cite{Feldmann:1998vh}, which is based on one single mixing angle $\theta_{\rm FKS}$. In the FKS-scheme, the flavor states are related to the physical states $\eta$, $\eta'$ according to \cite{Beneke:2002jn}
\begin{equation}
\left(
\begin{array}{cc}
    \ket{\eta} \\
    \ket{\eta'}  \\
   \end{array}
\right)   
 = 
\left( 
   \begin{array}{cc}
    \cos\theta_{\rm FKS} & -\sin\theta_{\rm FKS} \\
    \sin\theta_{\rm FKS} & \cos\theta_{\rm FKS}  \\
   \end{array} 
 \right)
 \left(
\begin{array}{cc}
    \ket{\eta_{q}} \\
    \ket{\eta_{s}}  \\
   \end{array}
\right),
\end{equation}
which leads to the following equations for the decay constants
\begin{align}
f^q_{\eta}&=f_q\cos\theta_{\rm FKS},&  f^q_{\eta'}&=f_q \sin\theta_{\rm FKS},\nonumber\\
f^s_{\eta}&=-f_s\sin\theta_{\rm FKS},& f^s_{\eta'}&=f_s \cos\theta_{\rm FKS}.
\end{align}
Therefore,  for the amplitudes of the processes $B\rightarrow M \eta$, $B\rightarrow M \eta'$ with $M\neq \eta, \eta'$ we obtain
\begin{eqnarray}
A(B\rightarrow M \eta)&=& 
\tilde{A}(B\rightarrow M \eta_q)\cos(\theta_{\rm FKS})
-
\tilde{A}(B\rightarrow M \eta_s)\sin(\theta_{\rm FKS}),\nonumber\\
A(B\rightarrow M \eta')&=& 
\tilde{A}(B\rightarrow M \eta_q)\sin(\theta_{\rm FKS})
+
\tilde{A}(B\rightarrow M \eta_s)\cos(\theta_{\rm FKS}),
\end{eqnarray}
where $\tilde{A}(B\rightarrow M \eta_{q})\propto F^{B\rightarrow M}_0(0) f_q$ and $\tilde{A}(B\rightarrow M \eta_{s})\propto F^{B\rightarrow M}_0(0) f_s$.\\

Since the form factors for the transitions $B\rightarrow \eta$ and $B\rightarrow \eta'$ can be parameterized as \cite{Beneke:2002jn}
\begin{eqnarray}
F^{B\rightarrow \eta^{(\prime)}}_{0}&=&
F_1\frac{f^q_{\eta^{(\prime)}}}{f_{\pi}}
+F_{2}\frac{\sqrt{2}f^{q}_{\eta^{(\prime)}} +f^{s}_{\eta^{(\prime)}}}{\sqrt{3}f_{\pi}},
\label{eq:Feta}
\end{eqnarray}
and considering $F_1=F_0^{B\rightarrow \pi}(0)$ and $F_2=0$, just as in \cite{Beneke:2002jn}, we write for the amplitudes of $B$-meson decays with final states consisting of pairs $\eta$, $\eta'$
\begin{eqnarray}
A(B\rightarrow \eta \eta)&=&
\tilde{A}(B\rightarrow \eta_q\eta_q)\cos^2\theta_{\rm FKS} +\tilde{A}(B\rightarrow \eta_s\eta_s)\sin^2\theta_{\rm FKS}\nonumber\\
&&- 2\tilde{A}(B\rightarrow \eta_q\eta_s)\cos\theta_{\rm FKS}\sin\theta_{\rm FKS},\nonumber\\
A(B\rightarrow \eta' \eta')&=&
\tilde{A}(B\rightarrow \eta'_q\eta'_q)\cos^2\theta_{\rm FKS} +\tilde{A}(B\rightarrow \eta'_s\eta'_s)\sin^2\theta_{\rm FKS}\nonumber\\
&&+ 2\tilde{A}(B\rightarrow \eta'_q\eta'_s)\cos\theta_{\rm FKS}\sin\theta_{\rm FKS},\nonumber\\
A(B\rightarrow \eta \eta')&=&\tilde{A}(B\rightarrow \eta_q \eta'_q)\cos\theta_{\rm FKS}\sin\theta_{\rm FKS}+\tilde{A}(B\rightarrow \eta_q \eta'_s)\cos^2\theta_{\rm FKS}\nonumber\\
&&-\tilde{A}(B\rightarrow \eta_s \eta'_q) \sin^2\theta_{\rm FKS} -\tilde{A}(B\rightarrow \eta_s\eta'_s)\sin\theta_{\rm FKS}\cos\theta_{\rm FKS}.
\end{eqnarray}

To perform our numerical estimates we need the value of the mixing angle $\theta_{\rm FKS}$, together with the values for the decay constants $f_q$ and $f_s$. They have been obtained from fits to data involving the processes $J/\psi\rightarrow \eta (\eta') \rho$, $\eta'\rightarrow \rho \gamma$, $\rho \rightarrow \eta \gamma$, $p\bar{p}\rightarrow \eta(\eta')M$, $J/\psi\rightarrow \eta\gamma$, $\eta(\eta')\rightarrow \gamma \gamma$ and $T\rightarrow P_1 P_2$, where $T$ is a $2^{++}$ tensor meson and $P_i$ is a pseudoscalar meson \cite{Feldmann:1998vh}, 
\begin{eqnarray}
\theta_{\rm FKS}=39.3^{\circ}\pm 1.0^{\circ},
\quad
f_{q}=(1.07\pm 0.02) f_{\pi},
\quad
f_{s}=(1.34\pm 0.06) f_{\pi}.
\label{eq:decay_constants}
\end{eqnarray}

For the purpose of our numerical determinations we calculate the coefficient $A_{M_1 M_2}$ in Eq.~(\ref{eq:AM1M2}) considering the form factors and decay constants included in Table~\ref{tab:FormFactors}. Moreover, the quantities relevant to the $\eta$ and $\eta'$ contributions are evaluated using Eq.~(\ref{eq:Feta}) and the inputs given in Eq.~(\ref{eq:decay_constants}). Finally, we find
\begin{eqnarray}
A_{M_1 M_2}=(1.25\pm 0.02)~{\rm GeV}^3.
\label{eq:AM1M@Average}
\end{eqnarray}

\begin{table}[htp]
\begin{center}
\begin{tabular}{ |c|c|c||c|c|c| } 
\hline
&&&&&\\
Parameter& Value & Reference & Parameter & Value & Reference\\ 
&&&&&\\
 \hline
  & $0.258\pm 0.031$ & \cite{Ball:2004ye} &&&\\
  & $0.25\pm 0.05$  & \cite{Khodjamirian:2006st} &&$0.331\pm 0.041$& \cite{Ball:2004ye} \\
  & $0.28\pm 0.05$ & \cite{Khodjamirian:2011ub} &&$0.31\pm 0.04$& \cite{Khodjamirian:2006st}\\  
 $F^{B\rightarrow \pi}_{0}(0)$ & $0.21\pm 0.07$  &  \cite{Gubernari:2018wyi}& $F^{B\rightarrow K}_{0}(0)$& $0.27\pm 0.08$&\cite{Gubernari:2018wyi}\\ 
  & $0.31\pm 0.02$  & \cite{SentitemsuImsong:2014plu}&& $0.395\pm 0.033$&\cite{Khodjamirian:2017fxg}\\ 
  &  $0.281\pm 0.038$ & \cite{Wang:2015vgv}&&$0.364\pm 0.05$&  \cite{Lu:2018cfc}\\
  &  $0.301\pm 0.023$& \cite{Khodjamirian:2017fxg}&&& \\
 \hline
 &&&&&\\
 $f_{\pi}~[\rm{GeV}]$  
 &
 $0.1302\pm 0.0008$
 &
 \cite{FlavourLatticeAveragingGroup:2019iem}
 &
 $f_{K}~[\rm{GeV}]$&$0.1557\pm 0.0007$& \cite{FlavourLatticeAveragingGroup:2019iem}\\
 &&&&&\\
 \hline
\end{tabular}
\end{center}
\caption{Form factors for the transitions $B\rightarrow \pi$ and $B\rightarrow K$ and decay constants $f_{\pi}$ and $f_{K}$. 
\label{tab:FormFactors}}
\end{table}


 \section{Details on the $\chi^2$-fit}
\label{sec:fit}

 To assess the set of values for the $SU(3)$-invariant amplitudes compatible with experimental data we perform a $\chi^2$-fit.
Thus, we establish the compatibility between the experimental data and our theoretical determinations by evaluating the function
 \begin{eqnarray}
 \chi^2=\sum_{i} \Bigl(\frac{\mathcal{O}_i^{\rm Theo}-\mathcal{O}_i^{\rm Exp}}{\sigma_i^{\rm Exp}}\Bigl)^2,
 \label{eq:chi2}
 \end{eqnarray}
 where $\mathcal{O}$ is either a branching fraction or a CP-asymmetry and the labels   ``Theo''  and  ``Exp'' make reference to the corresponding theoretical or experimental determinations. Moreover the sum runs over all the physical observables which have been measured. Our list of experimental inputs are presented in Tables \ref{tab:tableBr_combined} and~\ref{tab:tableAs_combined}. We exclude from the $\chi^2$-function quantities which either are not experimentally available, or where only an upper bound exists. Actually, the upper bounds will be considered as constraints during our fitting procedure, more details on this below.

We start by determining our relevant degrees of freedom. As discussed before there are in principle $10$ complex tree-like amplitudes and $10$ complex penguin-like amplitudes leading to a total of 40 real quantities, however $A^T_6$ can be absorbed into $B^T_6$ and $C^T_6$,  correspondingly $A^P_6$ can be absorbed into $B^P_6$ and $C^P_6$. Moreover we can get rid of a global phase by taking $C^P_3$ as a real parameter. Finally,  the study of $\eta-\eta'$ mixing requires an additional real quantity, the mixing angle $\theta_{FKS}$. Consequently, our parameter space consist of 36-dimensional vectors.
 
 There are two stages when it comes to determine the values of the $SU(3)$-invariant quantities from experimental data. First, we extract our best fit-point and second we will calculate the allowed regions about this particular point at a given confidence level.

  During the first stage we follow the standard approach and parameterize the tree and penguin amplitudes in terms of their modulus $|A_3|$, $|C_3|$, $|C_6|$, $|A_{15}|$, $|C_{15}|$, $|B_3|$, $|B_6|$, $|B_{15}|$, $|D_3|$ and their corresponding phases $\delta_{A_3}$, $\delta_{C_3}$, $\delta_{C_6}$, $\delta_{A_{15}}$, $\delta_{C_{15}}$, $\delta_{B_3}$, $\delta_{B_6}$, $\delta_{B_{15}}$, $\delta_{D_3}$ plus the $\eta-\eta'$ mixing angle $\theta_{FKS}$.  To determine the best fit point  we use random sampling, i.e.\ we generate randomly  $10^{9}$ points in our 36-dimensional space obeying a flat probability distribution. We then calculate the  $\chi^2$ function for each one of these points. We divide our $10^{9}$ evaluations into  $1000$ individual runs of $10^{6}$-points each. For each individual slot of $10^{6}$ points we select the top 5 points with the lowest $\chi^2$ determined previously. Then for each one of our  $1000$ slots, we apply the Sequential Least Square Programming algorithm, SLSQP, available in python, taking as starting points those with the corresponding  5 minimal $\chi^2$. Finally, the overall minimum of all the SLSQP-minimizations is taken  as our best-fit point.
 
 Based on different numerical tests we find that the random sampling procedure converges quickly to the region of minimum values if we generate points for $\{|C^{T}_3|,$
 $|C^{T}_6|, |C^{T}_{15}|,$ $|B^{T}_{6}|, |B^{T}_{15}|, |D^{T}_{3}|,  |C^{P}_{3}|, |C^{P}_{6}|, |C^{P}_{15}|,  |B^{P}_6|, |B^{P}_{15}|, |D^{P}_3|\}$
 inside the interval $[0, 0.3]$ and for $\{|A^{T}_{3}|, |A^{T}_{15}|,|B^{T}_{3}|, |A^{P}_{3}|, |A^{P}_{15}|, |B^{P}_{3}|\}$ inside the region $[0, 0.1]$. Finally,  for the phases we sample random values within $[-\pi, \pi]$. 
 
 Performing the $\chi^2$-fit based only on the experimental data
 leads to a rather unconstrained scenario with a large number of fit parameters. To improve the situation we opt for introducing a minimal set of extra conditions arising within the context of QCDF, while at the same time keeping the theoretical bias from that approach as marginal as possible. Currently, the color-allowed tree amplitude $\alpha_1$ has been calculated including corrections up to NNLO leading for the $B\rightarrow \pi \pi$ process to \cite{Beneke:2009ek}
 \begin{eqnarray}
 \alpha_{1}(\pi\pi)=1.000^{+0.029}_{-0.069}+(0.011^{+0.023}_{-0.050})i.
 \label{eq:alpha1}
 \end{eqnarray}
It can be seen from Eq.~(\ref{eq:alpha1}) that the real part is known with a precision better than $7\%$, and the study in~\cite{Beneke:2009ek} further reveals that it is robust against large higher-order perturbative corrections. Hence, during our statistical analysis we impose that our determination for the real part of $\alpha_{1}(\pi\pi)$ from the $SU(3)$ parameters is constrained to lie within the region
 \begin{eqnarray}
 \Re(\alpha_{1})=1.000^{+0.138}_{-0.138},
 \label{eq:alpharegions}
 \end{eqnarray}
 where we have adopted a conservative approach by considering a symmetric uncertainty which amounts to twice the theoretical value presented  in Eq.~(\ref{eq:alpha1}).
 
 According to Eq.~(\ref{eq:SU£toTop}), to map out the QCDF amplitude $\alpha_{1}$ into the $SU(3)$ evaluations we have to take into account the extra factor $A_{M_1 M_2}$ whose numerical value is presented in Eq.~(\ref{eq:AM1M@Average}). Thus, our full constraint reads
  \begin{eqnarray}
 \Re(A_{M_1 M_2}\alpha_1)=(1.245\pm 0.173)~{\rm GeV}^3.
 \label{eq:consalpha1}
 \end{eqnarray}
 
 Additionally, based on state of the art experimental determinations,  we require that the following bounds are obeyed during the minimization procedure
 \begin{eqnarray}
 Br(B_s\rightarrow \pi^0 \pi^0)<2.10\times 10^{-4},&~~~&
 Br(B_s\rightarrow \eta \pi^0)< 10^{-3},\nonumber\\
 Br(B^0\rightarrow \eta\eta)<10^{-6},&~~~&
 Br(B^0\rightarrow \eta'\eta')<1.7\times 10^{-6},\nonumber\\
 Br(B^0\rightarrow \eta' \eta)<1.2\times 10^{-6},&~~~&A_{CP}(B_s\rightarrow \eta K^0)<10^{-3}.
 \label{eq:inequalities}
 \end{eqnarray}
 
 Our main goal is to evaluate the size of the QCDF amplitudes $\beta_{i}$ as allowed by data, without introducing too much a priori information on the size of these quantities. We then perform a study including as constraints Eqs.~(\ref{eq:alpharegions})~--~(\ref{eq:inequalities}). In addition, based on the results in Eq.~(\ref{eq:TopologicalQCDF}) we impose the following conditions (dimensionful quantities are given in units of GeV${}^3$)
     \begin{eqnarray}
     T_{PA}=T_{SS}=T_{S}=0,&&|T_{P}|<0.1.
     \label{eq:constraints}
     \end{eqnarray}
 With the help of Eqs.~(\ref{eq:ToptoSU3}) we can easily translate Eqs.~(\ref{eq:constraints}) into constraints for the $SU(3)$-invariant parameters which are the quantities that we are actually fitting to experimental data. We would like to highlight the fact that we are being conservative with respect to the bound imposed over $T_{P}$ by enlarging the upper limit established by the magnitudes of the differences $\alpha^c_4-\alpha^u_4$ and $\alpha^c_{4, EW}-\alpha^u_{4, EW}$ by a factor of five.

 For the determination of the confidence intervals of our $SU(3)$ parameters, we perform a likelihood ratio test and determine the $p$  value applying Wilk's theorem.  In principle we can establish the confidence intervals for the modulus and phases of the relevant $SU(3)$-invariant quantities. However, we find that it is more informative to study the allowed regions in the space expanded by the real and the imaginary components of each one of the $SU(3)$-invariant amplitudes. Hence, the confidence level is estimated considering 
 \begin{eqnarray}
 1-\frac{\Gamma(\Delta\chi^2/2, \nu/2 )}{\Gamma(\nu/2)}  = 1-p,
 \end{eqnarray}
 where $\Gamma$ is the normalised upper incomplete Gamma function, $\nu$ represents the number of degrees of freedom and
 \begin{eqnarray}
 \Delta \chi^2=\chi^2_{0}-\chi_{min}^2.
 \label{eq:Deltachi}
 \end{eqnarray}
 In Eq.~(\ref{eq:Deltachi})  $\chi^2_{min}$ is the global minimum for the $\chi^2$ function and $\chi^2_{0}$ is the minimum subjected to the null-hypothesis over the value of one of the $SU(3)$-invariant parameters.

Since we are fitting for the real and imaginary parts of a complex quantity simultaneously $\nu=2$. Moreover, we take $1-p=0.68$, implying that for our confidence regions
\begin{eqnarray}
\Delta \chi^2 \leq 2.3.   
\end{eqnarray}
 
 \section{Results}
 \label{sec:results}
 
 \subsection{$SU(3)$-irreducible amplitudes}
 
 Let us now give the results of our $\chi^2$-fit. We start with the best-fit point of the $SU(3)$-irreducible amplitudes in polar coordinates (the modulus are given in units of GeV${}^3$),
 \begin{align}
 |A^T_{3}|&=0.029, \qquad &
 \delta_{A^T_3}&=-3.083, \qquad\qquad&
 |C^T_{3}|&=0.258, \qquad &
 \delta_{C^T_{3}}&=-0.105,\nonumber\\
 |C^T_{6}|&=0.235, &
 \delta_{C^T_{6}}&=-0.079, &
 |A^T_{15}|&=0.029, &
 \delta_{A^T_{15}}&=-3.083,\nonumber\\
 |C^T_{15}|&=0.151, &
 \delta_{C^T_{15}}&=0.061, &
 |B^T_3|&=0.034,&
 \delta_{B^T_3}&=3.087,\nonumber\\
 |B^T_6|&=0.033, &
 \delta_{B^T_6}&=-0.286, &
 |B^T_{15}|&=0.008, &
 \delta_{B^T_{15}}&=-1.892,\nonumber\\
 |D^T_3|&=0.055, &
 \delta_{D^T_3}&=2.942, & &&&
 \end{align}
 \begin{align}
 |A^P_{3}|&=0.014,&\qquad
 \delta_{A^P_3}&=-1.328, \qquad\qquad &
 |C^P_{6}|&=0.145, \qquad&
 \delta_{C^P_{6}}&=-2.881,\nonumber\\
 |A^P_{15}|&=0.003, &
 \delta_{A^P_{15}}&=2.234, &
 |C^P_{15}|&=0.003, &
 \delta_{C^P_{15}}&=-0.608,\nonumber\\
 |B^P_3|&=0.043,&
 \delta_{B^P_3}&=2.367,&
 |B^P_6|&=0.099,&
 \delta_{B^P_6}&=0.353,\nonumber\\
 |B^P_{15}|&=0.031,&
 \delta_{B^P_{15}}&=-0.690,&
 |D^P_{3}|&=0.030,&
 \delta_{D^P_{3}}&=0.477,\nonumber\\
 |C^P_3|&=0.008,&
 \theta_{FKS}&=0.628. & &&&
 \end{align}
 \begin{figure}[htp]
    \centering
    \includegraphics[width=7cm]{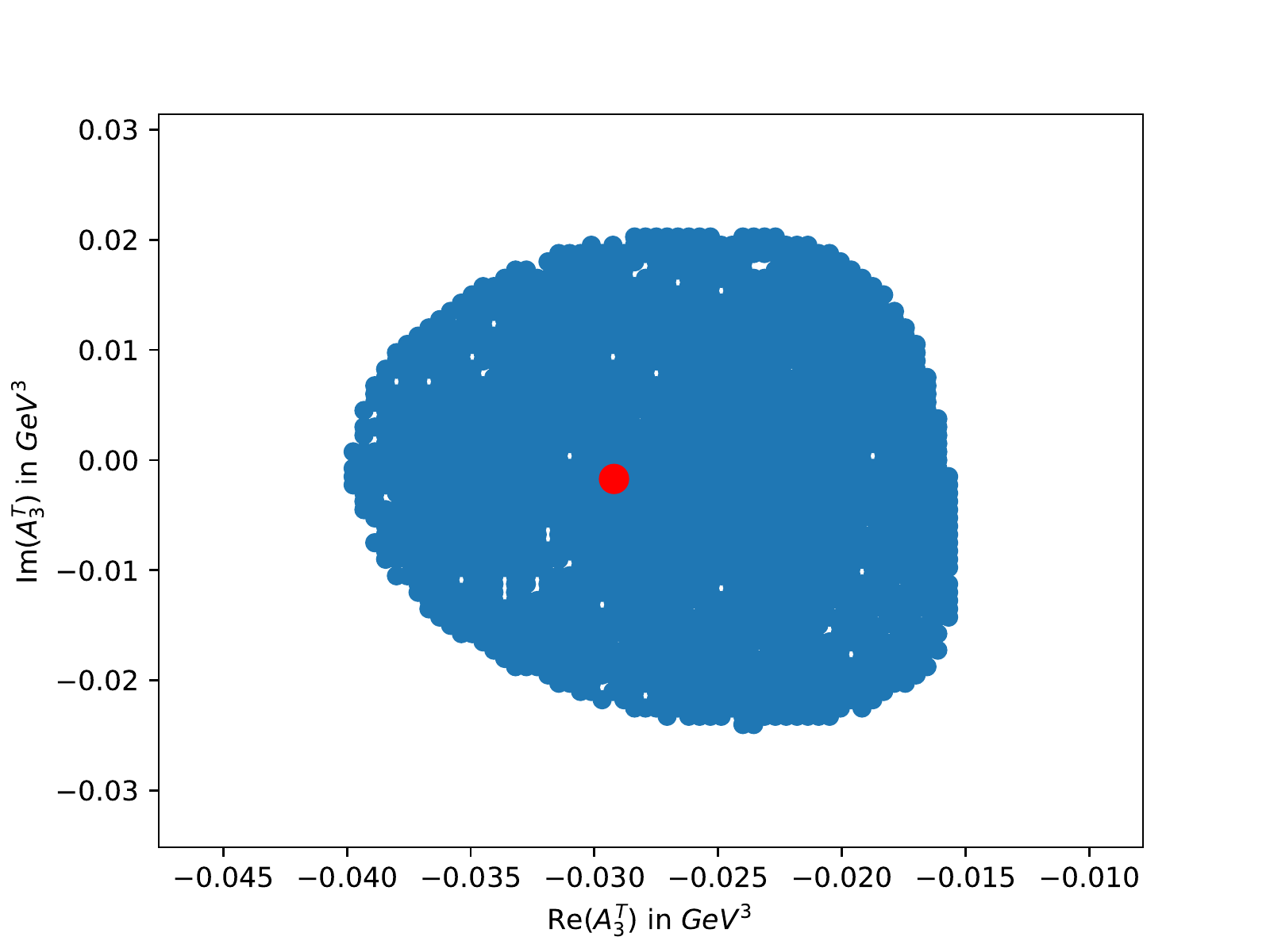}
    \includegraphics[width=7cm]{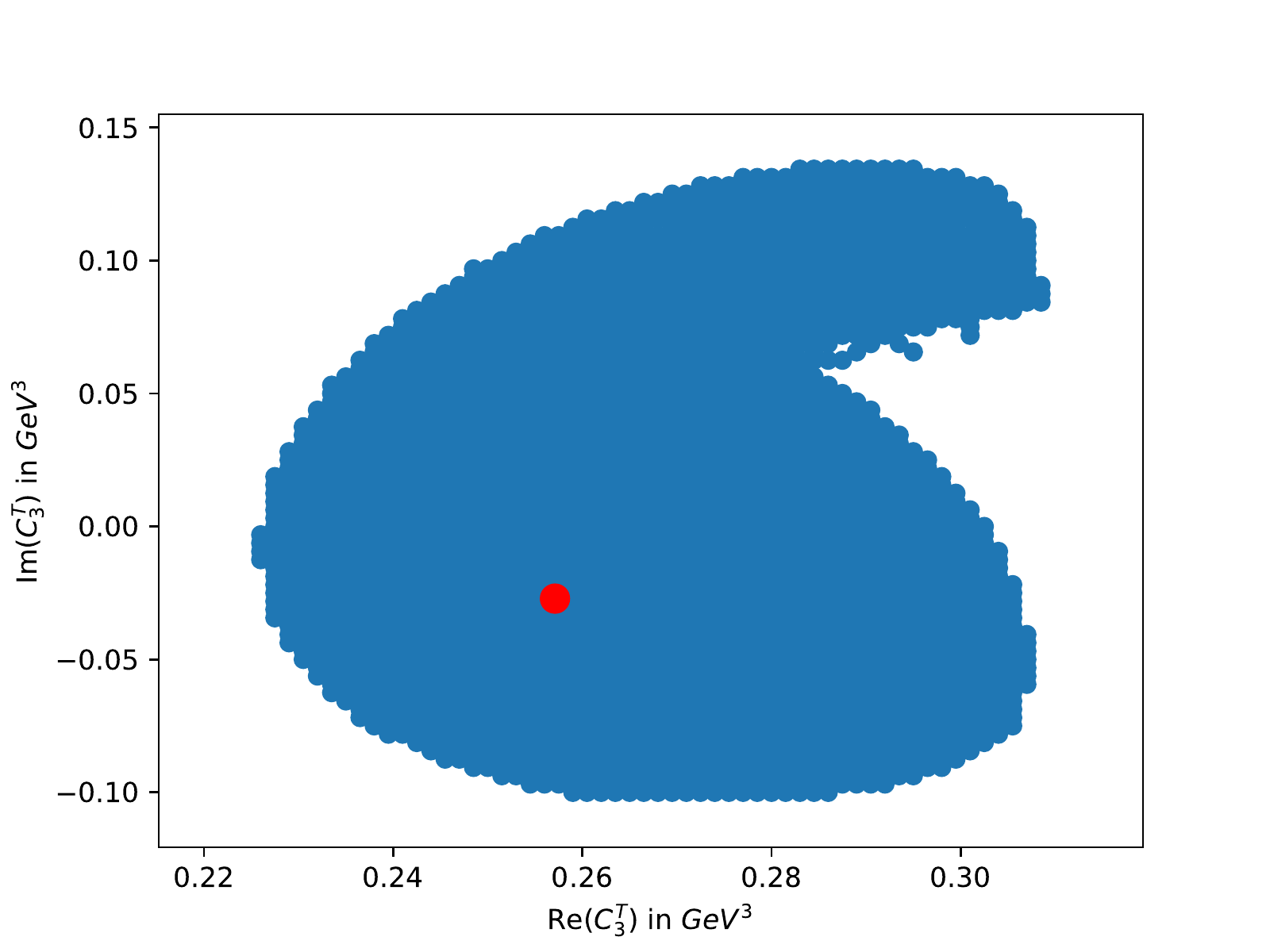}
    \includegraphics[width=7cm]{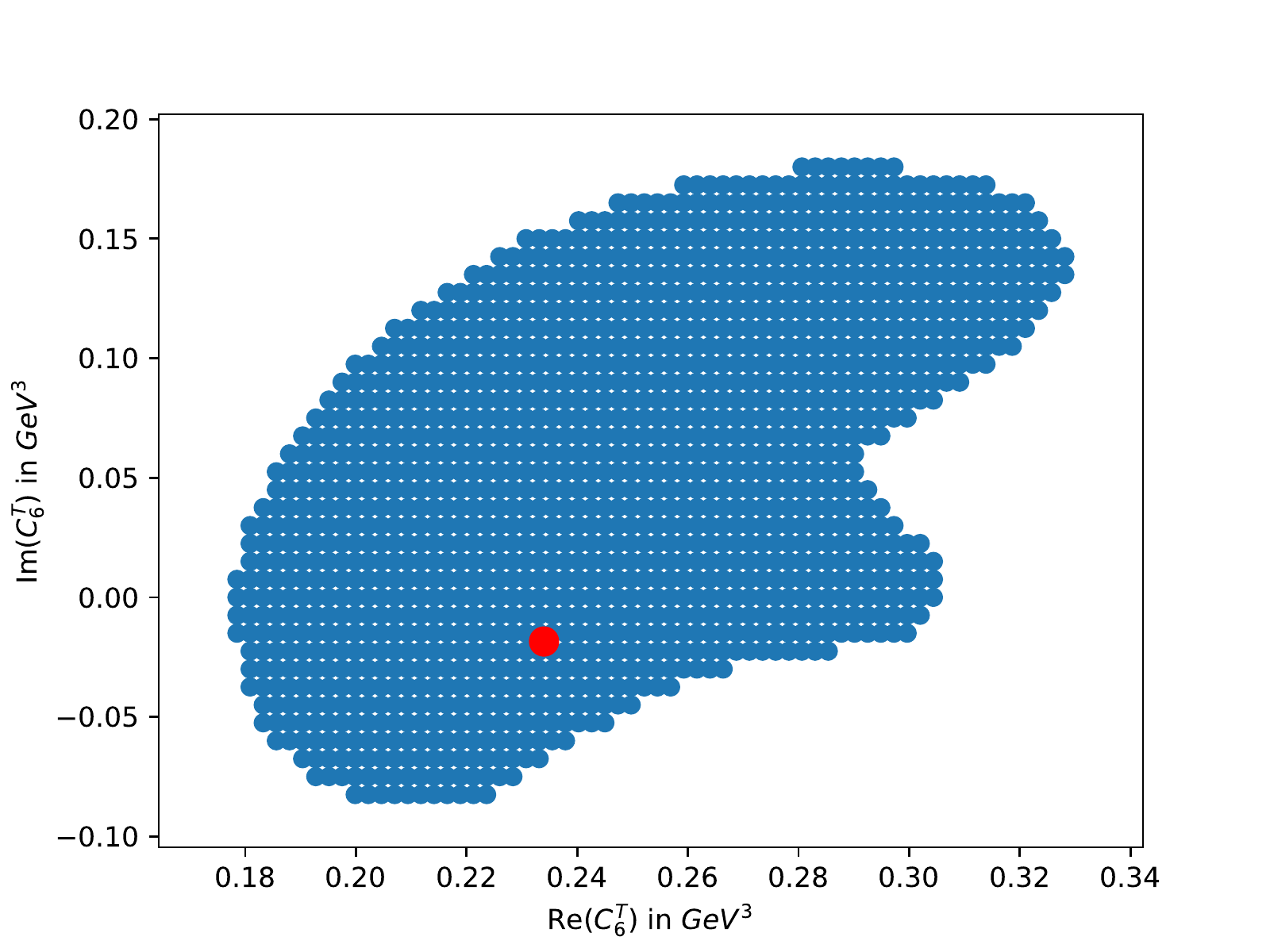}
    \includegraphics[width=7cm]{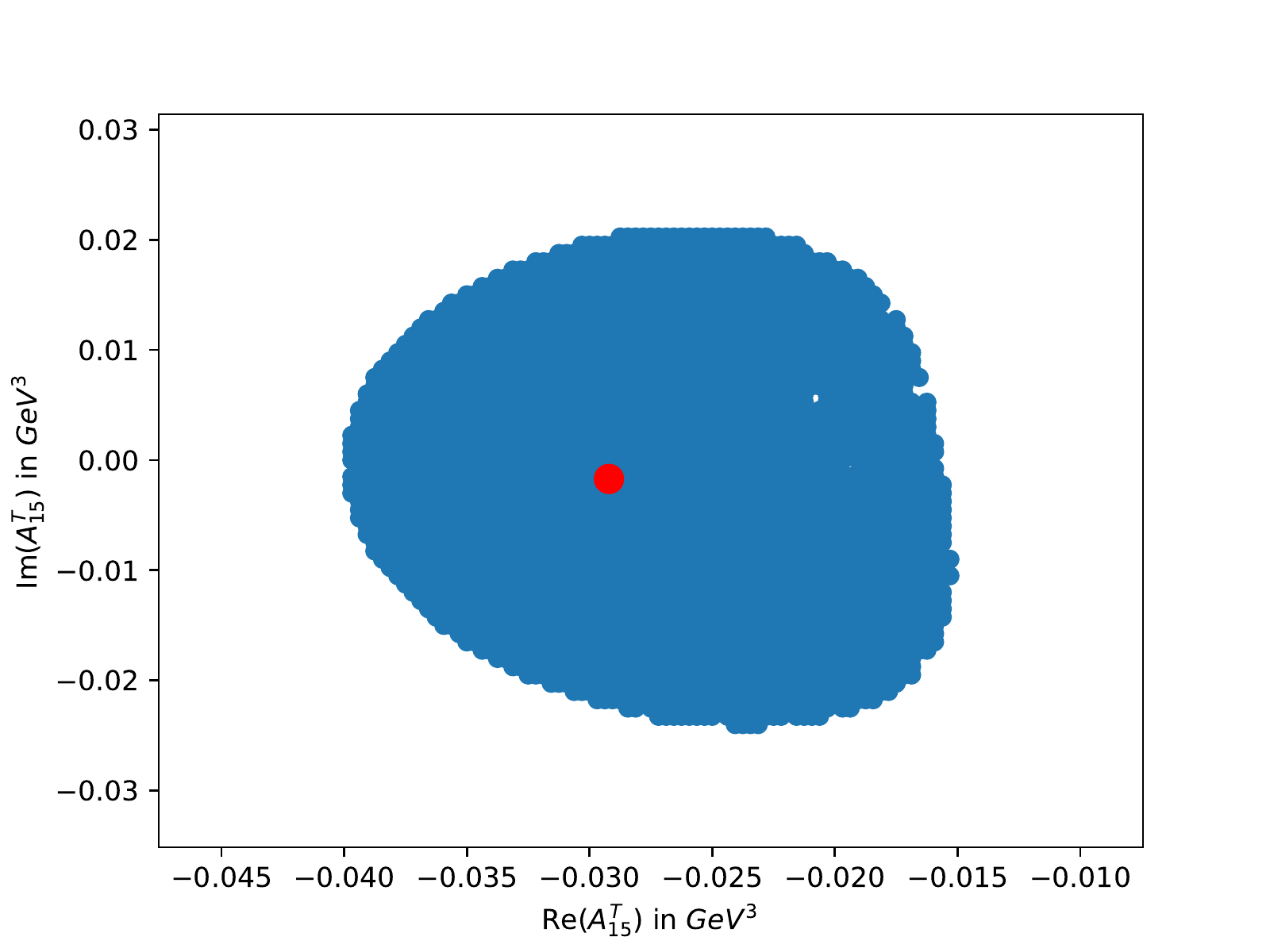}
    \includegraphics[width=7cm]{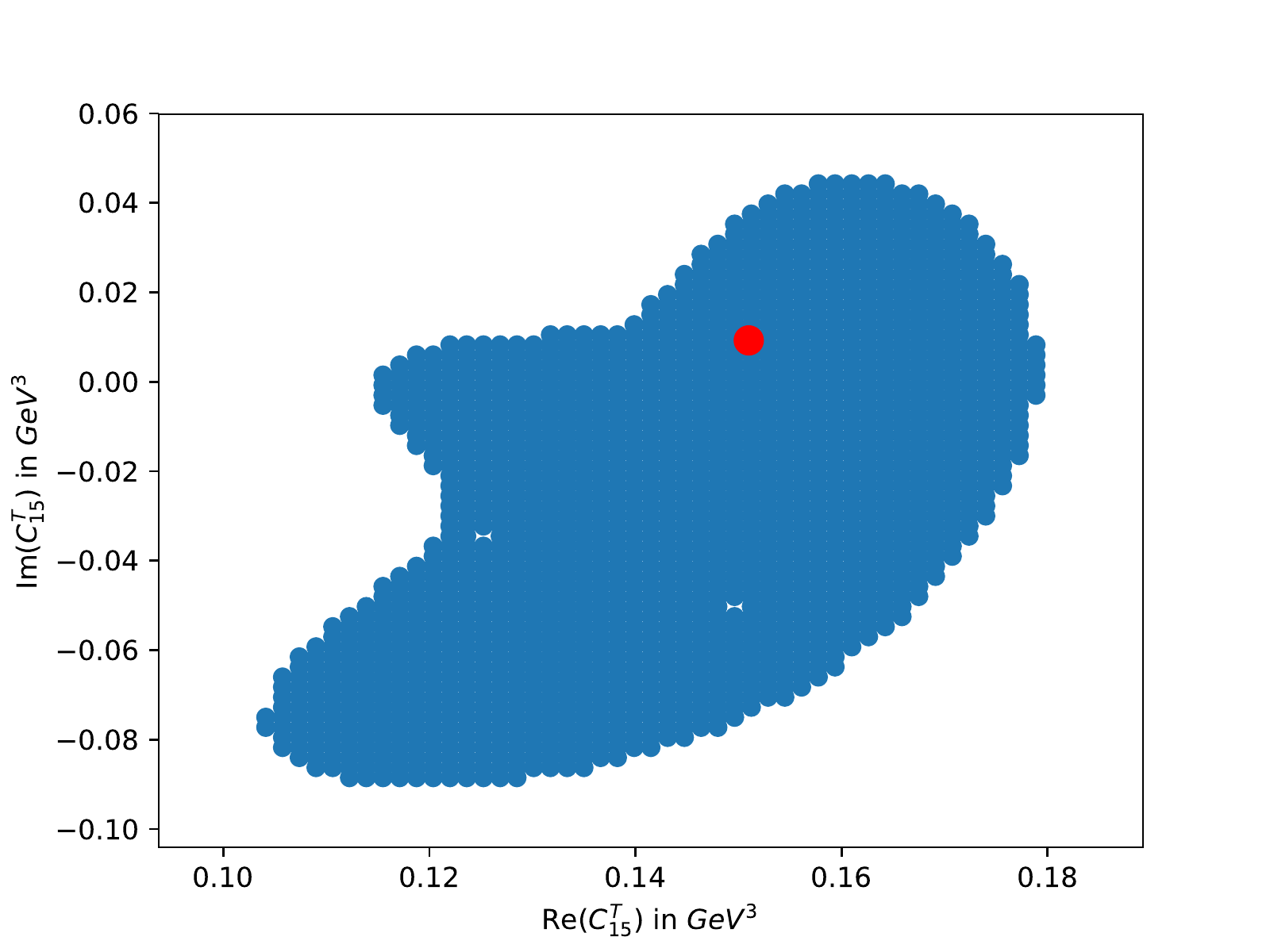}
    \includegraphics[width=7cm]{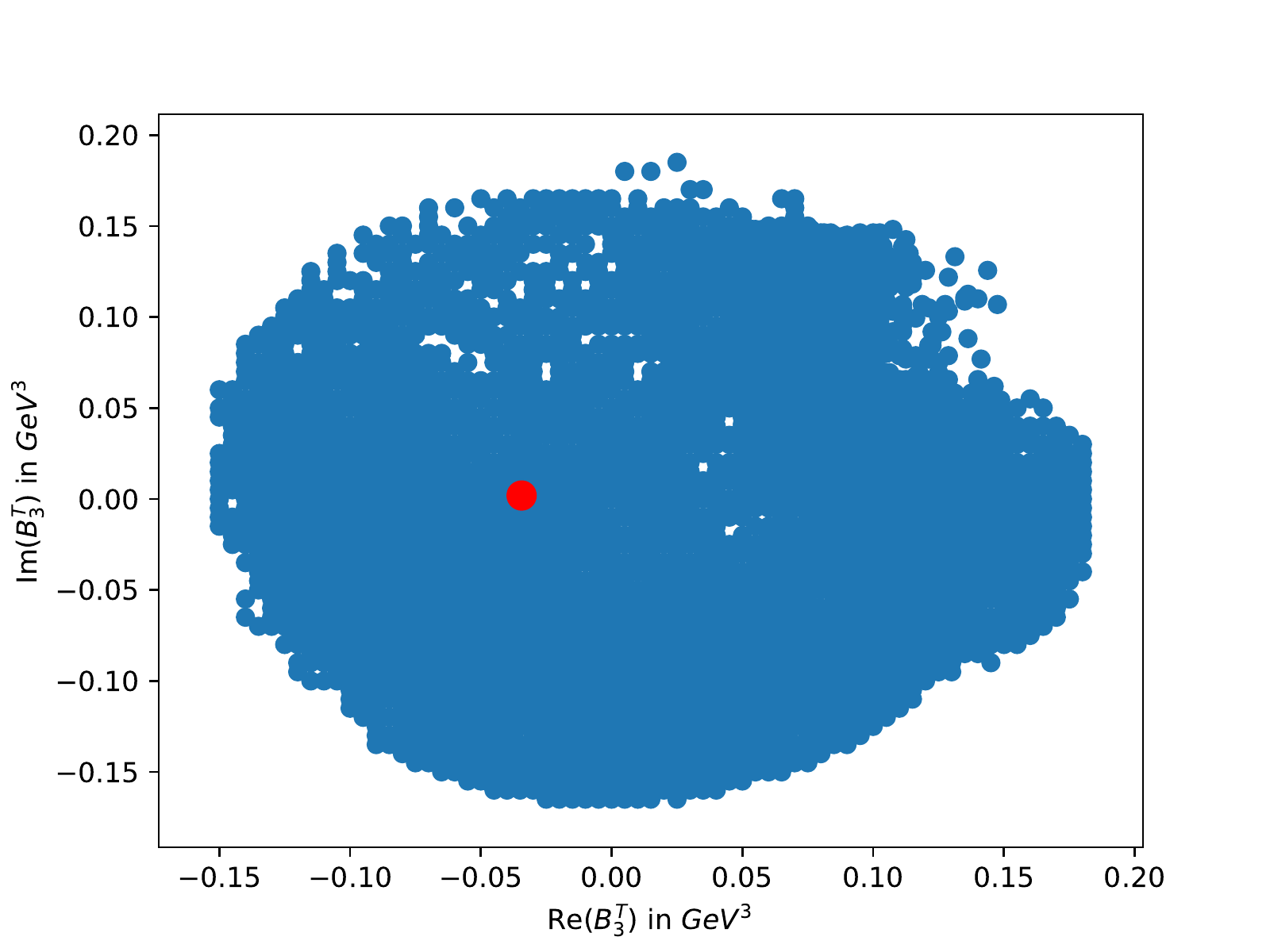}
    \includegraphics[width=7cm]{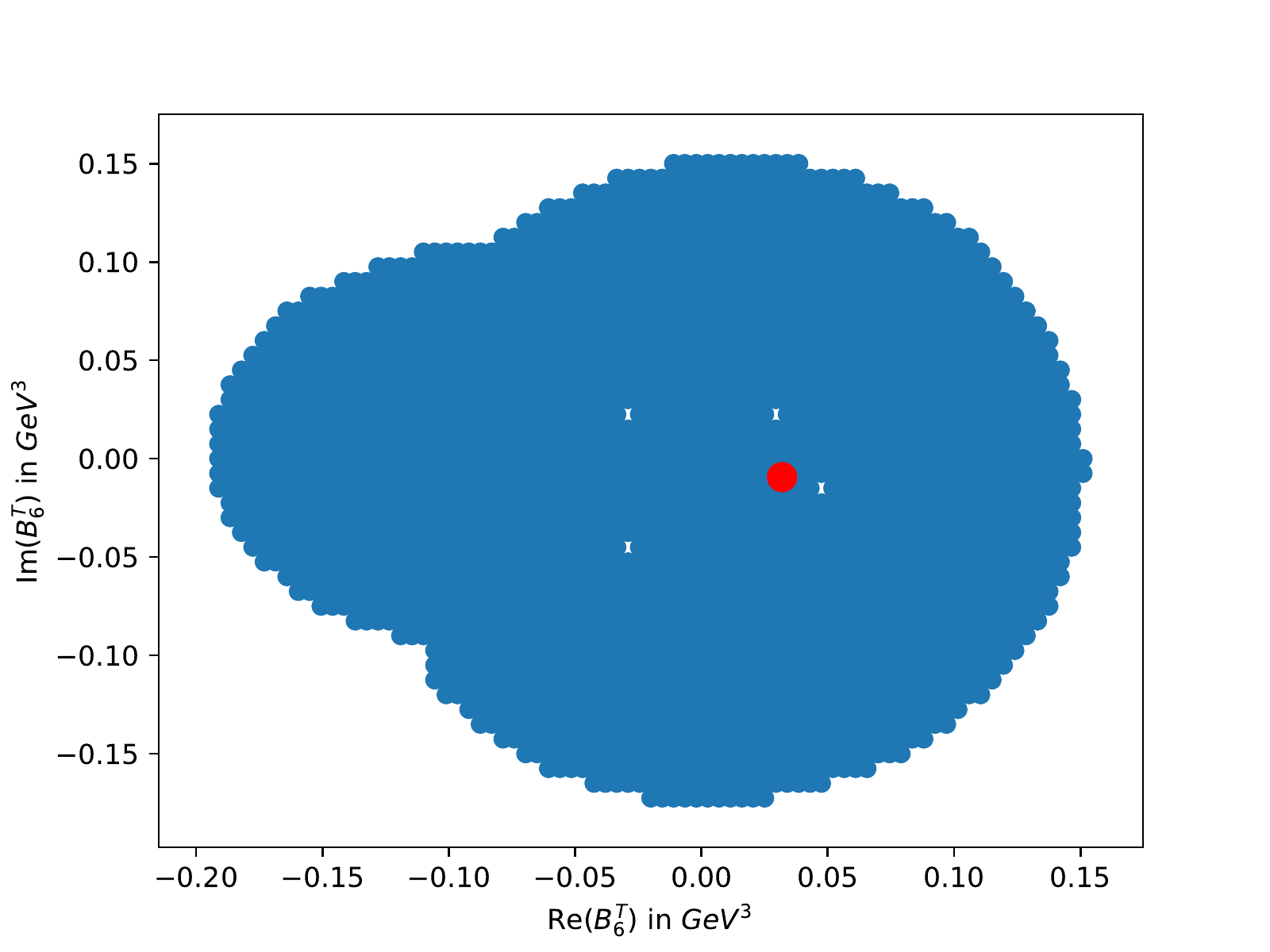}
    \includegraphics[width=7cm]{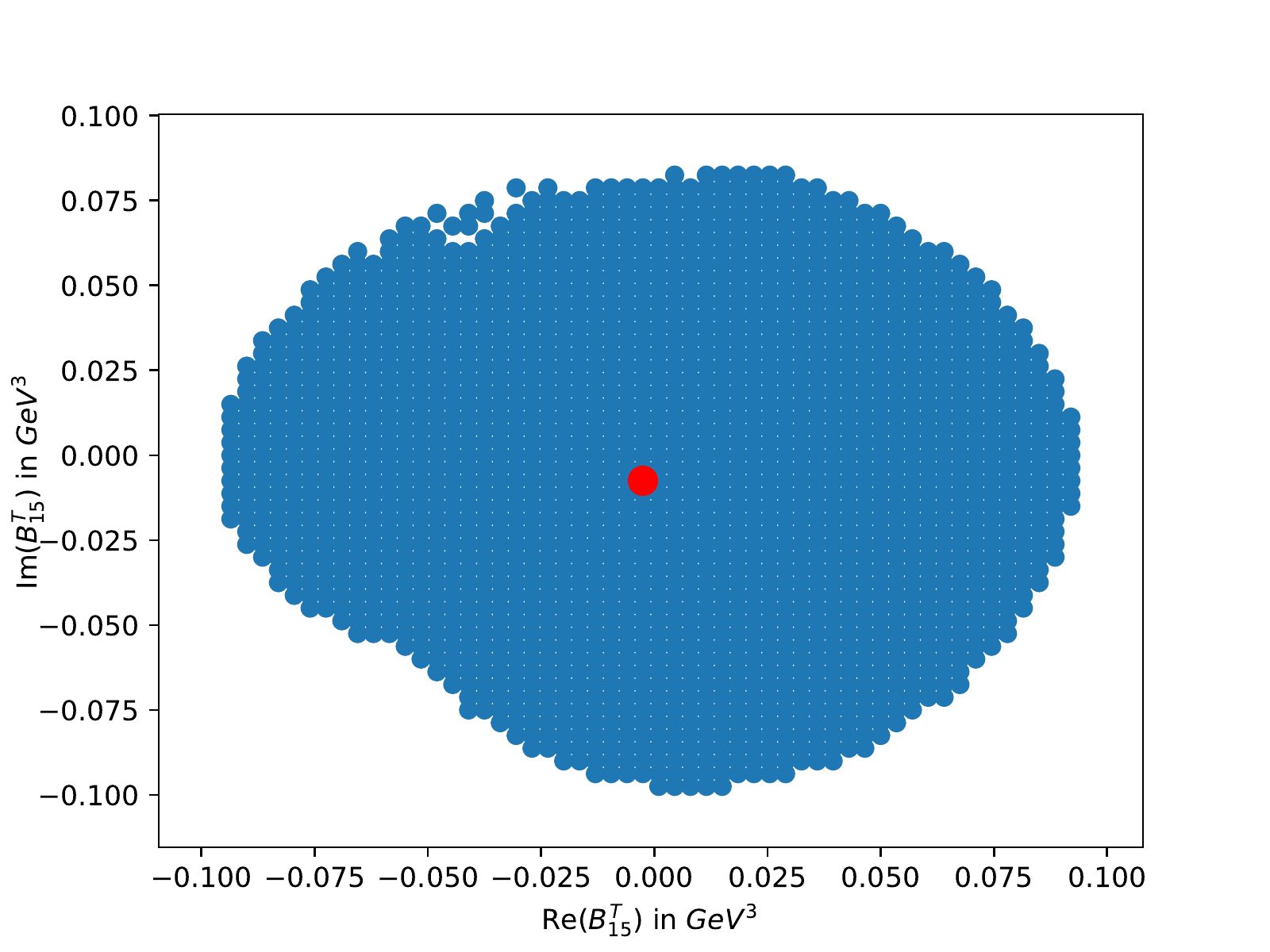}
    \caption{$SU(3)$ fit, part1. Amplitudes carry units of GeV${}^3$.\label{fig:SU3_1_1}}
\end{figure}
\begin{figure}[htp]
    \centering
    \includegraphics[width=7cm]{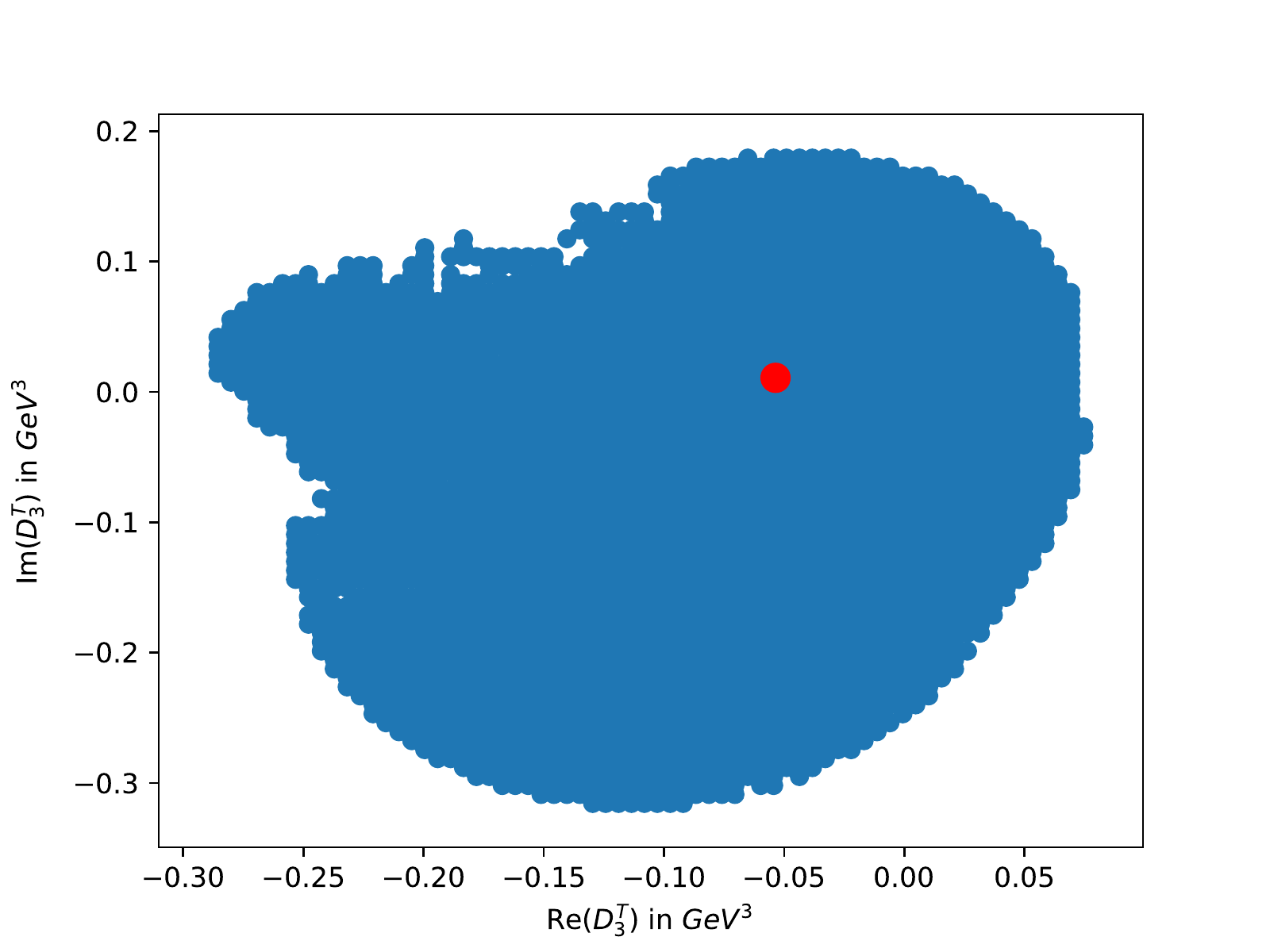}
    \includegraphics[width=7cm]{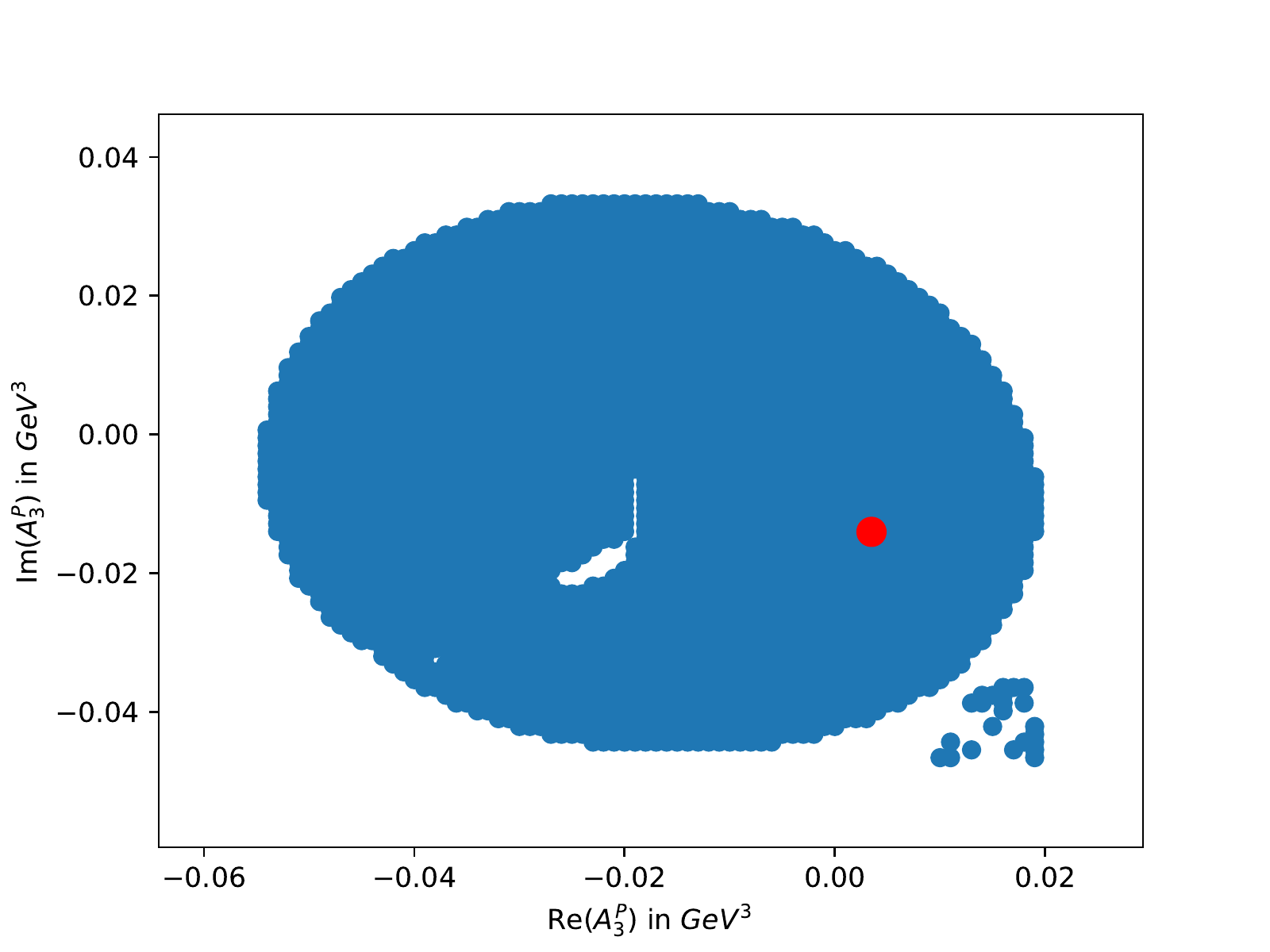}
    \includegraphics[width=7cm]{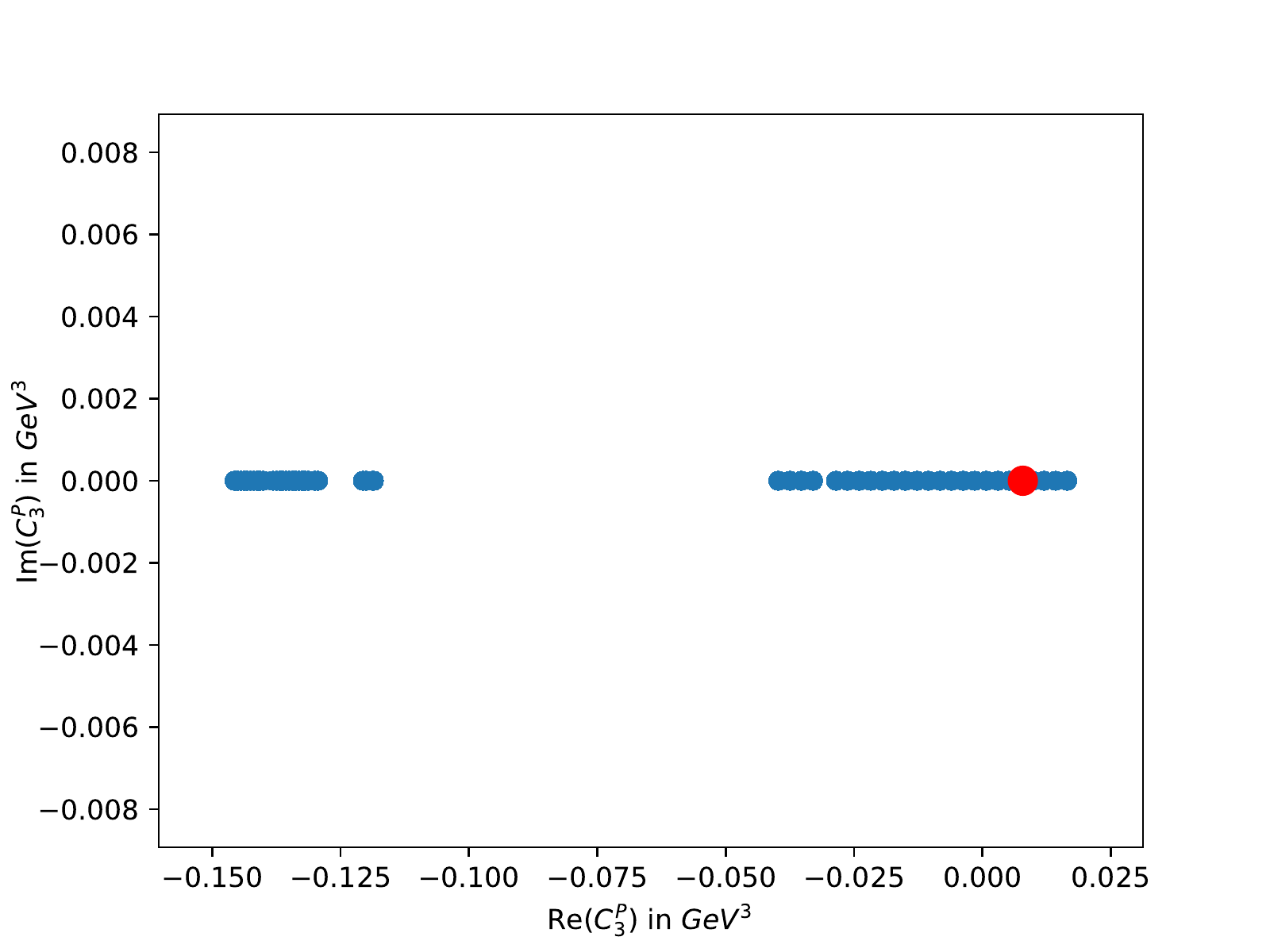}
    \includegraphics[width=7cm]{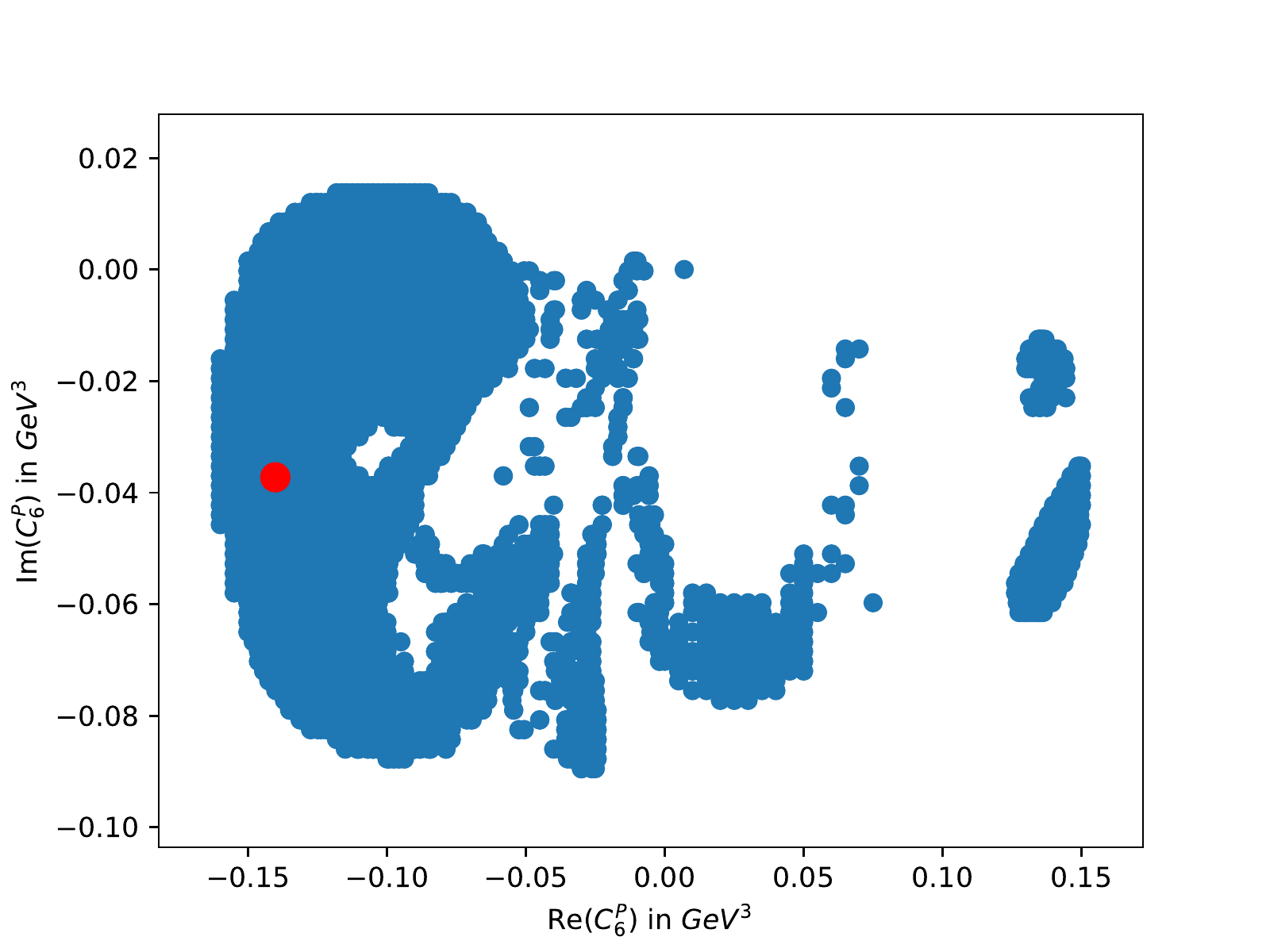}
    \includegraphics[width=7cm]{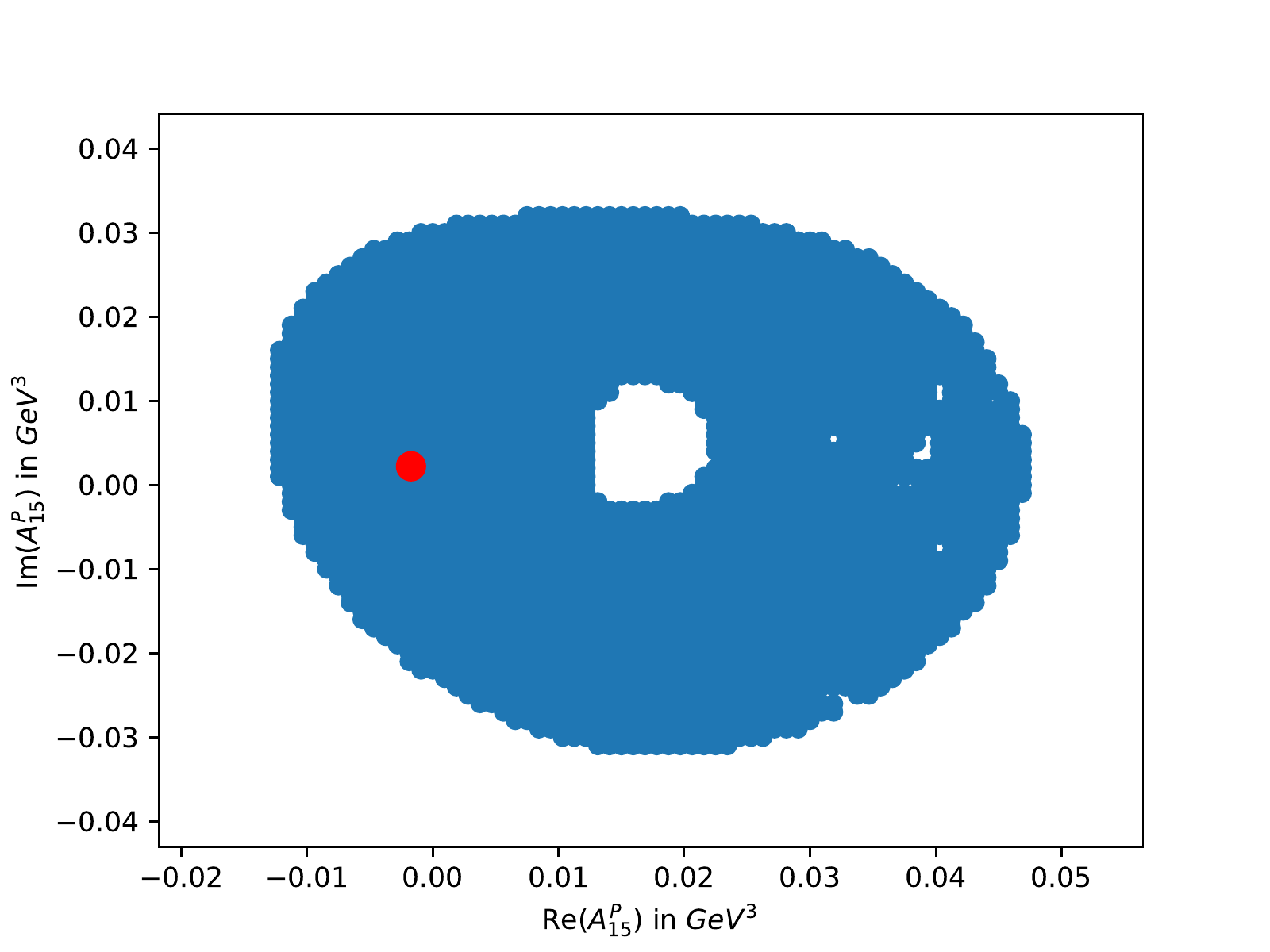}
    \includegraphics[width=7cm]{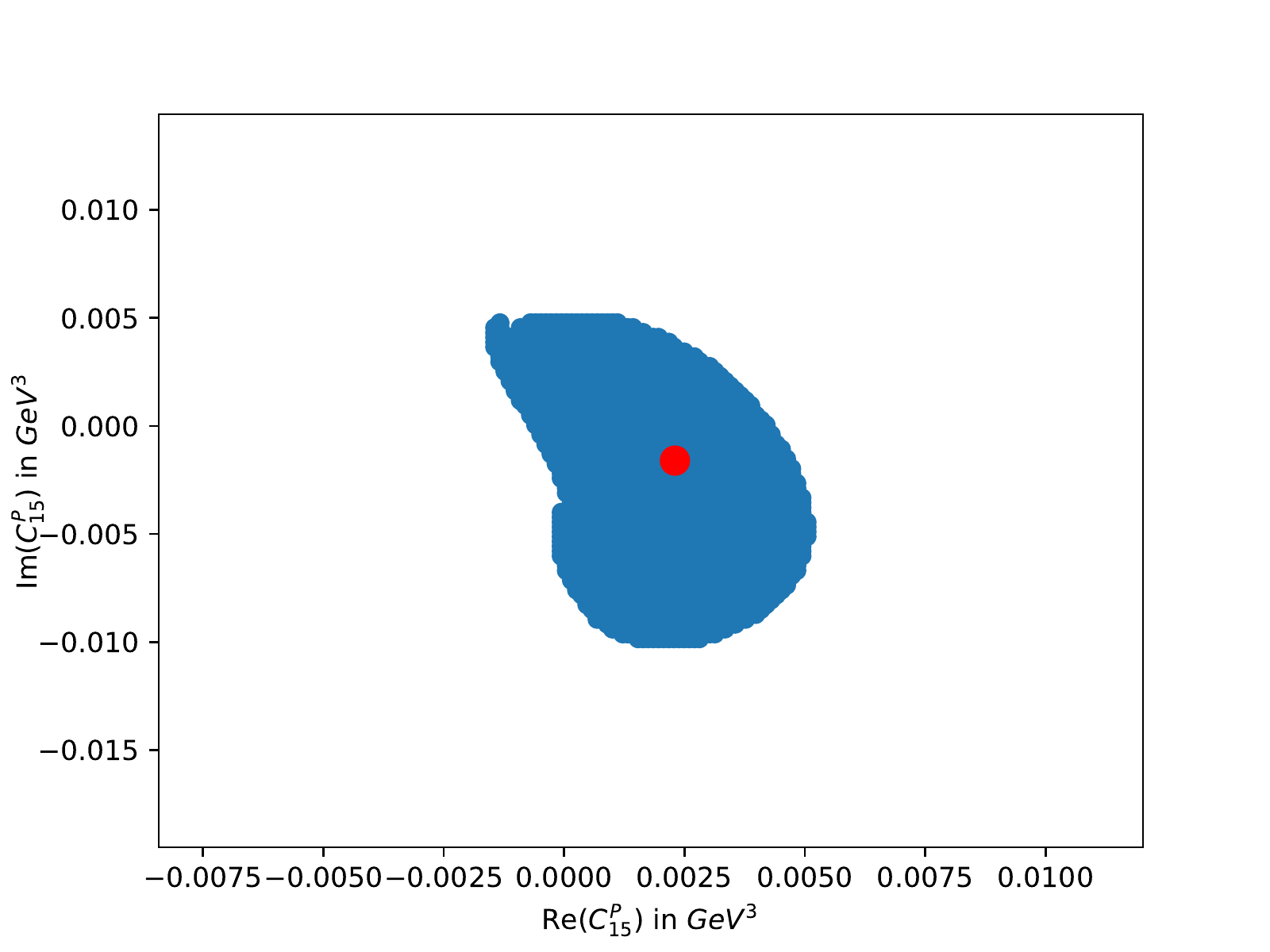}
    \includegraphics[width=7cm]{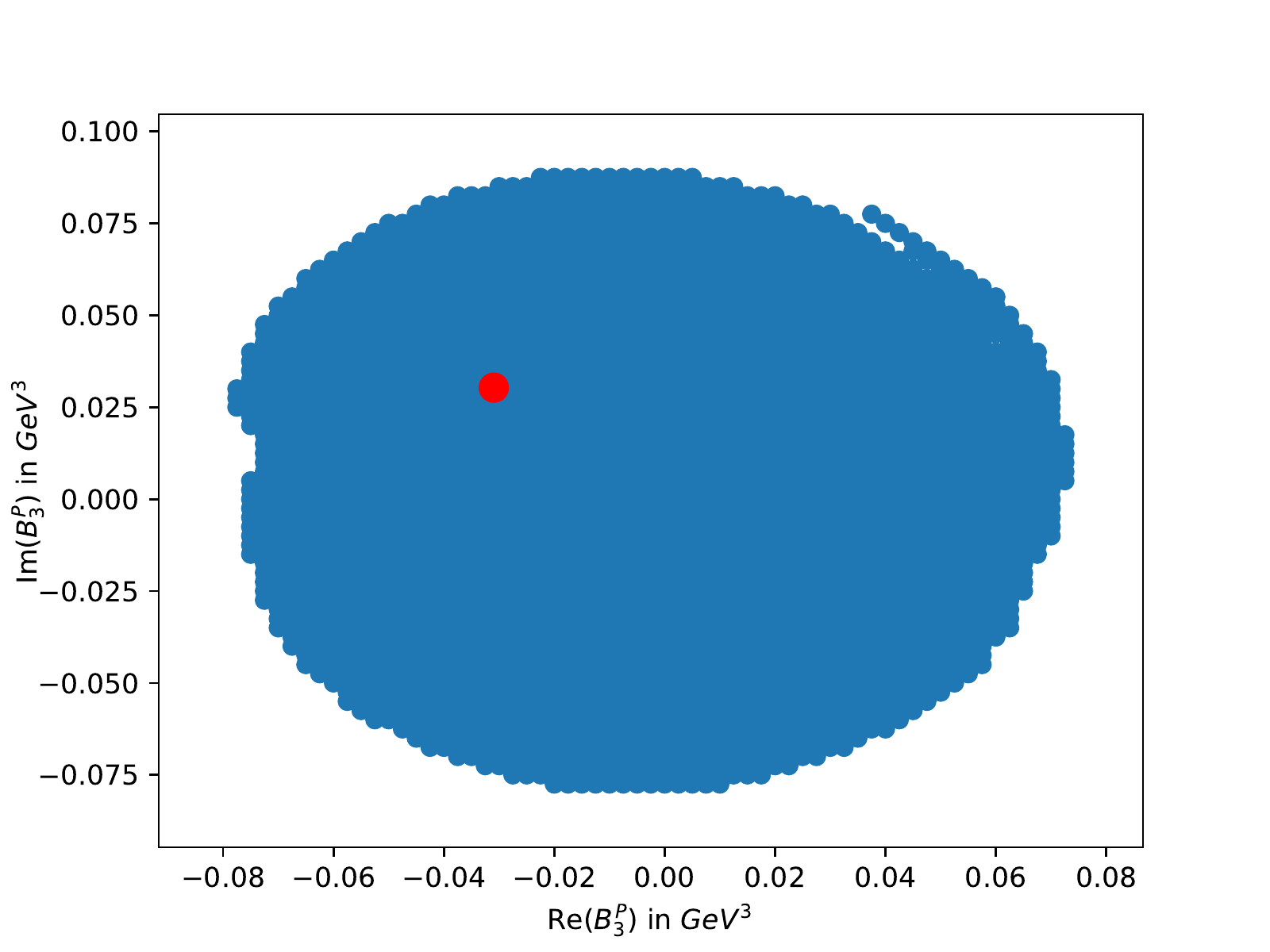}
    \includegraphics[width=7cm]{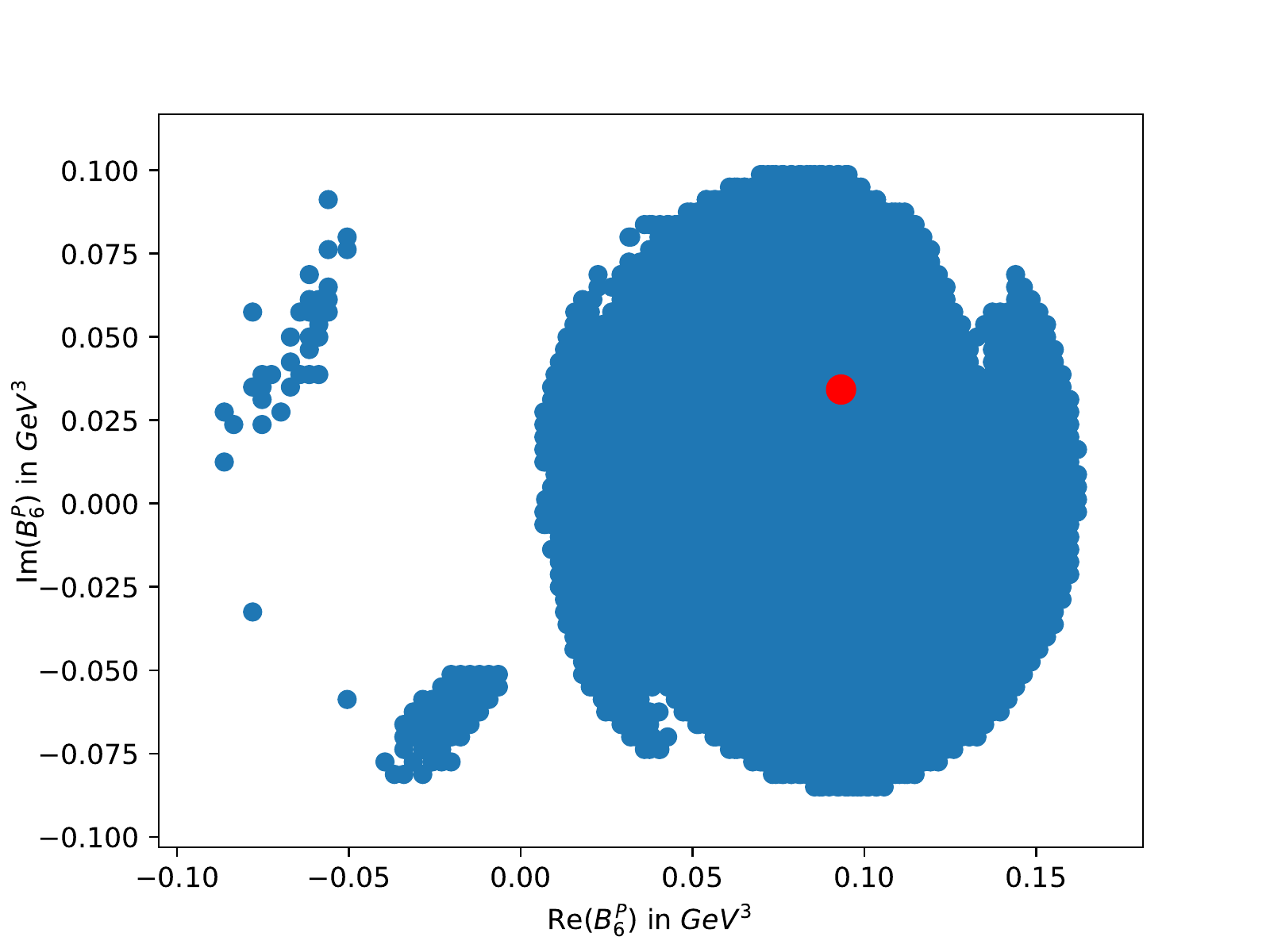}
    \caption{$SU(3)$ fit, part2. Amplitudes carry units of GeV${}^3$.\label{fig:SU3_2_1}}
\end{figure}
\begin{figure}[htp]
    \centering
    \includegraphics[width=7cm]{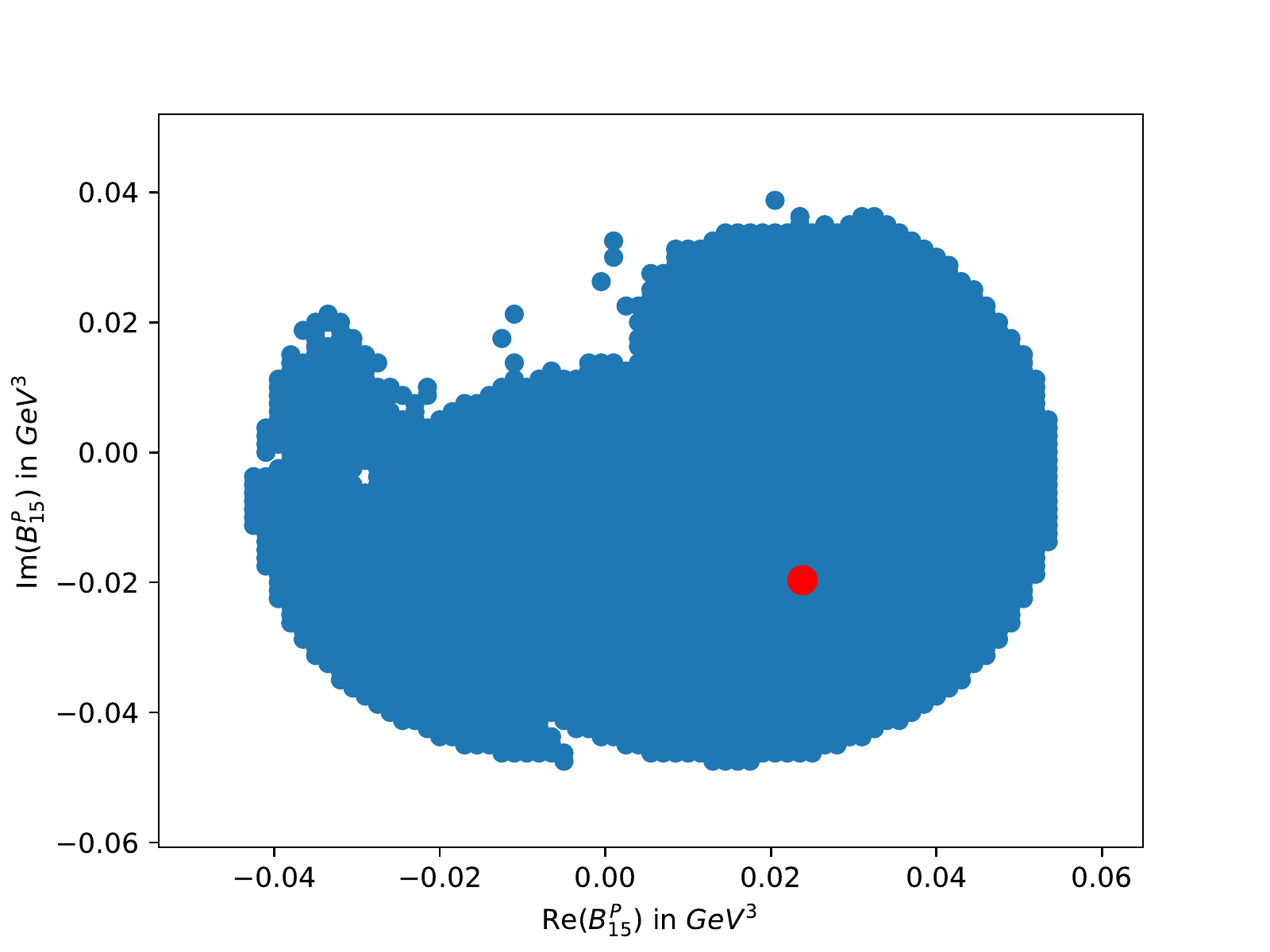}
    \includegraphics[width=7cm]{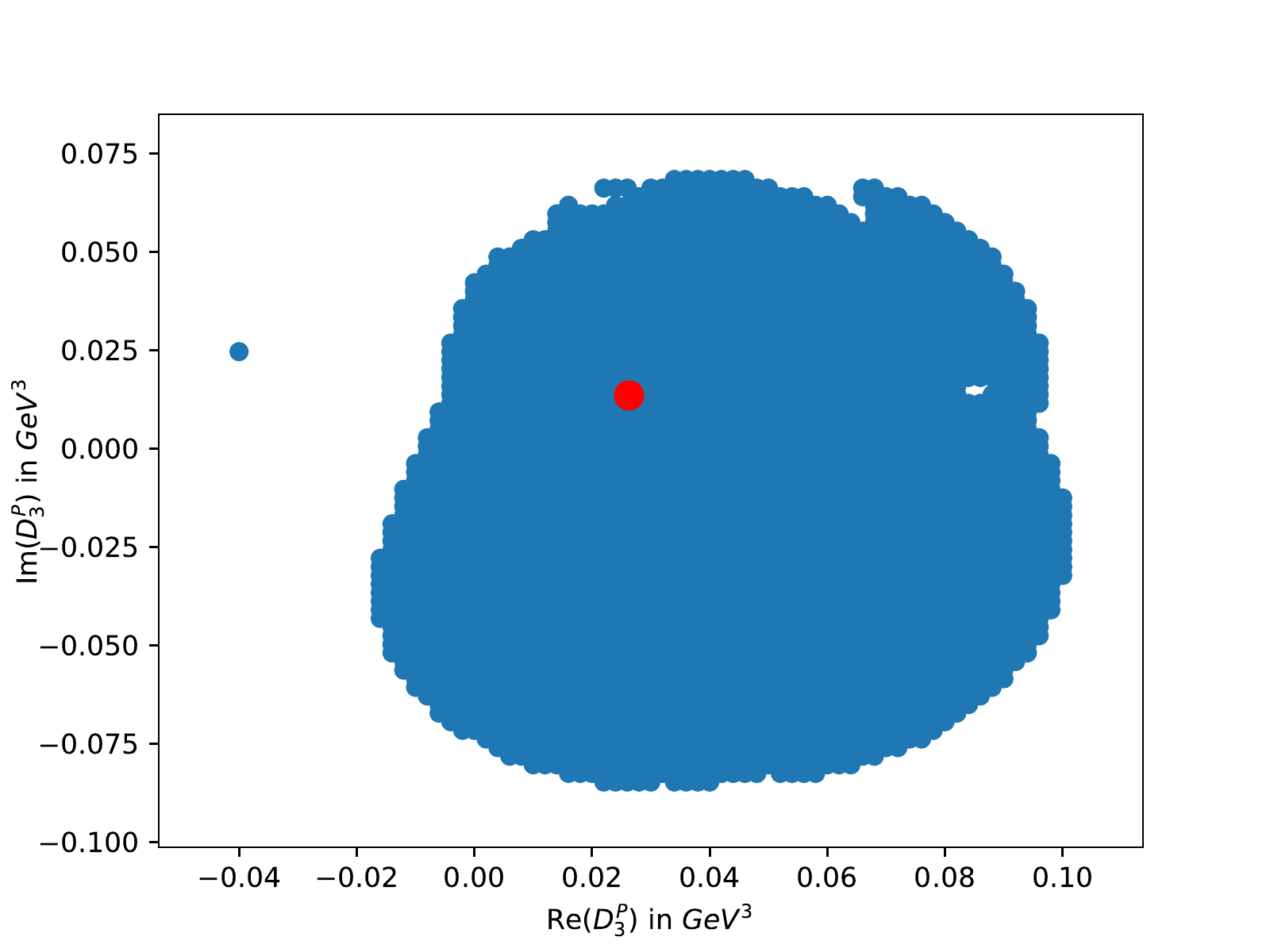}
    \caption{$SU(3)$ fit, part3. Amplitudes carry units of GeV${}^3$.\label{fig:SU3_3_1}}
\end{figure}
Additionally in Figs.~\ref{fig:SU3_1_1}, \ref{fig:SU3_2_1} and~\ref{fig:SU3_3_1} we present the two-dimensional confidence regions for the real and imaginary parts of the complex $SU(3)$-invariant quantities. Note that whereas our best-fit point is determined in polar coordinates, the two-dimensional confidence regions for each complex amplitude are obtained in the Cartesian basis.

In consistency with previous studies \cite{Hsiao:2015iiu,Fu:2003fy}, we see that the absolute value of the real and imaginary components of the annihilation contributions $A_i$ are always below $0.05$. On the other hand, the absolute values of the complex components for the rest of the amplitudes $C_i$, $B_i$ and $D_i$ are at most $0.32$. Notice however, that our estimation of the confidence regions are done simultaneously for the real and imaginary components of the individual amplitudes, i.e.\ they are performed for two degrees of freedom. This is a central difference with respect to the analyses in~\cite{Hsiao:2015iiu, Fu:2003fy} where to the best of our understanding the confidence regions are delivered taking into account uni-dimensional intervals for the magnitudes and phases of the different parameters involved. To assess the goodness of our fit we compute the $\chi^2$ per degree of freedom obtaining 
  \begin{eqnarray}
 \chi^2/d.o.f.&=&0.851.
 \end{eqnarray}
 
 \begin{table}[htp]
  \begin{center}
    \scalebox{0.85}{\begin{tabular}{|c|c|ccc||c|c|ccc|}
      \hline 
      \multirow{4}{*}{Channel} & \multicolumn{4}{|c||}{\textbf{Branching ratio}}&
      \multirow{4}{*}{Channel} & \multicolumn{4}{|c|}{\textbf{Branching ratio}}\\
      &\multicolumn{4}{|c||}{\textbf{in units of $10^{-6}$}} &&
      \multicolumn{4}{|c|}{\textbf{in units of $10^{-6}$}}\\
      \cline{2-5}\cline{7-10}
      &Experimental& \multicolumn{3}{|c||}{Theoretical} &
      &Experimental& \multicolumn{3}{|c|}{Theoretical}\\
      \hline
      \rule{0pt}{14pt}$B^-\rightarrow \pi^0\pi^-$ & $5.5\pm 0.4$ & &$6.04^{+2.42}_{-2.51}$ &&
      $B^-\rightarrow \eta \pi^-$& $4.02\pm 0.27$&&$3.80^{+1.25}_{-1.55}$ &\\[0.4em]
      $B^-\rightarrow K^0 K^-$ & $1.31\pm 0.17$ &&$1.36^{+0.17}_{-0.16}$&&
      $B^-\rightarrow \eta' \pi^-$&$2.7 \pm 0.9$&&$3.55^{+4.49}_{-1.67}$&\\[0.4em]
      $\bar{B}^0\rightarrow \pi^+ \pi^-$& $5.12\pm 0.19$&&$6.31^{+0.61}_{-0.50}$&&
      $\bar{B}^0\rightarrow \eta \pi^0$& $0.41\pm 0.17$&&$0.41^{+8.90}_{-4.08}$&\\[0.4em]
      $\bar{B}^0\rightarrow \pi^0 \pi^0$& $1.59\pm 0.26$&&$1.01^{+1.30}_{-0.51}$&&
      $\bar{B}^0\rightarrow \eta' \pi^0$& $1.2\pm 0.6$&&$1.20^{+3.62}_{-1.19}$&\\[0.4em]
      $\bar{B}^0\rightarrow K^+ K^-$& $0.078\pm 0.015$&&$0.13^{+0.08}_{-0.07}$&&
      $\bar{B}_s\rightarrow \eta K^0$& Not available&&$0.13^{+0.11}_{-0.08}$&\\[0.4em]
      $\bar{B}^0\rightarrow K^0 \bar{K}^0$& $1.21 \pm 0.16$&&$1.13^{+0.83}_{-0.91}$&&
      $\bar{B}_s\rightarrow \eta' K^0$& Not available &&$6.65^{+1.48}_{-1.65}$&\\[0.4em]
      $\bar{B}_s\rightarrow \pi^- K^+$& $5.8\pm 0.7$&&$7.75^{+0.63}_{-0.09}$&&
      $B^-\rightarrow \eta K^-$&$2.4\pm 0.4$&&$2.34^{+1.39}_{-1.67}$&\\[0.4em]
      $B^-\rightarrow \pi^0 K^-$& $12.9\pm 0.5$&&$12.78^{+1.75}_{-1.94}$&&
      $B^-\rightarrow \eta' K^-$&$70.4\pm 2.5$&&$70.82^{+11.16}_{-11.53}$&\\[0.4em]
      $B^-\rightarrow \pi^- \bar{K}^0$& $23.7\pm 0.8$&&$23.85^{+2.23}_{-2.31}$&&
      $\bar{B}^0\rightarrow \eta K^0$&$1.23\pm 0.27$&&$1.38^{+1.15}_{-0.36}$&\\[0.4em]
      $\bar{B}^0\rightarrow \pi^+ K^-$& $19.6 \pm 0.5$&&$19.47^{+1.72}_{-2.24}$&&
      $\bar{B}^0\rightarrow \eta' K^0$& $6.6\pm 0.4$&&$6.65^{+1.48}_{-1.65}$&\\[0.4em]
      $\bar{B}^0\rightarrow \pi^0 \bar{K}^0$& $9.9 \pm 0.5$&&$10.17^{+2.00}_{-2.30}$&&
      $\bar{B}_s\rightarrow \eta \pi^0$& $<10^{3}$&&$31.15^{+39.05}_{-31.14}$&\\[0.4em]
      $\bar{B}_s\rightarrow \pi^+ \pi^-$& $0.7 \pm 0.1$&&$0.57^{+0.40}_{-0.42}$&&
      $\bar{B}_s\rightarrow \eta' \pi^0$& Not available&&$11.13^{+74.75}_{-11.12}$&\\[0.4em]
      $\bar{B}_s\rightarrow \pi^0\pi^0$& $< 210$&&$0.28^{+0.20}_{-0.21}$&&
      $\bar{B}^0\rightarrow \eta \eta$&$<1$&&$0.30^{+0.70}_{-0.30}$&\\[0.2em]
      $\bar{B}_s\rightarrow K^+ K^-$& $26.6\pm 2.2$&&$20.63^{+6.80}_{-8.09}$&&
      $\bar{B}_s\rightarrow \eta \eta$& $<1.5\times 10^{3}$&&$2.58^{+36.53}_{-2.57}$&\\[0.4em]
      $\bar{B}_s\rightarrow K^0 \bar{K}^0$& $20\pm 6$&&$24.64^{+18.84}_{-21.14}$&&
      $\bar{B}^0\rightarrow \eta'\eta'$&$<1.7$&&$1.14^{+0.57}_{-1.07}$&\\[0.4em]
      $\bar{B}_s\rightarrow \pi^0 K^0$& Not available&&$0.71^{+1.47}_{-0.27}$&&
      $\bar{B}_s\rightarrow \eta'\eta'$& $33\pm 7$&&$33.00^{+24.52}_{-31.74}$&\\[0.4em]
      &  & & &&
      $\bar{B}^0\rightarrow \eta'\eta$& $<1.2$&&$0.61^{+0.59}_{-0.60}$&\\[0.4em]
      &  & & &&
      $\bar{B}_s\rightarrow \eta'\eta$& Not available&&$0.61^{+0.59}_{-0.60}$&\\[0.4em]
      \hline
    \end{tabular}}
     \caption{Experimental input and fit results for CP-averaged $B\to PP$ branching fractions.\label{tab:tableBr_combined}}
  \end{center}
\end{table}
\begin{table}[htp]
  \begin{center}
\scalebox{0.85}{\begin{tabular}{|c|c|ccc||c|c|ccc|}
      \hline 
      \multirow{4}{*}{Channel} & \multicolumn{4}{|c||}{\textbf{CP asymmetries}}&
      \multirow{4}{*}{Channel} & \multicolumn{4}{|c|}{\textbf{CP asymmetries}}  \\
      &\multicolumn{4}{|c||}{\textbf{in percent}}&&
      \multicolumn{4}{|c|}{\textbf{in percent}}\\
      \cline{2-5}\cline{7-10}
      &Experimental& \multicolumn{3}{|c||}{Theoretical}&
      &Experimental& \multicolumn{3}{|c|}{Theoretical}\\
      \hline
      \rule{0pt}{14pt}$B^-\rightarrow \pi^0\pi^-$ &  $3\pm 4$ &&$5.45^{+22.02}_{-20.60}$&&
      $B^-\rightarrow \eta \pi^-$& $-14\pm 7$&&$-11.37^{+14.49}_{-26.90}$&\\[0.4em]
      $B^-\rightarrow K^0 K^-$ & $4\pm 14$&&$18.82^{+36.93}_{-30.83}$&&
      $B^-\rightarrow \eta' \pi^-$&$6\pm 16$&&$4.71^{+59.79}_{-57.97}$&\\[0.4em]
      $\bar{B}^0\rightarrow \pi^+ \pi^-$& $32\pm 4$&&$35.01^{+3.19}_{-22.29}$&&
      $\bar{B}_s\rightarrow \eta K^0$&$<0.1$&&$0.10^{+0.00}_{-100.07}$&\\[0.4em]
      $\bar{B}^0\rightarrow \pi^0 \pi^0$& $33\pm 22$&&$-10.58^{+40.69}_{-89.40}$&&
      $\bar{B}_s\rightarrow \eta' K^0$&Not available&&$-0.58^{+100.57}_{-79.58}$&\\[0.4em]
     $\bar{B}^0\rightarrow K^0 \bar{K}^0$&$-60\pm 70$&&$-6.88^{+85.39}_{-81.37}$&&
      $B^-\rightarrow \eta K^-$&$-37\pm 8$&&$-42.23^{+42.23}_{-16.00}$&\\[0.4em]
      $\bar{B}_s\rightarrow \pi^- K^+$& $22.1\pm 1.5 $&&$20.84^{+2.39}_{-2.57}$&&
      $B^-\rightarrow \eta' K^-$&$0.4\pm 1.1$&&$0.63^{+3.98}_{-4.30}$&\\[0.4em]
      $B^-\rightarrow \pi^0 K^-$& $3.7\pm 2.1$&&$3.72^{+7.19}_{-4.35}$&&
      $\bar{B}^0\rightarrow \eta K^0$&Not available&&$-0.01^{+40.07}_{-0.02}$&\\[0.4em]
      $B^-\rightarrow \pi^- K^0$&$-1.7\pm 1.6$&&$-1.08^{+1.76}_{-2.32}$&&
      $\bar{B}^0\rightarrow \eta' K^0$&$-6\pm 4$ &&$0.03^{+4.82}_{-11.69}$&\\[0.4em]
      $\bar{B}^0\rightarrow \pi^+ K^-$& $-8.3\pm 0.4$&&$-8.38^{+8.38}_{-1.01}$&&
      $\bar{B}^0\rightarrow \eta \pi^0$&Not available&&$-27.39^{+127.11}_{-72.58}$&\\[0.4em]
      $\bar{B}^0\rightarrow \pi^0 \bar{K}^0$& $0\pm 13$&&$-0.97^{+19.35}_{-3.20}$&&
      $\bar{B}^0\rightarrow \eta' \pi^0$&Not available&&$-43.67^{+143.63}_{-56.33}$&\\[0.4em]
      $\bar{B}_s\rightarrow K^+ K^-$& $-14\pm 11$&&$-10.58^{+10.58}_{-3.60}$&&
      $\bar{B}_s\rightarrow \eta \pi^0$& Not available&&$0.88^{+94.98}_{-98.70}$&\\[0.4em]
       $\bar{B}_s\rightarrow \pi^+ \pi^-$& Not available&&$17.56^{+11.84}_{-38.25}$&&
      $\bar{B}_s\rightarrow \eta' \pi^0$& Not available&&$1.57^{+77.56}_{-95.66}$&\\[0.4em]
      $\bar{B}_s\rightarrow \pi^0\pi^0$& Not available&&$17.56^{+11.84}_{-38.25}$&&
      $\bar{B}^0\rightarrow \eta \eta$&Not available&&$3.46^{+96.50}_{-103.45}$&\\[0.4em]
      $\bar{B}_s\rightarrow K^0 \bar{K}^0$& Not available&&$0.31^{+5.07}_{-4.59}$&&
      $\bar{B}_s\rightarrow \eta \eta$& Not available&&$14.29^{+76.81}_{-113.09}$&\\[0.4em]
      $\bar{B}^0\rightarrow K^+ K^-$& Not available&&$-78.45^{+161.99}_{-20.78}$&&
      $\bar{B}^0\rightarrow \eta'\eta'$&Not available&&$42.41^{+57.55}_{-142.41}$&\\[0.4em]
      $\bar{B}_s\rightarrow \pi^0 K^0$&  Not available&&$13.74^{+29.49}_{-113.73}$&&
      $\bar{B}_s\rightarrow \eta'\eta'$&Not available&&$-2.05^{+15.29}_{-13.44}$&\\[0.4em]
      &  & & &&
      $\bar{B}^0\rightarrow \eta'\eta$&Not available&&$-12.32^{+112.32}_{-87.67}$&\\[0.4em]
      &  & & &&
      $\bar{B}_s\rightarrow \eta'\eta$&Not available&&$3.43^{+96.36}_{-103.22}$&\\[0.4em]
      \hline
    \end{tabular}}
     \caption{Experimental input and fit results for $B\to PP$ CP asymmetries.
     \label{tab:tableAs_combined}}
  \end{center}
\end{table}
 
 \begin{figure}[htp]
    \centering
    \includegraphics[width=7cm]{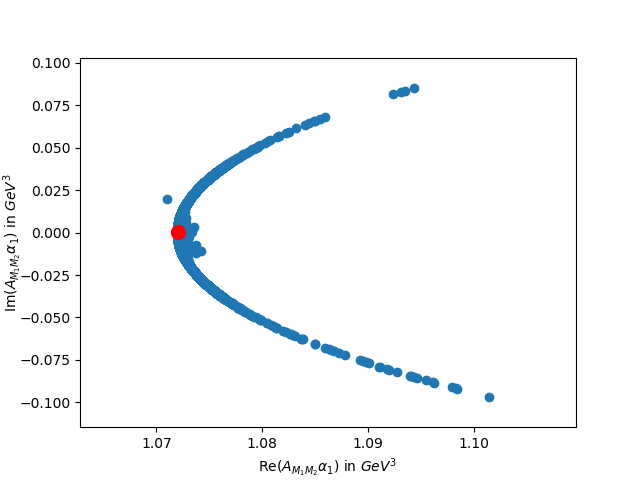}
    \includegraphics[width=7cm]{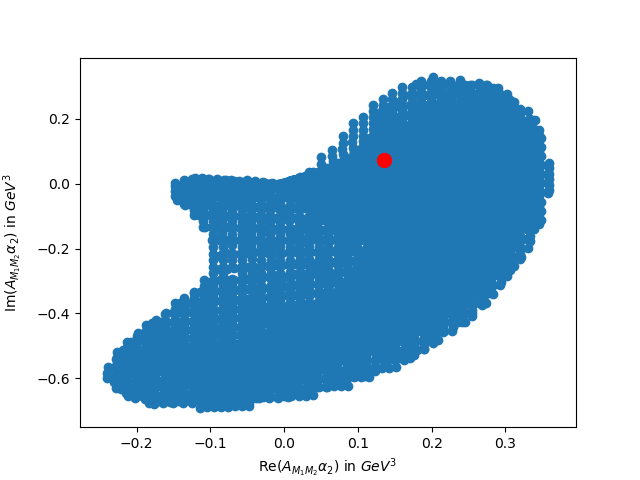}
    \includegraphics[width=7cm]{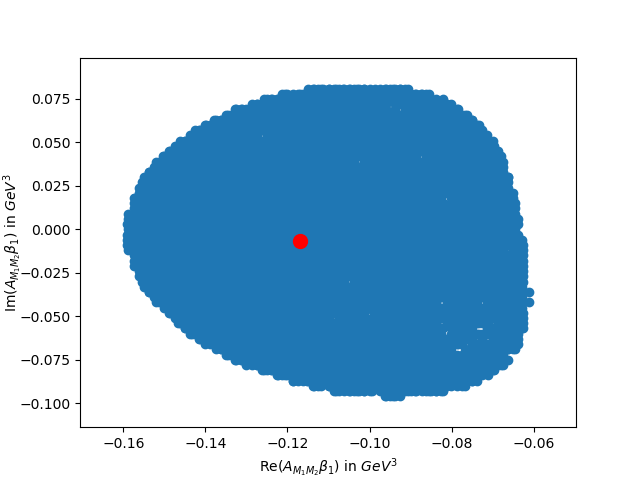}
    \includegraphics[width=7cm]{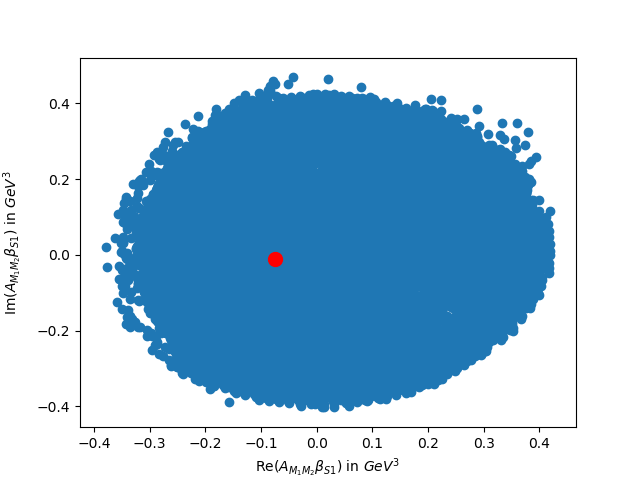}
    \includegraphics[width=7cm]{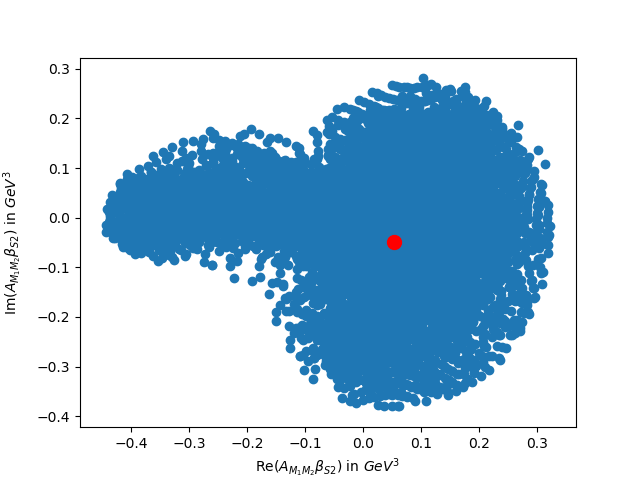}
    \includegraphics[width=7cm]{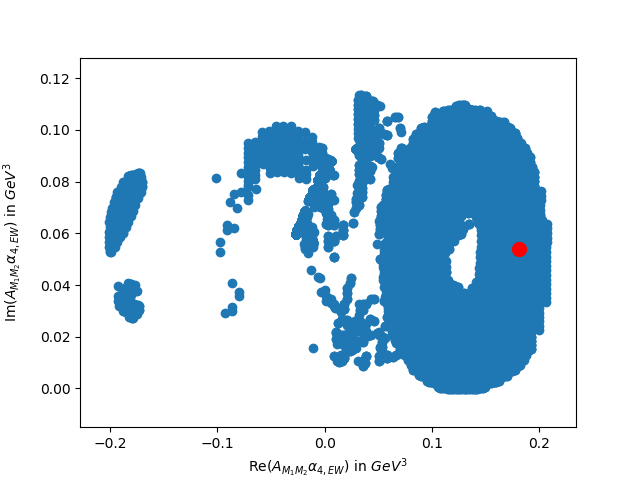}
    \includegraphics[width=7cm]{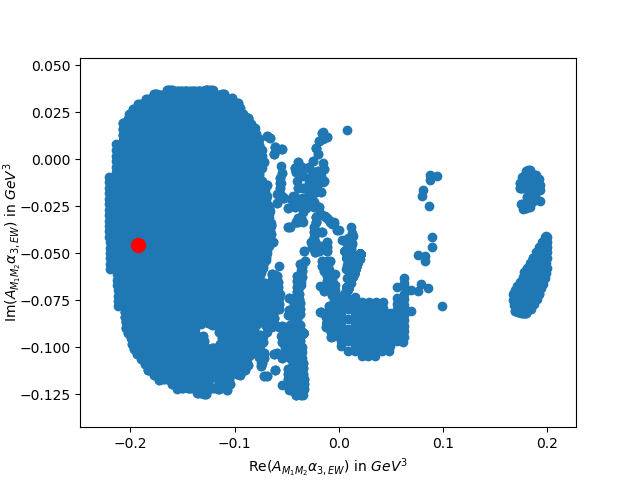}
    \includegraphics[width=7cm]{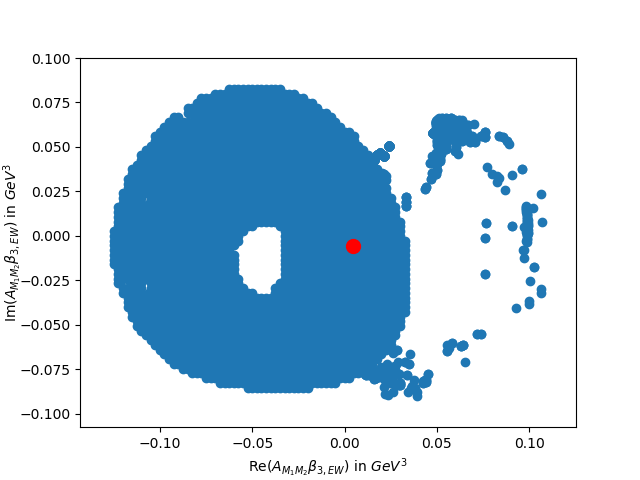}
    \caption{QCDF fit, part1. Amplitudes carry units of GeV${}^3$.\label{fig:QCDF_1_1}}
\end{figure}
\begin{figure}[htp]
    \centering
    \includegraphics[width=7cm]{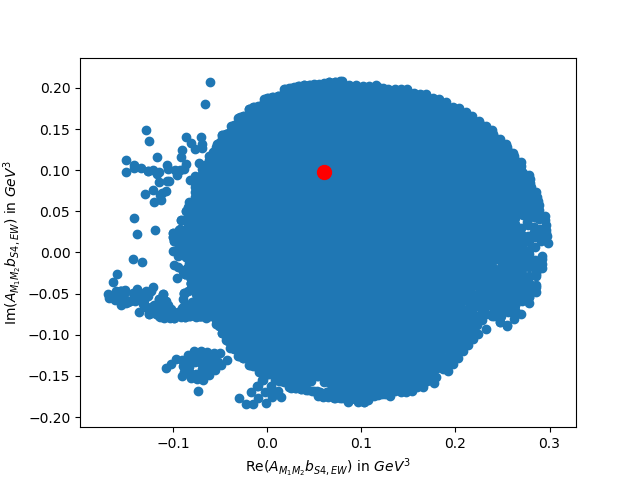}
    \includegraphics[width=7cm]{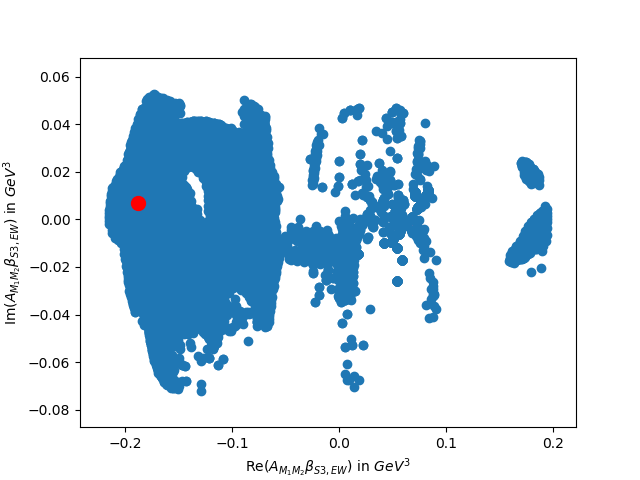}
    \includegraphics[width=7cm]{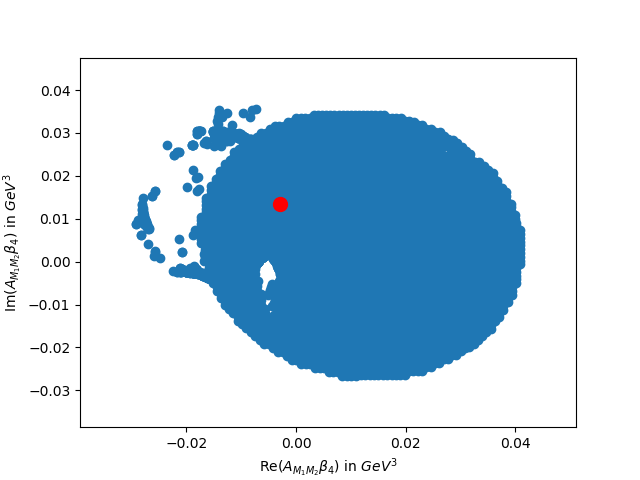}
    \includegraphics[width=7cm]{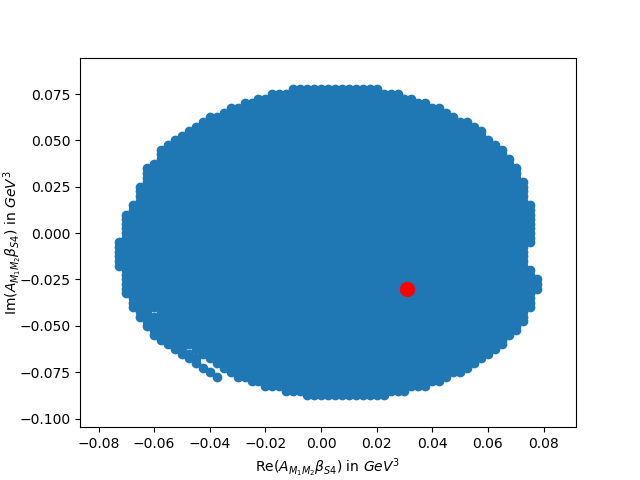}
    \includegraphics[width=7cm]{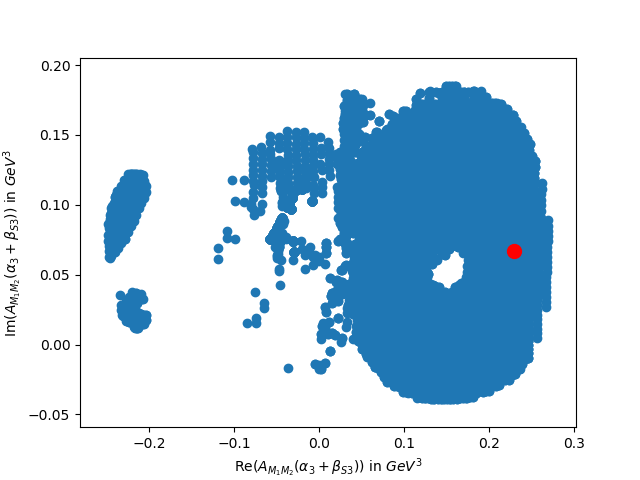}
    \includegraphics[width=7cm]{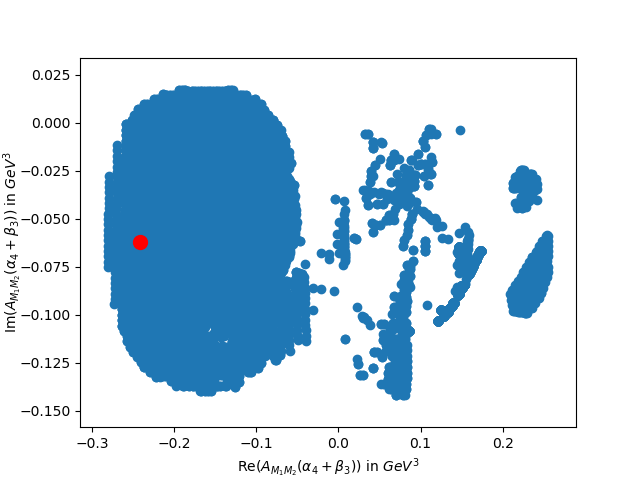}
    \caption{QCDF fit, part2. Amplitudes carry units of GeV${}^3$.\label{fig:QCDF_2_0}}
\end{figure}

 One of the consequences of the relatively large confidence intervals are bigger uncertainties for the evaluations of the branching fractions and CP asymmetries, which we show in Tables~\ref{tab:tableBr_combined} and~\ref{tab:tableAs_combined}. We can see that most of our results agree with the experimental determinations within the $1\sigma$ range. There are, however, two exceptions: $\mathcal{B}(B^0\rightarrow \pi^+ \pi^-)$ and $\mathcal{B}(B_s\rightarrow \pi^- K^+)$ where the discrepancies are at the level of $2\sigma$ and $2.5\sigma$ respectively. We can identify the source of these discrepancies in the relatively high value of the pre-factor $A_{M_1 M_2}$ appearing in Eq.~(\ref{eq:AM1M@Average}), which is in turn driven by the $B \to \eta^{(\prime)}$ form factors. Moreover, the value of $A_{M_1 M_2}$ has an additional impact on the fit through the constraint from Eq.~(\ref{eq:consalpha1}). As a matter of fact, data prefers a relatively low value for the product $A_{M_1 M_2}\alpha_1\sim 1.0$~GeV${}^3$. We checked this by performing the $\chi^2$-fit  with an increased uncertainty to the number in Eq.~(\ref{eq:consalpha1}), and also by determining $A_{M_1 M_2}$ including only  the form factors and decay constants of the transitions $B\to \pi$ and $B\to K$. Both scenarios lead to a reduced value of the $\chi^2$ per degree of freedom. However, we prefer to keep the setup of the analysis in the way described, since we consider that determining $A_{M_1 M_2}$ from all $B \to P$ form factors is the most transparent way of carrying out our analysis which consistently relies on the $SU(3)$ symmetry.

 \subsection{QCDF amplitudes}
 
  Translating our best-fit point to the QCDF amplitudes we obtain (all dimensionful numbers are given in GeV${}^3$)
 \begin{align}
 A_{M_1 M_2}\alpha_1&=1.072+5.596\times 10^{-5} i,
 & A_{M_1 M_2}\alpha_2&=0.136+0.073i, \nonumber\\
A_{M_1 M_2} \beta_1&=-0.117-0.007 i, &
A_{M_1 M_2} \beta_2&=A_{M_1 M_2}\beta_1, \nonumber\\
A_{M_1 M_2} \beta_{S1}&=-0.074-0.0112i, &
 A_{M_1 M_2} \beta_{S2}&=0.054-0.049 i, \nonumber\\
 A_{M_1 M_2} \alpha_{3, EW}&= -0.193-0.045 i, &
 A_{M_1 M_2} \alpha^{c}_{4, EW}&=0.181+0.053 i,\nonumber\\
 A_{M_1 M_2} \beta_{3, EW}&= 0.005-0.006 i, &
 A_{M_1 M_2} b_{4, EW}&=A_{M_1 M_2} \beta_{3, EW},\nonumber\\
 A_{M_1 M_2} \beta_{S3, EW}&=-0.188+0.007 i, &
 A_{M_1 M_2} b_{S4, EW}&=0.061+0.098 i, \nonumber\\
 A_{M_1 M_2}\beta_4&=-0.003+0.013 i, &
 A_{M_1 M_2}\beta_{S4}&=0.031-0.030 i, \nonumber\\
 A_{M_1 M_2}\Bigl(\alpha_3 +\beta_{S3}\Bigl)&=0.230+0.067 i, &
 A_{M_1 M_2}\Bigl(\alpha_4 +\beta_{3}\Bigl)&=-0.242-0.062 i.
 \end{align}
 
 Using the relations in~(\ref{eq:TopologicalQCDF}) we can map out the confidence regions of the  $SU(3)$-irreducible amplitudes into the QCDF ones presented in Figs.~\ref{fig:QCDF_1_1} and~\ref{fig:QCDF_2_0}. Note that in those figures we plot the product of $A_{M_1M_2}$ times the QCDF amplitudes, whereas in the lines to follow we refer to the QCDF amplitudes themselves. One of our main results is the size of the annihilation amplitudes $\beta_i$ and $b_i$. While some of them such as  $\beta_4$ and  $\beta_{S4}$ get constrained around or below $0.1$, others can be up to $\sim 0.3$ as in the case of $\beta_{S1}$ and $\beta_{S2}$. Concerning $\alpha_1$, the best fit for the real part is at the lower end of the interval defined by Eq.~(\ref{eq:alpharegions}), and its imaginary component can be up to $0.08$.  The imaginary part of $\alpha_2$ can grow to values of up to $-0.5$. Within our analysis we can only evaluate the combinations $\alpha_3 + \beta_{S3}$ and  $\alpha_4 + \beta_{3}$, and is not possible to isolate any of the amplitudes independently  without introducing extra information. While many of the uncertainty bands are still rather sizeable, our results can act as a valuable guide to get a handle on the size of annihilation contributions in foreseen phenomenological studies on two-body charmless non-leptonic $B_{(s)}$ decays, and be considered as a first step towards improving the treatment of annihilation amplitudes with respect to the early QCDF studies~\cite{Beneke:2001ev,Beneke:2003zv}.
 
 The extraction of the annihilation contributions can in principle also be carried out by fitting to data Eqs.~(\ref{eq:QCDF1}) and (\ref{eq:QCDF2}), however the confidence regions are in general bigger than those obtained in Figs.~\ref{fig:QCDF_1_1} and~\ref{fig:QCDF_2_0} since,
 as discussed before, the cancellations between the different up and charm penguin contributions 
 arising in the $\Lambda_u-\Lambda_t$ parameterization does not occur in this
 particular case, leading to the coefficients $c^{{\rm QCDF}, p}_i$ being expressed as linear combinations of the QCDF amplitudes $\{\alpha_i,\alpha_i^{p},\beta_i,b_i\}$, which makes it relatively difficult to disentangle the pure annihilation contribution.

\subsection{Estimating the size of $SU(3)$ breaking}

So far our discussion has been limited to the situation where the flavour $SU(3)$-symmetry holds. A proper treatment of the $SU(3)$ breaking situation requires a full and careful analysis where this effect is incorporated in the theoretical formulas leading to the extraction of information from data. This will correspond to an entire new study on which we comment briefly in the outlook section. We can, however, carry out a preliminary estimation of the implications of giving up the $SU(3)$-flavour symmetry, in particular on the weak annihilation amplitudes. One way of accounting for $SU(3)$-breaking is by estimating the coefficient $A_{M_1 M_2}$ introduced in Eq.~(\ref{eq:AM1M2}) in a channel-dependent fashion. Consider for instance the transitions $B\rightarrow K K$ for which the global factor extracted from $KK$ final states only reads
\begin{eqnarray}
A_{K K}&=&1.54\pm 0.09,
\end{eqnarray}
and consequently, with the $SU(3)$-symmetric $A_{M_1 M_2}$ from~(\ref{eq:AM1M@Average}),
\begin{eqnarray}
A_{K K}/A_{M_1 M_2}&=&1.23 \pm 0.07. 
\end{eqnarray}

\begin{figure}[t]
    \centering
    \includegraphics[width=10cm]{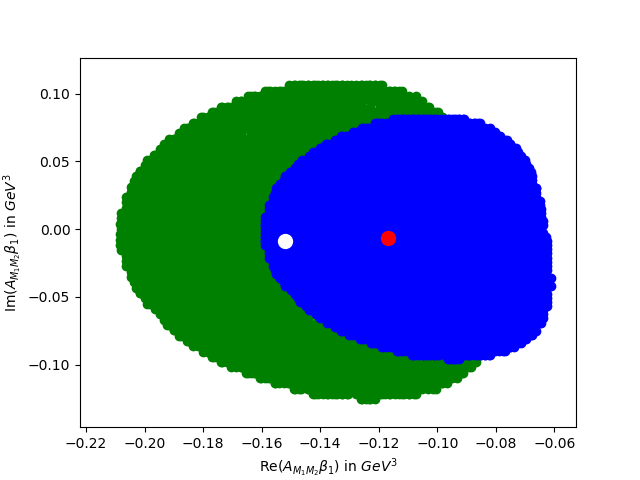}
    \caption{Comparison of the $SU(3)$ symmetric (blue, best fit point in red) and $SU(3)$ breaking (green, best fit point in white) cases for the QCDF amplitude $\beta_1$. \label{fig:SU3_breaking_comparison}}
\end{figure}

Therefore, the maximum deviation with respect to the $SU(3)$ symmetric case is $30\%$ for this particular situation, a value that is within the ballpark of numbers quoted in the literature. We can then study the impact of this deviation on the regions for the weak annihilation amplitudes. Consider for example the case of $A_{M_1 M_2}\beta_1$ shown in Fig.~\ref{fig:QCDF_1_1} which now gets modified according to Fig.~\ref{fig:SU3_breaking_comparison}. We can see how the imaginary part grows up to values slightly above $\pm 0.1$ and the real part shifts to values in the range $[-0.21, -0.08]$. As stated before this is a very preliminary study of the $SU(3)$-flavour symmetry breaking effects which will be properly extended in a forthcoming analysis.

 
 \section{Conclusion and Outlook}
 \label{sec:conclusion}
 There are two well-known model independent descriptions of non-leptonic $B$ decays, the topological and the $SU(3)$-irreducible one. In addition, the QCDF framework offers a systematic way of parameterizing pertubative and non-perturbative contributions. Even though QCDF offers a theoretically solid technique to address the perturbative quantities, the weak annihilation amplitudes, which belong to the non-perturbative category, are challenging to estimate in quantum field theory. In the present work we have determined the possible regions of weak annihilation amplitudes from a data-driven approach.
 
 We first establish the transformation rules between the topological and the QCDF decomposition. We then perform a global fit to the $SU(3)$-irreducible amplitudes with a few additional mild assumptions, and give predictions for CP-averaged branching fractions and direct CP asymmetries. Then, assuming flavour $SU(3)$ to be unbroken, we exploit the connection between the QCDF and the $SU(3)$-irreducible decomposition to translate the fit results into $1\sigma$ confidence regions of the real and imaginary parts of the QCDF amplitudes. 

One of our main results is that the size of some of the annihilation amplitudes (such as  $\beta_4$ and  $\beta_{S4}$) get constrained very well around or below $0.1$, others can take values up to $\sim 0.3$ (such as $\beta_{S1}$ and $\beta_{S2}$). While the confidence regions obtained from the fit are in many cases still sizeable they can be regarded as an upper bound, and provide valuable information on non-perturbative input parameters in future phenomenological studies on two-body charmless non-leptonic $B_{(s)}$ decays. Moreover, with improved experimental data the situation will further improve.

We would like to highlight the fact that we consider an $SU(3)$-invariant approach because this reduces the number of parameters to be fitted to a manageable level. However, in future studies a first step towards a more complete description should include an implementation of $SU(3)$ breaking. This can, for instance, be done by taking the leading-power results from QCDF, where the $SU(3)$ breaking is incorporated naturally in form factors, decay constants and light-cone distribution amplitudes. This will allow for a direct fit to the QCDF weak annihilation amplitudes. We leave such a study to future work.

 
 \section*{Acknowledgments}

We would like to thank Thorsten Feldmann and Thomas Mannel for useful discussions. This research was supported by the Deutsche Forschungsgemeinschaft (DFG, German Research Foundation) under grant 396021762 — TRR 257 ``Particle Physics Phenomenology after the Higgs Discovery''. We also acknowledge the use of the TTP computer cluster at KIT in Karlsruhe.

\bibliographystyle{utphys}
\bibliography{main}

\end{document}